\documentclass[a4paper,11pt]{article}
\pdfoutput=1 

\usepackage{jheppub2} 
\usepackage{bbm,bm,graphicx,mathtools,color,hyperref,slashed}
\usepackage{amssymb,amsmath}

\usepackage{comment}
\usepackage[all]{xy}
\usepackage[dvipsnames]{xcolor}

\graphicspath{{./Figures/}}

\newcommand{\diff}{\mathrm{d}}
\newcommand{\p}{\partial}
\newcommand{\ve}{\varepsilon}
\newcommand{\Diff}{{\mathcal{D}}}

\newcommand{\tr}{\mathrm{tr}}

\newcommand{\im}{\mathrm{i}}

\newcommand{\calA}{\mathcal{A}}

\newcommand{\calR}{\mathcal{R}}

\newcommand{\calV}{\mathcal{V}}

\newcommand{\rmd}{\mathrm{d}}
\newcommand{\rme}{\mathrm{e}}
\newcommand{\rmf}{\mathrm{f}}
\newcommand{\rmq}{\mathrm{q}}

\newcommand{\rmB}{\mathrm{B}}
\newcommand{\rmL}{\mathrm{L}}
\newcommand{\rmR}{\mathrm{R}}
\newcommand{\rmV}{\mathrm{V}}

\def\Z{{\mathbb Z}}
\def\R{{\mathbb R}}

\preprint{YITP-22-04}

\title{Center vortex and confinement in Yang-Mills theory and QCD with anomaly-preserving compactifications}

\author[1]{Yuya Tanizaki,}

\affiliation[1]{Yukawa Institute for Theoretical Physics,  Kyoto University, Kyoto 606-8502, Japan}

\emailAdd{yuya.tanizaki@yukawa.kyoto-u.ac.jp}

\author[2]{Mithat \"{U}nsal}

\affiliation[2]{Department of Physics, North Carolina State University, Raleigh, NC 27607, USA}

\emailAdd{unsal.mithat@gmail.com}

\abstract{
We construct an anomaly-preserving compactification of 4d gauge theories, including $SU(N)$ Yang-Mills theory, $\mathcal{N}=1$ supersymmetric Yang-Mills theory, and QCD, down to 2d by turning on 't Hooft flux through $T^2$. 
It provides a new framework to analytically calculate nonperturbative properties such as confinement, chiral symmetry breaking, and multi-branch structure of vacua.  
We give the semiclassical description of these phenomena based on the center vortex and show that it enjoys the same anomaly matching condition with the original $4$d gauge theory. 
We conjecture that the weak-coupling vacuum structure on small $T^2 \times \mathbb{R}^2$ is adiabatically connected to the strong-coupling regime on $\mathbb{R}^4$ without any phase transitions.  
In QCD with fundamental quarks as well, we can turn on 't Hooft flux either by activating $SU(N_f)_{\mathrm{V}}$ symmetry twist for $N_f=N$ flavors or by introducing a magnetic flux of baryon number $U(1)_{\mathrm{B}}$ for arbitrary $N_f$ flavors.  
In both cases, the weak-coupling center-vortex theory gives the prediction consistent with chiral Lagrangian of $4$d QCD. 
}

\begin{document}
\maketitle
%--------------------------------------------------------------------------------------

\section{Introduction and Summary}\label{sec:introduction}

Color confinement~\cite{Wilson:1974sk} is the fundamental property of $4$d pure Yang-Mills theory. 
It has been suspected for a long time that topological defects play important roles to understand confinement, such as condensation of monopoles and proliferation of center vortices~\cite{Nambu:1974zg, Mandelstam:1974pi,  Polyakov:1975rs, tHooft:1977nqb, Cornwall:1979hz, Nielsen:1979xu, Ambjorn:1980ms}. 
Monopoles and center vortices are particle-like and string-like objects in $4$d spacetime, respectively, and they both carry color-magnetic charges. 
Their condensation or proliferation means liberation of magnetic charges, which leads to confinement of color-electric charges. 
Despite its beauty, it is not easy to incorporate this idea with the actual calculation to demonstrate confinement of $4$d gauge theories. 
From the practical point of view, the difficulty reflects the fact that they are extended objects in the Euclidean path integral. 
Their action densities are inversely proportional to the gauge coupling, $\sim O(1/g^2)$, and thus the entropy for such configurations, $\sim O(1)$, cannot overcome the suppression by the action density in the weak-coupling regime. 
In the strong-coupling regime of $U(1)$ lattice gauge theories~\cite{Banks:1977cc, Savit:1977fw,Creutz:1979kf, Creutz:1979zg}, this relation is reversed and the confinement occurs due to the proliferation of monopole worldlines as confirmed numerically. 
Currently, its analytical success in $4$d continuum field theory is limited to the case with sufficiently many supersymmetries, in which we can control strongly-coupled regimes~\cite{Seiberg:1994rs}, and it demonstrates the confinement by monopole (or dyon) condensation for softly-broken $\mathcal{N}=2$ supersymmetry. 

However, this does not rule out the weak-coupling semiclassical analysis with monopoles or center vortices to describe confinement. 
Indeed, when we consider Yang-Mills theory on $\mathbb{R}^3\times S^1$ with suitable deformations, many important features of the confinement phase can be obtained by the semiclassical analysis using monopoles~\cite{Unsal:2007vu,Unsal:2007jx,Unsal:2008ch,Shifman:2008ja, Davies:2000nw}.
In this case, the monopole worldline can wrap the compactified $S^1$ direction, and thus monopoles behave as instantons, or point-like objects, in $3$d effective field theory on $\mathbb{R}^3$. 
It gives us a chance to circumvent the arguments on action density versus entropy when $S^1$ is sufficiently small. 
This program has revealed many examples of confinement, chiral symmetry breaking, and the multi-branch structure of vacua at weak couplings. An important lesson is that these nonperturbative phenomena are not necessarily strong coupling effects.

\begin{figure}[t]
\vspace{-1.5cm}
\begin{center}
\hspace{-1cm}
\includegraphics[width = 1.0\textwidth]{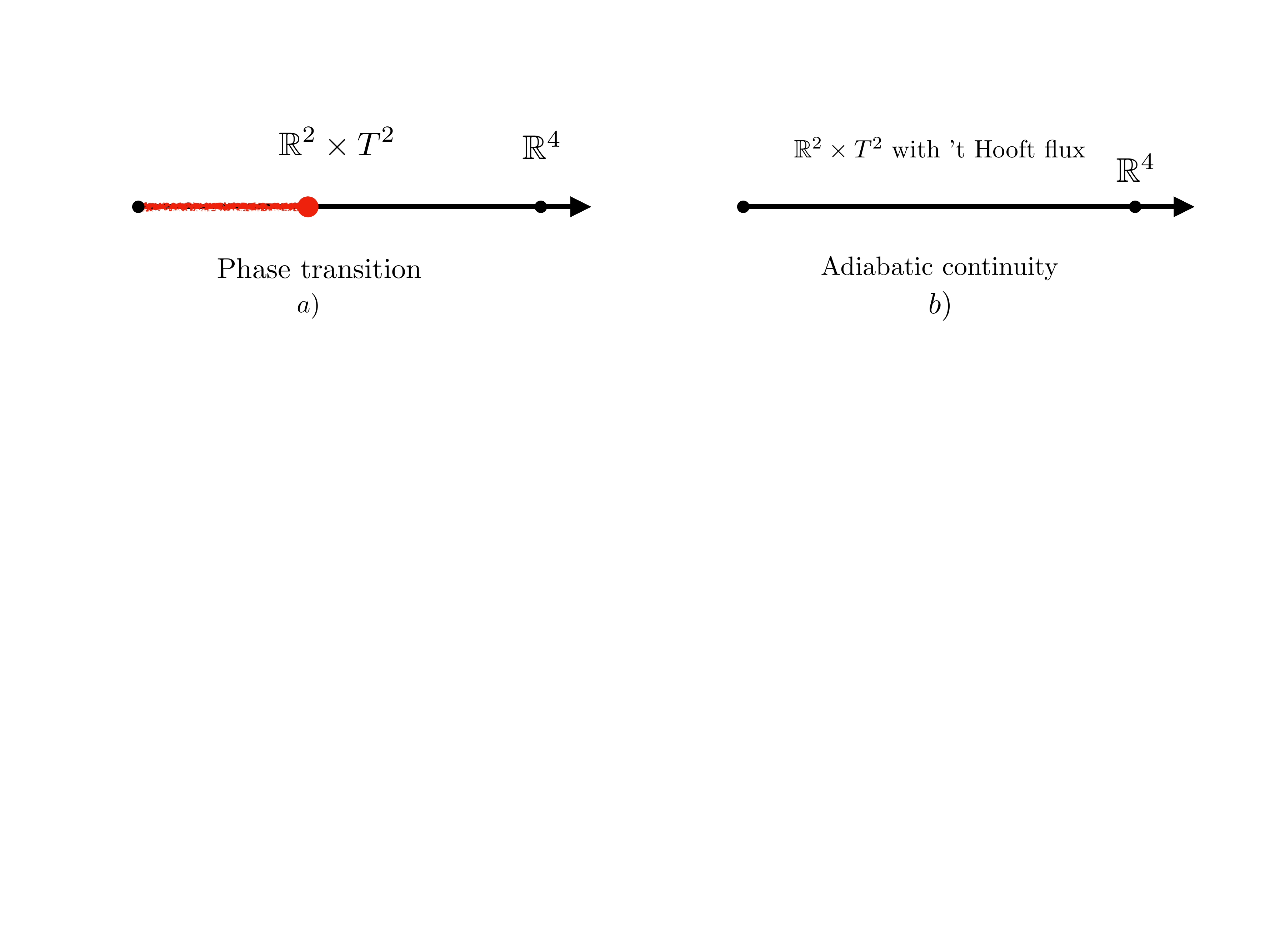}
\vspace{-8.0cm}
\caption{
Expected phase diagrams as a function of the compactification size $L$ of $T^2$. 
(a)~In a $T^2$ compactification of 4d gauge theories (YM, QCD) without 't Hooft flux, there is generically a  phase transition such as center-symmetry breaking and chiral restoration.  
(b)~In 't Hooft flux background, the phase transitions can be avoided. 
Dynamics of small $T^2 \times \R^2$  theory can be adiabatically  connected to the strongly-coupled dynamics on $\R^4$, and it is semi-classically calculable. Small $T^2$ regime provides a  window in which center-vortex mechanism is operative. 
 }
\label{fig:conjecture}
\end{center}
\end{figure}

In this work, we propose a new calculable framework for $4$d gauge theories by compactifying the spacetime to small  $\mathbb{R}^2\times T^2$ while preserving anomalies. This construction gives a new semiclassical description of confinement and chiral symmetry realization based on center vortices. 
Our analysis of this paper includes Yang-Mills (YM) theory, $\mathcal{N}=1$ supersymmetric Yang-Mills (SYM) theory, and quantum chromodynamics (QCD) with fundamental quarks. 
When we consider the $T^2$ compactification, the center-vortex worldsheet can wrap around it, and then it can be regarded as a point-like object in $\mathbb{R}^2$. 
For this purpose, we have to perform a specific $T^2$ compactification so that such a point-like center vortex exists as a local minimum of the YM action. 
It turns out that we can achieve it without adding any deformations or extra matter fields to the theories, and we only need to take the 't~Hooft twisted boundary condition along the $T^2$ direction~\cite{tHooft:1979rtg,tHooft:1981sps}. 
Remarkably, these analytical computations on small $\mathbb{R}^2\times T^2$ with 't~Hooft flux can reproduce important qualitative features of $4$d confinement for all these theories, YM, $\mathcal{N}=1$ SYM, and QCD. 
We uncover the kinematical reasoning behind this matching by computing 't~Hooft anomalies of both $2$d and $4$d field theories, and we show that they are intimately related under the presence of 't~Hooft flux~\cite{Tanizaki:2017qhf,Yamazaki:2017dra}. 
In other words, this $T^2$ compactification preserves 't~Hooft anomaly, and thus both $2$d and $4$d dynamics are constrained by the same anomaly matching condition. 
Based on these observations, we cannot help but conjecture that our semiclassical center-vortex theory on $\mathbb{R}^2\times T^2$ with 't~Hooft flux is adiabatically connected to the strongly-coupled dynamics of confining gauge theories on $\mathbb{R}^4$ without having phase transitions (see Fig.~\ref{fig:conjecture}). 
Of course, this is quite a nontrivial conjecture on dynamics of $4$d gauge theories, and we should carefully examine and improve it in future studies. 

In the following, let us summarize the main results. 

\begin{figure}[t]
%\vspace{-1.5cm}
\begin{center}
\hspace{-1cm}
\includegraphics[width = 1.0\textwidth]{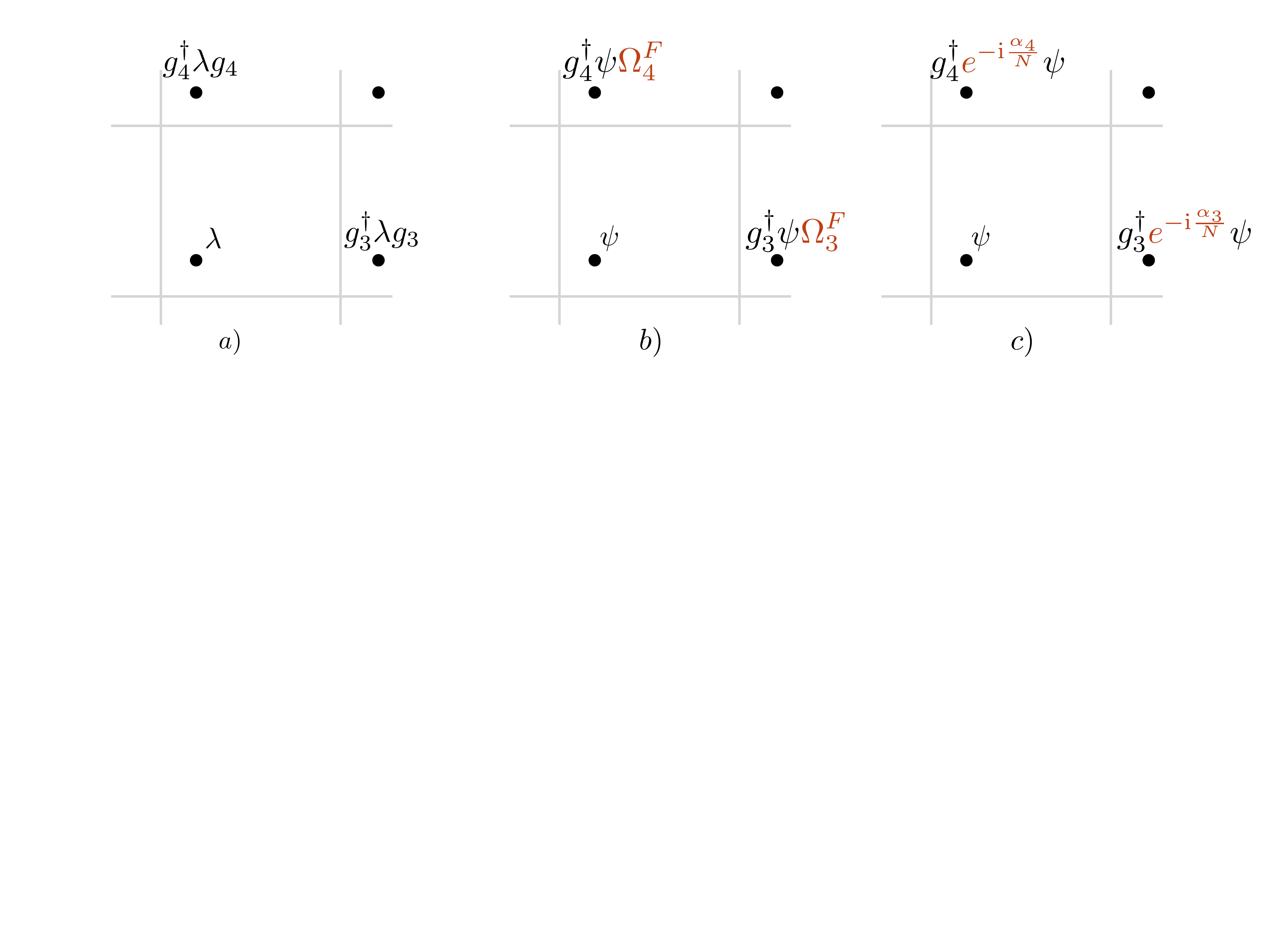}
\vspace{-7.5cm}
\caption{ \textbf{(a)} YM theory (with adjoint matters $\lambda$): 't~Hooft flux  through $T^2\ni (x_3,x_4)$ can be realized by the twisted boundary condition, which is the mild violation of the cocycle condition, $g_3(L)^\dagger g_4(0)^\dagger =\rme^{2\pi \im n_{34}/N} g_4(L)^\dagger g_3(0)^\dagger$, for the $SU(N)$ transition functions, $g_3(x_4),g_4(x_3)$. \\
\textbf{(b,c)} QCD with fundamental quarks $\psi$: In this case, we cannot naively introduce  't~Hooft flux in the gauge sector. 
We have two ways to go around this obstacle using symmetry of QCD. 
(b)~'t~Hooft flux in QCD with $N_f = N$ by using non-commuting 
$SU(N_f)_{\rm V}$ flavor twist, with $\Omega_3^F\Omega_4^F=\rme^{2\pi\im/N_f}\Omega_4^F \Omega_3^F$. (c)~'t~Hooft flux in QCD with any $N_f $ by using $U(1)_{\rm B} =  U(1)_{\rm q}/{\mathbb Z}_N$ magnetic flux, $\int_{T^2}\diff A_\rmB=2\pi$, which requires fractionalized $U(1)$ transition functions, $\rme^{\im \alpha_3(x_4)/N}, \rme^{\im \alpha_4(x_3)/N}$, for quark fields.  
 }
\label{fig:twist}
\end{center}
\end{figure}

In Sec.~\ref{sec:YM}, we consider the $SU(N)$ pure YM theory on $M_4=M_2\times T^2$, where $T^2\ni (x_3,x_4)$ is much smaller than the strong scale $\Lambda^{-1}$ and the minimal 't~Hooft flux, $n_{34}=1 \bmod N$, is inserted in this small $T^2$ (see Fig.~\ref{fig:twist}~(a)). 
The YM theory has the center symmetry, or the $\mathbb{Z}_N$ $1$-form symmetry~\cite{Gaiotto:2014kfa}, denoted as $\mathbb{Z}_N^{[1]}$. 
Under the $T^2$ compactification, this $4$d $1$-form symmetry splits into  $1$-form symmetry and two $0$-form symmetries, and the $2$d effective theory on $M_2$ enjoys $(\mathbb{Z}_N^{[1]})_{2\rmd}\times \mathbb{Z}_N^{[0]}\times \mathbb{Z}_N^{[0]}$. 
Under the presence of 't~Hooft flux, the $0$-form center symmetries are unbroken at the classical level, and thus the deconfinement for Polyakov loops, $P_3, P_4$, does not occur. 
This already gives us a hint that the adiabatic continuity may be possible in this compactification. 
However, the perturbative spectrum of $2$d gauge fields is completely gapped due to the 't~Hooft flux, and thus the Wilson loop in $M_2$ obeys the perimeter law within the perturbation theory, which seems to be problematic. 
It is the center vortex that resolves this issue in the semiclassical analysis. 

As we will discuss in Sec.~\ref{sec:center_vortex_theory}, the center vortex realizes a self-dual configuration of YM theory on $\mathbb{R}^2\times T^2$ with the 't~Hooft flux. 
It carries the fractional topological charge, $Q_{\mathrm{top}}=\pm {1\over N}$, and its classical action is given by  $\mathrm{Re}(S_{\mathrm{YM}})={8\pi^2\over N g^2}$. 
It is slightly unfortunate that there do not yet exist analytic solutions, but its presence is numerically well established~\cite{Gonzalez-Arroyo:1998hjb, Montero:1999by, Montero:2000pb}. 
Since it has a fixed size along the $\mathbb{R}^2$ direction, we can perform the dilute gas approximation without suffering from infrared divergence. 
As a result, we obtain the multi-branch structure of the ground states, 
\begin{equation}
    E_k(\theta)\sim -\Lambda^2 (\Lambda L)^{5/3} \cos\left({\theta-2\pi k\over N}\right), 
\end{equation}
where $\Lambda$ is the strong scale, $L \,(\ll \Lambda^{-1})$ is the length of small $T^2$, $\theta$ is the topological vacuum angle, and $k=0,1,\ldots, N-1$ is the branch label. 
This multi-branch structure is expected also for the confinement dynamics of $4$d YM theory~\cite{Witten:1980sp, DiVecchia:1980yfw, Witten:1998uka}, and we here obtain it in the weak-coupling semiclassical analysis for the $T^2$-compactified setup. 
This fractional $\theta$ dependence is also a key for the area law of the Wilson loop. When $-\pi<\theta<\pi$, the string tension for the Wilson loop of the representation $\mathcal{R}$ is given by 
\begin{equation}
    T_\calR(\theta)=E_0(\theta+2\pi |\calR|)-E_0(\theta), 
\end{equation}
where $|\calR|$ is the $N$-ality of the representation. Therefore, for quarks with non-zero $N$-ality, there is a linear confinement as $T_\calR(\theta)>0$ for generic values of $\theta$. 
It is recently found that the multi-branch nature of vacua is controlled by the discrete 't~Hooft anomaly of $4$d YM theory at $\theta=\pi$~\cite{Gaiotto:2017yup, Gaiotto:2017tne}, and we shall see that this anomaly persists under this $T^2$ compactification in Sec.~\ref{sec:anomaly_compact}. 
Therefore, our result is not a coincidence, but comes from a fact that $2$d and $4$d dynamics is constrained by the same 't~Hooft anomaly.

In Sec.~\ref{sec:SYM}, we move on to the $\mathcal{N}=1$ SYM theory on $M_2\times T^2$ with the 't~Hooft flux along small $T^2$. 
We take the supersymmetry-preserving boundary condition for gluinos, $\lambda$, and thus the Witten index, $\tr(-1)^F=N$, suggests that there are $N$ vacua as a consequence of discrete chiral symmetry breaking, $\mathbb{Z}_{2N}\xrightarrow{\mathrm{SSB}}\mathbb{Z}_2$, \cite{Witten:1982df}. 
We confirm it by applying the semiclassical center-vortex theory, and we obtain that 
\begin{equation}
    \langle \tr(\lambda\lambda)\rangle_k\sim \Lambda^3 \rme^{\im (\theta-2\pi k)/N},
\end{equation}
where $k=0,1,\ldots, N-1$. 
In this case, the chiral condensate does not depend on the compactification size $L$ at the leading order of semiclassical analysis. 
Again, we can reproduce the qualitative feature of $4$d dynamics of SYM theory by considering the theory on small $\mathbb{R}^2\times T^2$ with 't~Hooft flux. 

Armed with this semiclassical center-vortex theory on confinement, we study QCD with fundamental quarks in Sec.~\ref{sec:QCD_fund}. 
Under the presence of fundamental matters, however, one cannot naively impose the 't~Hooft twisted boundary condition as pointed out by 't~Hooft himself~\cite{tHooft:1981sps}, since the wave function of fundamental matter cannot be single-valued with the 't~Hooft flux. 
We find that there are at least two ways to go around this obstacle, and then we can use the center-vortex theory to study QCD. 
The first one is to use the $SU(N_f)_\rmV$ flavor symmetry when $N_f=N$, and the second one is to use the $U(1)_\rmB$ magnetic flux and this is useful for any number of flavors~(see Fig.~\ref{fig:twist}~(b) and (c), respectively).   

In Sec.~\ref{sec:QCD_flavor_twist}, we consider $N_f=N$-flavor massless QCD on small $T^2$ with the non-commuting $SU(N)_\rmV$ flavor twist (Fig.~\ref{fig:twist}~(b)). 
This boundary condition explicitly breaks the non-Abelian chiral symmetry, and thus the $2$d effective theory only has the baryon-number and discrete chiral symmetries, $U(1)_\rmB\times (\mathbb{Z}_N)_\rmL$. 
$4$d massless QCD has the discrete anomaly that involves $SU(N)_\rmV/\mathbb{Z}_N$ and this $2$d symmetry~\cite{Tanizaki:2018wtg}, and it is preserved under this flavor-twisted $T^2$ compactification. 
Anomaly matching predicts $N$ degenerate gapped vacua as a consequence of discrete chiral symmetry breaking. 
Within the perturbative analysis, we show that there exists massless Dirac fermion $\tr(\psi)$ that comes from the diagonal component in the color-flavor space and other $N^2-1$ components are gapped. 
Applying the center-vortex theory, we find that the $2$d massless fermion develops the chiral condensate,
\begin{equation}
    \langle \tr(\overline{\psi_\rmL})\tr(\psi_\rmR)\rangle_k\sim \Lambda^3 \rme^{-\im (\theta-2\pi k)/N},
\end{equation}
which satisfies the anomaly matching. 
Moreover, we also compute the large-$T^2$ case with this boundary condition using the chiral Lagrangian, and we find the same result: Nambu-Goldstone bosons become massive due to the flavor twist, and there are $N$ gapped vacua. 

In Sec.~\ref{sec:QCD_baryon_flux}, we consider massless QCD at arbitrary flavors $N_f$ with the minimal $U(1)_\rmB$ magnetic flux, $\int_{T^2}\diff A_\rmB=2\pi$ (Fig.~\ref{fig:twist}~(c)). 
In this case, we can preserve the $U(1)_\rmB$-$SU(N_f)_\rmL$-$SU(N_f)_\rmL$ perturbative anomaly through $T^2$ compactification, and the $2$d effective theory enjoys the $SU(N_f)_\rmL$-$SU(N_f)_\rmL$ 't~Hooft anomaly. 
This $2$d anomaly can be matched by the $SU(N_f)$ level-$1$ Wess-Zumino-Witten ($SU(N_f)_1$ WZW) model, and we explicitly derive it from the $4$d chiral Lagrangian for large $T^2$ with the $U(1)_\rmB$ flux.
On small $T^2$, we first solve the Dirac zero modes and find that each flavor gives a single $2$d massless Dirac fermion at the perturbative level, so we have $N_f$ massless Dirac fermions in $2$d. 
By non-Abelian bosonization, we obtain the $U(N_f)_1$ WZW model instead of $SU(N_f)$, but the center-vortex gives a mass to one of the massless modes and resolves the discrepancy. 
This center-vortex induced mass term can be understood as the $\eta'$ mass term in the $4$d chiral Lagrangian. 

It is quite surprising that a simple idea based on $T^2$ compactification and the center vortex works so nicely just by introducing 't~Hooft flux. 
If the adiabatic continuity really works for a wide class of $4$d gauge theories, then this provides a new way to analyze the strongly-coupled dynamics through the weakly-coupled semiclassical regime. 
It is intriguing that there now exist two different semiclassical limits of Yang-Mills theory, on $\R^3 \times  S^1$ and on $\R^2 \times T^2$,  both of which seem to be adiabatically connected to the strong dynamics on $\R^4$ but exhibit two different confinement mechanisms, via  monopole-instantons and center vortices,  respectively.

%--------------------------------------------------------
%--------------------------------------------------------
\section{\texorpdfstring{$SU(N)$}{SU(N)} Yang-Mills theory on \texorpdfstring{$M_2\times T^2$}{M2xT2} without and with 't~Hooft flux}
\label{sec:YM}

In this section, we study $4$-dimensional $SU(N)$ Yang-Mills (YM) theory on small $T^2$ compactification, so that the total $4$d spacetime is $M_4=M_2\times T^2$ and we consider the $2$d effective theory on $M_2$.  
The classical action of the theory is given by 
\begin{align}
S_{\mathrm{YM}}={1\over g^2}\int \tr[F(a)\wedge \star F(a)]+{\im\, \theta\over 8\pi^2}\int \tr[F(a)\wedge F(a)]. 
\end{align}
where $a$ is the $SU(N)$ gauge field, $F(a)=\diff a + \im a\wedge a$ is the field strength, $\theta$ is  topological theta angle, and 
${1\over 8\pi^2}\int \tr[F(a)\wedge F(a)]\in \Z$ is the instanton number.

$SU(N)$ Yang-Mills theory enjoys $\mathbb{Z}_N$ $1$-form symmetry~\cite{Gaiotto:2014kfa}, and we denote it as $\mathbb{Z}_N^{[1]}$. 
Under the compactification with $2$-torus $T^2$, we can choose whether we insert an 't~Hooft flux. 
In both ways of $T^2$ compactification, the $\mathbb{Z}_N^{[1]}$ symmetry in $4$d decomposes into $\mathbb{Z}_N^{[1]}$ in $2$d and $\mathbb{Z}_N\times \mathbb{Z}_N$ ordinary center symmetry:
\begin{align}
 \label{oneform-red}
 \left(\Z_N^{[1]}\right)_{\rm 4d} \xrightarrow{T^2\,\, \mbox{compact.}}   \left( \Z_N^{[1]}\right)_{\rm 2d} \times \Z_N^{[0]} \times \Z_N^{[0]}.
\end{align} 
Here, the $\Z_N^{[1]}$ symmetry acts on Wilson loops inside $M_2$,  and  $\Z_N^{[0]} \times \Z_N^{[0]}$  acts on Polyakov loops wrapping along $T^2$. 
When $T^2$ is large, the strong dynamics in $4$d should be recovered, and thus these center symmetries are unbroken. 
We would like to find a semiclassical description that has the same feature even for small $T^2$. 

We  first briefly discuss this theory on $M_2 \times T^2$ without 't Hooft flux. 
However, our main emphasis is the case with the nontrivial 't~Hooft flux on $T^2$. 
Introducing a nontrivial 't~Hooft flux, we will show that the $2$d effective theory shows confinement and the ground states have the multi-branch structure.
As the 4d YM theory is expected to have the same feature, we can conjecture that the $2$d theory is adiabatically connected to large volume limit of pure Yang-Mills theory in the sense of gauge invariant order parameters.

%-----------------------------------------------------
\subsection{Periodic \texorpdfstring{$T^2$}{T2} compactification and absence of adiabatic continuity}
\label{sec:periodic_T2}

Consider compactification of pure YM theory on $\R^2 \times T^2$, where $T^2$ is symmetric and its size  $L$  is much smaller than  strong length scale, $\Lambda^{-1}$.  
Here, we take the periodic boundary condition for the gauge field, so the 't~Hooft flux is not inserted. 
Let us denote the compactified direction as $x_3$ and $x_4$, and the holonomies along those directions as 
\begin{equation}
    P_3=\mathcal{P}\exp\left(\im \int_{0}^{L}a_3\diff x_3\right),\quad 
    P_4=\mathcal{P}\exp\left(\im \int_{0}^{L}a_4\diff x_4\right). 
\end{equation}
As the classical Yang-Mills action includes the term $ \frac{1}{g^2} \tr (F_{34})^2$, the classical minima is given by flat connections, $F_{34} =0$.  
This shows that $P_3$ and $P_4$ commutes with each other, and thus they are simultaneously diagonalizable by a gauge transformation: 
\begin{align}
\label{com-pair}
P_3 &= {\rm diag}  ( \rme^{\im \alpha_1}, \rme^{\im \alpha_2}, \ldots, \rme^{\im \alpha_N} ),  \cr 
P_4 &= {\rm diag} ( \rme^{\im \beta_1},  \rme^{\im \beta_2},  \ldots, \rme^{\im \beta_N} ) . 
\end{align} 
Each diagonalized holonomy takes its value in the maximal torus ${\bf T}_N\simeq U(1)^{N-1}$ of $SU(N)$, and the remaining part of gauge transformations is the Weyl group $S_N$, which respects the diagonal form and permutes the eigenvalues of the Polyakov loops simultaneously. 
The classical moduli space is given by
\begin{align}
\label{moduli}
{\cal M}_{\rm cl} =  ({\bf T}_N)^2/S_N . 
\end{align}

This classical moduli space does not survive quantum mechanically. 
At the one-loop order, Gross-Pisarski-Yaffe type calculation~\cite{Gross:1980br,Luscher:1982ma,vanBaal:1986ag,Unsal:2010qh} generates 
a potential for the gauge holonomies  $P_3,  P_4$:
 \begin{align}
\label{pot}
V_{\mathrm{1-loop}} = -\frac{2}{\pi^2 L^4}  \sum_{ (n_3,n_4) \in  \Z^2\setminus \{\bm{0}\}}{\vphantom{\sum}} \frac{1}{(n_3^2 + n_4^2)^2} |\tr (P_3^{n_3}  P_4^{n_4})|^2. 
\end{align} 
This potential has $N^2$ isolated minima located at 
  \begin{align}
\label{pot-min}
 {1\over N} \langle \tr P_3\rangle = \rme^{\im \frac{2\pi k_3}{N}  }, \qquad  
  {1\over N}\langle \tr P_4\rangle = \rme^{\im \frac{2\pi k_4}{N}  }, 
\end{align} 
where $k_3,k_4=0,1,\ldots, N-1$. 
Therefore, the theory spontaneously breaks the $ \Z_N^{[0]} \times \Z_N^{[0]}$ center symmetry. 

On each vacuum that breaks $\mathbb{Z}_N^{[0]}\times \mathbb{Z}_N^{[0]}$, the $2$d effective theory is given by $2$d $SU(N)$ YM theory coupled to massive adjoint scalars $(a_3,a_4)$, or $(P_3, P_4)$. 
If we neglect those massive scalars, we find that the Wilson loop $W_{\calR}(C)$ inside $M_2$ obeys the area law with the Casimir scaling,
\begin{align} 
\langle W_{\calR} (C) \rangle   =  \exp\left( -  g_2^2  C_2 ( \calR ) {\rm Area}(C) \right),
\end{align} 
where  $C_2 ( {\calR} )$   is the quadratic Casimir element of the representation $\calR$, and the value of the tension is dictated by $g_2^2 = \frac{g^2}{L^2}$.  
This violates the $N$-ality rule, but we can expect that the $N$-ality rule of string tensions is obtained by taking into account the effect of massive scalars. 
In any case, the Wilson loop in $M_2$ with nontrivial $N$-ality obeys the area law, and thus $\mathbb{Z}_N^{[1]}$ in $2$d is unbroken. Therefore, when we consider the periodic $T^2$ compactification, we have 
\begin{equation}
    \left( \Z_N^{[1]}\right)_{\rm 2d} \times \Z_N^{[0]} \times \Z_N^{[0]}
    \xrightarrow{\mathrm{SSB}}
    \left( \Z_N^{[1]}\right)_{\rm 2d}
\end{equation}
for sufficiently small $T^2$. 

When the $T^2$ size is large,  the $ \Z_N^{[0]} \times \Z_N^{[0]}$ symmetry is not spontaneously broken as the $4$d YM theory shows confinement. 
As a result, there has to be a phase transition between the small and large  $T^2$ limits. 
This shows that periodic $T^2$ compactification cannot be used to study the $4$d confinement dynamics, and we should consider some other setups to achieve the adiabatic continuity between $2$d and $4$d theories.

%------------------------------------------------------
\subsection{\texorpdfstring{$T^2$}{T2} compactification with a nontrivial 't Hooft flux}

Since the periodic compactification turns out not to be suitable for adiabatic continuity, let us examine other boundary conditions. 
In the following, we take the 't~Hooft twisted boundary condition for the small $T^2$ direction~\cite{tHooft:1979rtg}, and we shall see how the classical vacuum is affected. 
As in the case of the periodic compactification, we should have $F_{34}=0$ to minimize the classical action, but the Polyakov loops $P_3, P_4$ does not commute with each other due to the 't~Hooft twist. Instead, they obey 
\begin{equation}
\label{eq:algebra_tHooftflux}
    P_3 P_4 = P_4 P_3\, \rme^{2\pi \im n_{34}/N}, 
\end{equation}
with some $n_{34}\in \mathbb{Z}_N$, and we specifically choose $n_{34}=1$ (For details of the 't~Hooft flux, see Appendix~\ref{app:tHooft_flux}). 

This algebra~\eqref{eq:algebra_tHooftflux} with $n_{34}=1$ can be satisfied by the shift and clock matrices,
\begin{align}
\label{unbroken}
    P_3=S,\quad  
    P_4= C, 
\end{align}
where $C\propto \mathrm{diag}(1,\omega,\ldots, \omega^{N-1})$ and $(S)_{i,j}\propto\delta_{i+1,j}$ with $\omega=\rme^{2\pi\im/N}$, and this gives the classical vacuum with the 't~Hooft flux. 
We note that there always exists a gauge transformation that brings a minimal action configuration to the above gauge configuration. 
For the detailed discussions, see Appendix~\ref{sec:tHooft_flux_2dtorus_continuum} and \ref{sec:tHooft_flux_2dtorus_lattice}. 
Since 
\begin{equation}
    \tr(P_3^{n_3} P_4^{n_4})=0
\end{equation}
for any $(n_3,n_4)\not=(0,0) \bmod N$, the $\mathbb{Z}_N^{[0]}\times \mathbb{Z}_N^{[0]}$ center symmetry is kept intact by this classical vacuum.\footnote{ One may wonder if quantum fluctuations can destabilize the  $\Z_N^{[0]} \times \Z_N^{[0]} $ symmetric minimum \eqref{unbroken}  to $\Z_N^{[0]} \times \Z_N^{[0]} $ broken minima \eqref{pot-min}. 
In fact, this problem is very important to know if twisted Eguchi-Kawai large-$N$ reduction works or not~\cite{Eguchi:1982nm, GonzalezArroyo:1982hz}. 
When $N$ is fixed, we can show that the center symmetry is unbroken at sufficiently weak couplings by the similar discussion. 
However, destabilization takes place on 1-site model at some intermediate couplings, and the critical value $g_*^2$ depends on $N$ as $1/(Ng^2_*)\sim N$~\cite{Bietenholz:2006cz, Teper:2006sp,Azeyanagi:2007su}. 
Later, it has been pointed out that stability can be maintained for judicious choice of $n_{\mu \nu}$ that depends on $N$ when taking the large-$N$ limit~\cite{GonzalezArroyo:2010ss} or by introducing  fermionic matter in adjoint representation \cite{Azeyanagi:2010ne}.   

We may also need to choose some nontrivial twists if we try to make a connection between our continuum theory and the large-$N$ volume independence. 
As another possibility, we may extend the applicability of adiabatic continuity by adding a massive adjoint fermion with the same boundary condition, which would prohibit the phase transitions at intermediate length scales. 
The adjoint fermion decouple for sufficiently small $T^2$ because of its mass term, and we can apply the following analysis without any changes.

As we shall see, our semiclassical center-vortex description of confinement gives predictions that are consistent with the adiabatic continuity. 
This suggests the existence of a continuous path connecting the confinement on small $T^2$ regimes and the $4$d confinement without phase transitions.
However, we must note that such a continuous path may exist in the enlarged theory space. 
}

Let us discuss the perturbative mass spectrum of $2$d gauge fields on $M_2$: 
\begin{equation}
    \sum_{m=1,2}a_m\diff x_m. 
\end{equation}
Because of the 't~Hooft flux, its boundary condition along $T^2$ is given by 
\begin{align}
    a_m(\bm{x},x_3+L,x_4)&=S^{-1} a_m(\bm{x},x_3,x_4) S,\nonumber\\
    a_m(\bm{x},x_3,x_4+L)&=C^{-1} a_m(\bm{x},x_3,x_4)C, 
    \label{eq:twistedBC_gauge}
\end{align}
where $\bm{x}\in M_2$ and $(x_3,x_4)\in T^2$. For performing Fourier expansion, it is convenient to introduce the following basis for the Lie algebra \cite{GonzalezArroyo:1982hz}:
\begin{equation}
J_{\bm{p}}=\omega^{-p_3 p_4/2} C^{-p_3} S^{p_4}, 
\label{basis1}
\end{equation}
with $\bm{p}=(p_3,p_4)\in (\mathbb{Z}_N)^2$. This satisfies $J_{\bm{p}}^\dagger=J_{-\bm{p}}$, and we can expand $a_m\in \mathfrak{su}(N)$ as 
\begin{equation}
    a_m(\bm{x},x_3,x_4)=\sum_{\bm{p}\not=0}a^{(\bm{p})}_m(\bm{x},x_3,x_4) J_{\bm{p}}, 
\end{equation}
with $(a_m^{(\bm{p})})^*=a_m^{(-\bm{p})}$. 
The twisted boundary condition~\eqref{eq:twistedBC_gauge} can be written as 
\begin{align}
    a_m^{(\bm{p})}(\bm{x},x_3+L,x_4)&=\omega^{p_3}a_m^{(\bm{p})}(\bm{x},x_3,x_4),\nonumber\\
    a_m^{(\bm{p})}(\bm{x},x_3,x_4+L)&=\omega^{p_4} a_m^{(\bm{p})}(\bm{x},x_3,x_4), 
    \label{tbc2}
\end{align}
and thus the Fourier expansion of $a_m^{(\bm{p})}$ is given by 
\begin{equation}
    a_m^{(\bm{p})}(\bm{x},x_3,x_4)=\sum_{k_3,k_4\in \mathbb{Z}}\tilde{a}_m^{(\bm{p},\bm{k})}(\bm{x})\exp\left({2\pi \im\over NL}\Bigl((Nk_3+p_3)x_3+(Nk_4+p_4)x_4\Bigr)\right). 
\end{equation}
A part of the $4$d kinetic term, $\tr(F_{m3}^2)+\tr(F_{m4}^2)$, gives the mass of each mode, $a_m^{(\bm{p},\bm{k})}$, as 
\begin{equation}
    M_{\bm{p},\bm{k}}^2=\left({2\pi\over NL}\right)^2\Bigl((Nk_3+p_3)^2+(Nk_4+p_4)^2\Bigr). 
    \label{gluonspectrum}
\end{equation}
We note that the traceless condition, $\tr(a)=0$, requires $\bm{p}\not=0$, and thus there is no zero mode. 
The perturbative fluctuations are completely gapped under the presence of the 't~Hooft flux, and thus semiclassical calculations do not suffer from infrared divergence.\footnote{Note that the spacing  in the perturbative spectrum is controlled by $\frac{2\pi}{NL}$, which is much finer than the usual Kaluza-Klein spacing $\frac{2\pi}{L}$. In particular, at finite $L$, if one takes $N \rightarrow \infty$ limit, the perturbative states form a continuum, despite the fact that the theory  is formulated on a finite space with size $L$.  This is a perturbative manifestation of the non-perturbative large-$N$ volume independence~\cite{GonzalezArroyo:1982hz,Unsal:2010qh}. } Note that the 4d coupling $Ng^2(\mu)$ runs according to standard renormalization group at distances much  smaller  than $\frac{NL}{2\pi}$ and freezes at longer length scales. As a result,   $Ng^2(\frac{NL}{2\pi} ) \ll 1$ at small $T^2$ and the theory  admits a weak-coupling calculable regime. 

One way to understand the perturbative mass gap is to regard $P_3$ and $P_4$ as the adjoint Higgs fields in $2$d effective theory. 
At the classical vacuum~\eqref{unbroken}, the Higgsing of the gauge group occurs as 
\begin{equation}
    SU(N)\xrightarrow{\mathrm{Higgsing}} \mathbb{Z}_N,
\end{equation}
and the low-energy effective theory becomes the discrete gauge theory at the perturbative level. 
This also suggests that the Wilson loop inside $M_2$ obeys the perimeter law within the perturbation theory. 
In the next section, we shall examine how the $\mathbb{Z}_N^{[1]}$ symmetry in $2$d is restored by taking into account nonperturbative effects.

%------------------------------------------------------
\subsection{Semiclassical confinement mechanism via center vortices}
\label{sec:center_vortex_theory}

We have seen that the $0$-form center symmetry is unbroken when we introduce the 't~Hooft flux on small $T^2$, but $\mathbb{Z}_N^{[1]}$ is spontaneously broken within the perturbation theory. 
We should address the question whether the $\mathbb{Z}_N^{[1]}$ symmetry is restored by some nonperturbative effects. 
In this section, we uncover that center vortex plays an important role for the restoration of $\mathbb{Z}_N^{[1]}$, and quantitative features of $4$d YM theory can be reproduced by the dilute gas approximation of center vortices. 

Center vortex is a dynamical codimension-$2$ object in Euclidean spacetime, and its liberation is one of the famous scenarios to explain quark confinement~\cite{tHooft:1977nqb, Cornwall:1979hz, Nielsen:1979xu, Ambjorn:1980ms, DelDebbio:1996lih,  Faber:1997rp, DelDebbio:1998luz, Langfeld:1998cz, Kovacs:1998xm, Engelhardt:1999fd, deForcrand:1999our}. 
See Ref.~\cite{Greensite:2003bk} for a review about the center-vortex induced confinement scenario. 
Although the center-vortex theory provides a promising scenario to understand various properties of confinement, it seems that it does not yet give analytically calculable semi-classical models of confinement to our best knowledge. 
We will see that the small $T^2$ compactification with the 't~Hooft flux gives a semiclassically calculable setup of confinement based on center vortices, while maintaining many quantitative features of $4$d confinement.\footnote{In the context of $4$d confinement, monopoles are also considered to play important roles. In our $T^2$-compactification setup, however, we have to explain the confinement of $2$d effective theory. 
Since monopoles are defined by its magnetic fluxes that penetrate the $2$-sphere at infinities, we need at least $3$ spacetime dimensions to have such configurations. This is why we consider center vortices as the leading candidate for the semiclassical description of confinement mechanism. }  

Most part of the following discussion turns out to be quite parallel with the semiclassical analysis of $2$d charge-$N$ Abelian Higgs model, and we discuss it in Appendix~\ref{sec:2dAbelianHiggs}. 

\subsubsection{Center vortex configurations}

Here, let us explain the structure of center vortex configurations. 
Unfortunately, self-dual solutions of Yang-Mills equation on the torus $T^4$ have not been found analytically, so we cannot use analytical solutions for our semiclassical analysis. 
Still, numerical studies~\cite{Gonzalez-Arroyo:1998hjb, Montero:1999by, Montero:2000pb} by A.~Gonz\'alez-Arroyo and A.~Montero have confirmed the existence of center vortex configuration 
as a self-dual solution, and let us briefly explain its properties.

We consider the $4$-torus $M_4=T^4$ as our spacetime, and the first two directions $\mu=1,2$ are large and the other two directions $\mu=3,4$ are small:
\begin{align}
 \underbrace{T^2}_{L_l  \gg \Lambda^{-1}} \times  \underbrace{{ T}^2}_{L_s \ll \Lambda^{-1} } \xrightarrow{L_l\to \infty}  \,\, \R^2\times T^2 .
\end{align}
We introduce the background $\mathbb{Z}_N$ $2$-form gauge field $B$ for the center symmetry. 
Following Refs.~\cite{Kapustin:2014gua,Kapustin:2013qsa}, we  introduce it by promoting the $SU(N)$ gauge field $a$ to a $U(N)$ gauge field $\tilde{a}$, and we regard $B$ as a $U(1)$ $2$-form gauge field with the constraint 
\begin{equation}
    NB=\tr(\tilde{F}), 
\end{equation}
where $\tilde{F}=F(\tilde{a})=\diff \tilde{a}+\im \tilde{a}\wedge \tilde{a}$ is the $U(N)$ field strength (For details, see Appendix~\ref{sec:tHooft_flux_gauging}). 
This is related to the 't~Hooft twist $n_{\mu\nu}$ by 
\begin{equation}
    {1\over N}n_{\mu\nu}={1\over 2\pi}\int_{(T^2)_{\mu\nu}}B \quad (\bmod 1). 
\end{equation}
Under the presence of $B$, the topological charge can take fractional values~\cite{vanBaal:1982ag}:
\begin{align}
    Q_{\mathrm{top}}&={1\over 8\pi^2}\int_{T^4} \tr\left((\tilde{F}-B)^2\right) \nonumber\\
    &={1\over 8\pi^2}\int_{T^4}\left(\tr(\tilde{F}^2)-NB^2\right)\nonumber\\
    &\in-{1\over N}{\ve_{\mu\nu\rho\sigma}n_{\mu\nu}n_{\rho\sigma}\over 8}+\mathbb{Z}. 
    \label{eq:Qtop_tHooft}
\end{align}
The fractional part comes from ${1\over 8\pi^2}\int NB^2$, and the integer part from ${1\over 8\pi^2}\int \tr(\tilde{F}^2)\in \mathbb{Z}$. 
We note that $\ve_{\mu\nu\rho\sigma}n_{\mu\nu}n_{\rho\sigma}/8\in\mathbb{Z}$, and thus  $Q_{\mathrm{top}}\in {1\over N}\mathbb{Z}$. The fractional part is completely determined by the 't~Hooft twist. 

Especially, by taking $n_{12}=1$ and $n_{34}=1$, we have $Q_{\mathrm{top}}=-1/N \bmod 1$. 
Then, Bogomol'nyi–Prasad–Sommerfield (BPS) bound~\cite{Bogomolny:1975de,Prasad:1975kr} tells that the Yang-Mills kinetic term is bounded from below as 
\begin{equation}
    \mathrm{Re}(S_{\mathrm{YM}})\ge {8\pi^2\over g^2}|Q_{\mathrm{top}}|={8\pi^2\over N g^2}, 
\end{equation}
and the lower bound is given by $1/N$-th of the instanton action $S_\mathrm{I}={8\pi^2/g^2}$. 
The inequality is saturated if and only if the (anti-)self-dual equation is satisfied, and not much is known for its analytic solutions with fractional topological charges (except for Abelian-like constant solutions on a  torus with certain aspect ratio~\cite{tHooft:1981nnx}). 
In Refs.~\cite{Gonzalez-Arroyo:1998hjb, Montero:1999by, Montero:2000pb}, the authors numerically searched the absolute minimum of lattice YM action with this setup by using the naive cooling technique~\cite{GarciaPerez:1989gt}. 
They found that the fractional instanton can achieve the BPS bound within the lattice discretization error, and the solution has the well-defined limit in $L_l\to \infty$. 
The self-dual solution has the localized action density along the $\mathbb{R}^2$ direction, so it has a vortex-like structure. 

Let us mention how this vortex-like solution affects the Wilson loop, $W_\calR(C)$, in $\mathbb{R}^2$. 
When the vortex is outside of the loop, the Wilson loop takes the trivial value, $W(C)=1$. 
When the vortex is inside of the loop, however, the phase of Wilson loop is rotated as 
\begin{equation}
    W_{\calR}(C)=\exp(2\pi\im |\calR|/N),  
    \label{eq:WilsonLoop_singleVortex}
\end{equation}
where $|\calR|$ is the $N$-ality of the representation $\calR$, i.e. the number of boxes of the Young tableaux mod~$N$. 
This is exactly what we expect for the center vortex~\cite{tHooft:1977nqb, Cornwall:1979hz, Nielsen:1979xu, Ambjorn:1980ms}, so the self-dual configuration with the above setup can be regarded as the center vortex.

To wrap-up,  it is known that the topological charge is quantized in units of  $Q_{\mathrm{top}}=1/N$ in 
background $\mathbb{Z}_N$ $2$-form gauge field $B$ (or 't~Hooft flux background).   
Although an analytic center-vortex solution is not yet known,   numerical simulations unambiguously  demonstrate that configurations which saturate BPS bound with action $\frac{1}{N}S_{\mathrm{I}}$  exist.  
We will establish semi-classical theories based on proliferation of these configurations in various gauge theories: Yang-Mills, ${\cal N}=1$ SYM, and QCD with fundamental quarks.

\subsubsection{Dilute gas of center vortices and confinement}
\label{sec:DCVGA}

We have seen that there exists the center vortex as the self-dual configuration on $\mathbb{R}^2\times T^2$ with the 't~Hooft flux. 
It carries the fractional topological charge, $Q_{\mathrm{top}}=\pm 1/N$, with the YM action $S_{\mathrm{vortex}}=S_\mathrm{I}/N=8\pi^2/(Ng^2)$. 
As it has the fixed radius, we can perform the dilute gas approximation without suffering from infrared divergence.\footnote{According to Ref.~\cite{Montero:2000pb}, the radius of the center vortex is proportional to $\sqrt{N}$. When we would like to take the large-$N$ limit, we should care about the order of infinite volume limit and large-$N$ limit. In this paper, we keep $N$ to be finite, and thus the dilute gas approximation is valid for sufficiently large $M_2$. } 
The Boltzmann weight for the single center vortex is given by 
\begin{equation}
    K\exp(-S_\mathrm{I}/N)\exp(\pm \im\theta/N), 
\end{equation}
where the $\pm$ sign depends on the topological charge $Q_{\mathrm{top}}=\pm 1/N$. Here, $K\sim 1/L_s^2$ is a prefactor with mass dimension $2$, which should be determined by careful analysis of the fluctuation determinant. 

When performing the dilute gas approximation, we must specify the details of the boundary condition. Our spacetime is given by $M_4=M_2\times T^2$. Here, $M_2$ is a large but compact $2$d manifold, and we do not insert any 't~Hooft flux along this direction. $T^2$ is a small $2$-torus with the nontrivial 't~Hooft flux. 
Then, we have $n_{34}=1$ but all the other components of $n_{\mu\nu}$ vanish, and in particular $n_{12}=0$. 
As a result, $\ve_{\mu\nu\rho\sigma}n_{\mu\nu}n_{\rho \sigma}=0$, and \eqref{eq:Qtop_tHooft} tells that $Q_{\mathrm{top}}\in \mathbb{Z}$, so single vortex configurations are not summed in the path integral. 
Let $n$ and $\overline{n}$ be the number of center vortices and anti-vortices, then 
\begin{equation}
    Q_{\mathrm{top}}={n-\overline{n}\over N}. 
\end{equation}
We must restrict the summation of dilute gas configurations so as to satisfy $n-\overline{n}\in N\mathbb{Z}$.

The partition function within the dilute gas approximation is then given by 
\begin{equation}
    Z(\theta)=\sum_{n,\overline{n}\ge 0}{V^{n+\overline{n}}\over n! \overline{n}!}K^{n+\overline{n}}\rme^{-(n+\overline{n})S_I/N}\rme^{\im (n-\overline{n})\theta/N}\delta_{n-\overline{n}\in N\mathbb{Z}}, 
\end{equation}
where $V$ is the volume of $M_2$. 
The Kronecker delta factor, $\delta_{n-\overline{n}\in N\mathbb{Z}}$, is introduced in order to impose the restriction, and it is convenient to write down the constraint as Fourier series, 
\begin{equation}
    \delta_{n-\overline{n}\in N\mathbb{Z}}=\sum_{k=0}^{N-1}\rme^{-{2\pi \im \,k\over N}(n-\overline{n})}.
\end{equation}
Substituting this expression and performing the summation over $n,\overline{n}$, we obtain
\begin{align}
    Z(\theta)&=\sum_{n,\overline{n}\ge 0}{V^{n+\overline{n}}\over n! \overline{n}!}K^{n+\overline{n}}\rme^{-(n+\overline{n})S_I/N}\rme^{\im (n-\overline{n})\theta/N}\sum_{k=0}^{N-1}\rme^{-{2\pi \im \,k\over N}(n-\overline{n})}\nonumber\\
    &=\sum_{k=0}^{N-1}\exp\left[V K \rme^{-S_{\mathrm{I}}/N+\im (\theta-2\pi k)/N}+V K \rme^{-S_{\mathrm{I}}/N-\im (\theta-2\pi k)/N}\right]\nonumber\\
    &=\sum_{k=0}^{N-1}\exp\left[-V\left(-2K \rme^{-S_I/N}\cos\left({\theta-2\pi k\over N}\right)\right)\right],
\end{align}
and thus we can reproduce the multi-branch structure of $4$d YM theory~\cite{Witten:1980sp, DiVecchia:1980yfw, Witten:1998uka}:
\begin{align}
    E_k(\theta)&=-2K\rme^{-S_I/N}\cos\left({\theta-2\pi k\over N}\right) \nonumber\\
    &\sim -\Lambda^2 (\Lambda L_s)^{5/3}\cos\left({\theta-2\pi k\over N}\right).
    \label{eq:multi_branch}
\end{align}
To find the last expression, we use the fact that the leading coefficient of the YM beta function is ${11\over 3}N$, so that  $\rme^{-S_{\mathrm{I}}/N} \sim (\Lambda L_s)^{\frac{11}{3}}$. 
Each branch of the ground states does not have $2\pi$ periodicity for its $\theta$ dependence, but the partition function is $2\pi$ periodic thanks to the level crossing between those branches. 
There is a first-order phase transition at $\theta=\pi$, at which the level crossing occurs.

\begin{figure}[t]
\vspace{-1.0cm}
\begin{center}
\includegraphics[width = 1.0\textwidth]{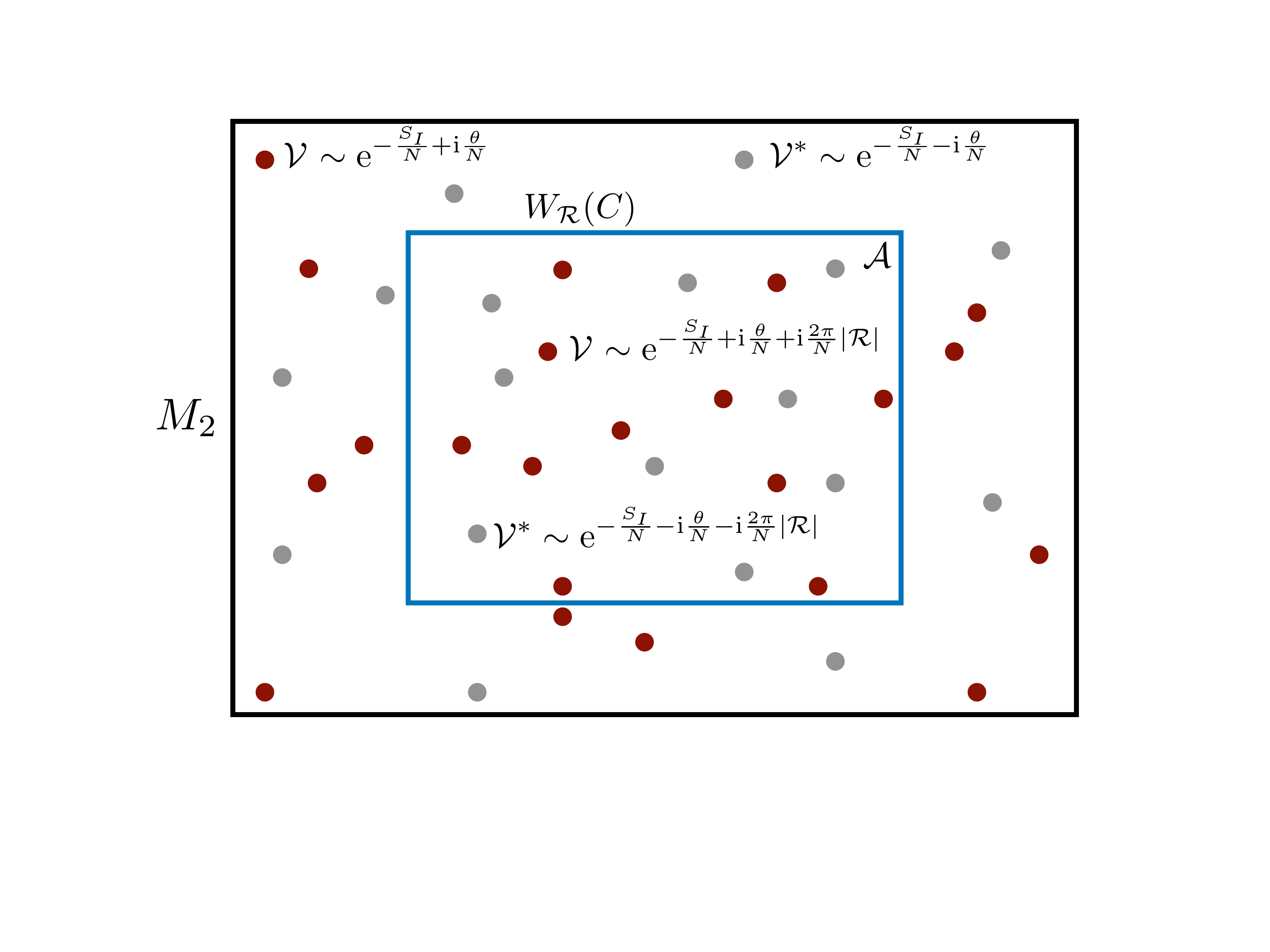}
\vspace{-2.5cm}
\caption{Center vortices are point-like objects on  $M_2$ and their fugacity is proportional to $\exp(-S_{\rm vortex})  = \exp(-{8\pi^2\over N g^2}\pm \mathrm{i}{\theta\over N})  $.  Due to their non-trivial mutual statistics with the Wilson loop, the center vortex acquires an extra phase $\exp(\pm{2\pi\im\over N}|\mathcal{R}|)$ when it is inside the Wilson loop of the representation $\mathcal{R}$. 
As a result, the proliferation of the center vortex leads to the area law of Wilson loops, and confinement is realized. }
\label{fig:prolif}
\end{center}
\end{figure}

\begin{figure}[t]
\vspace{-1.0cm}
\begin{center}
\includegraphics[width = 0.9\textwidth]{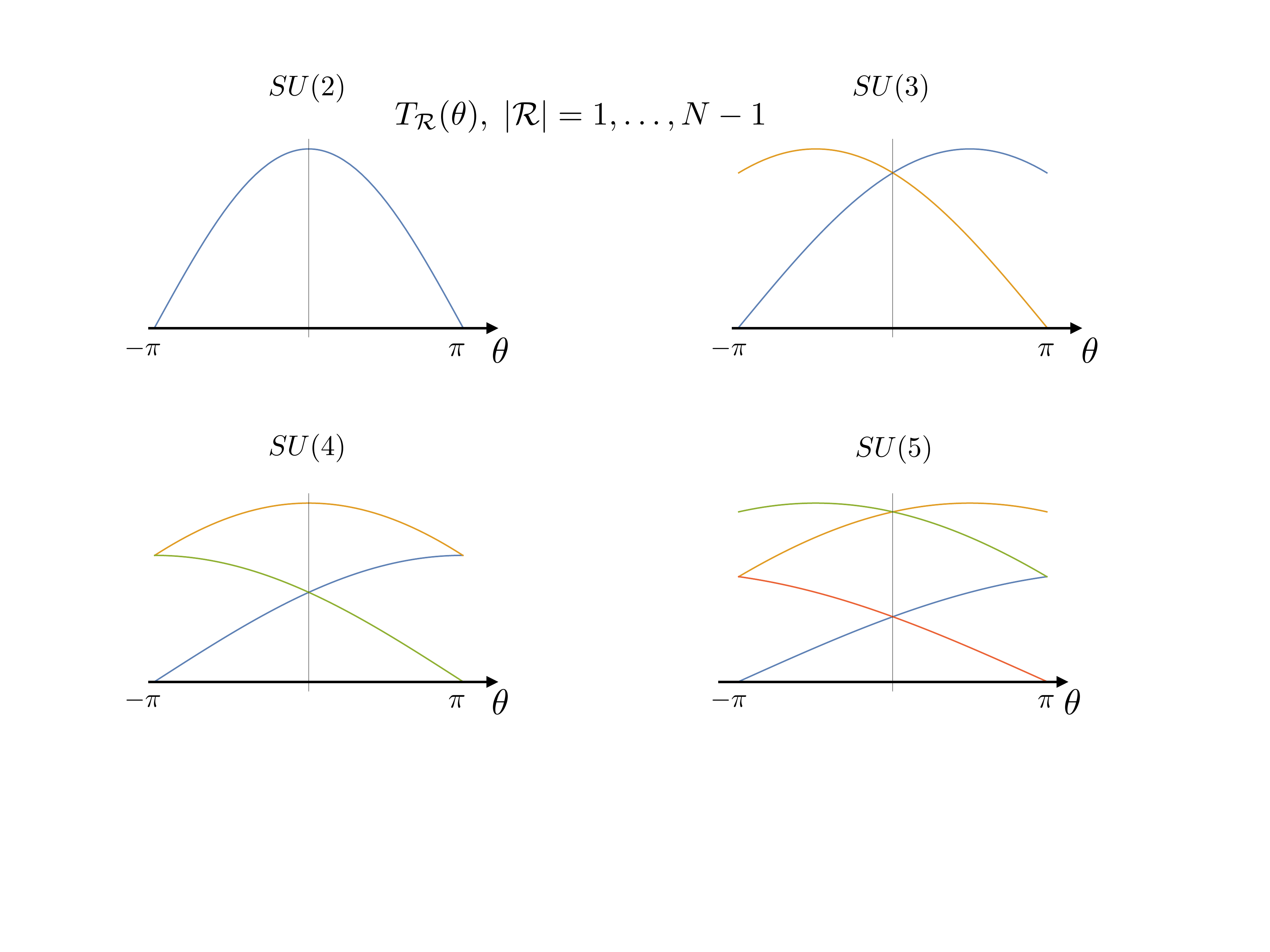}
\vspace{-2.5cm}
\caption{The $\theta$ angle dependence of the string tensions on $M_2 \times T^2$ with 't Hooft flux through small $T^2$. 
At $\theta=\pi$, one of the tensions vanishes, which is consistent with the requirement of $2$d 't~Hooft anomaly between the $1$-form symmetry and $CP$ at $\theta=\pi$. 
}
\label{fig:tensions}
\end{center}
\end{figure}

Next, let us derive the area law of the Wilson loop $W_{\calR}(C)$ inside $M_2$, and we denote the area enclosed by $C$ as $\calA$. 
When the vortex is inside the Wilson loop, it acquires an extra factor~\eqref{eq:WilsonLoop_singleVortex} (see Fig.~\ref{fig:prolif}), and we should take it into account in the dilute gas approximation. 
For this purpose, we split the number of vortices $n$ into $n_1$ and $n_2$, which represent the number of vortices inside and outside of the Wilson loop, respectively, and we do the same for $\overline{n}$. 
The expectation value of the Wilson loop can be obtained as 
\begin{align}
    \langle W_{\calR}(C)\rangle &= {1\over Z(\theta)}
    \sum_{n_1,n_2,\overline{n}_1,\overline{n}_2}{\calA^{n_1+\overline{n}_1}(V-\calA)^{n_2+\overline{n}_2}\over n_1!\, n_2!\, \overline{n}_1!\, \overline{n}_2!} \left(K\,\rme^{-S_I/N}\right)^{n_1+n_2+\overline{n}_1+\overline{n_2}}\nonumber\\
    &\qquad \times \rme^{\im (n_1+n_2-\overline{n}_1-\overline{n}_2)\theta/N} \, \rme^{2\pi\im (n_1-\overline{n}_1)|\calR|/N} \,  \delta_{n_1+n_2-\overline{n}_1-\overline{n}_2\in N\mathbb{Z}}\\
    &={1\over Z(\theta)}\sum_{k=0}^{N-1}\rme^{-V E_k(\theta)}\exp\Bigl(-\calA (E_k(\theta+2\pi|\calR|)-E_k(\theta))\Bigr). 
\end{align}
In order to find the familiar expression, let us set $-\pi<\theta<\pi$ and take the infinite volume limit $V\to \infty$ of $M_2$. The $k=0$ state is selected by this procedure, and we obtain the area law,
\begin{align}
\langle W_{\cal R}((C) \rangle =\exp\Bigl(-\calA \bigl(E_0(\theta+2\pi|\calR|)-E_0(\theta)\bigr)\Bigr). 
\end{align}
The string tension $T_{\calR}(\theta)$ for the representation $\calR$ is then given by 
\begin{align}
    T_{\calR}(\theta)&=E_0(\theta+2\pi|\calR|)-E_0(\theta)\nonumber\\
    &\sim \Lambda^2 (\Lambda L_s)^{5/3} \left(\cos{\theta\over N}-\cos{\theta+2\pi |\calR|\over N}\right). 
\label{thetadep}
\end{align}
We note that this expression is valid only for $|\theta|<\pi$. In Appendix~\ref{app:area_law}, we derive this formula~\eqref{thetadep} from a slightly different viewpoint. 
The $\theta$ angle dependence   of string tensions \eqref{thetadep} is shown in Fig.~\ref{fig:tensions}. 
Note that although we observe confinement as the string tensions are  non-zero at generic $\theta$, one of the tensions vanishes at $\theta=\pi$. 
This turns out to be a consequence of the mixed 't~Hooft anomaly between the $1$-form symmetry and $CP$ at $\theta=\pi$ in $2$d as we shall discuss in the next section.

For $\theta=0$, the string tensions reduce to 
\begin{align}
T_{\calR}  \sim \Lambda^2  (\Lambda L_s)^{5/3}  \sin^2  \frac{\pi |\calR|}{N}.  
\label{eq:tension_theta0}
\end{align}
We note that the string tensions depend only on $N$-ality $|\calR|$, not on the representation $\calR$ itself. 
This is a common feature in the center-vortex confinement scenario. 
For quarks with non-zero $N$-ality charges, $|\calR| \neq 0$ (mod $N$), there is a linear confinement, induced by vortices in the semi-classical domain, so we have unbroken $\mathbb{Z}_N^{[1]}$ symmetry. 
For  zero $N$-ality  representations,  the area term vanishes, and the leading behavior is given by the perimeter law.

Few other remarks are in order. 
The $N$-ality rule of the string tension is a desirable feature of the string tensions for sufficiently larger Wilson loops, which is believed to be true for $4$d pure Yang-Mills theory. 
However, the linear confinement appears in much shorter length scale for pure YM theory on $\mathbb{R}^4$, and the Casimir scaling controls the string tension in that regime, which predicts $T_k ={k(N-k)\over N-1}T_1 \sim k T_1 $ when $k\ll N$.
The dilute gas approximation of center vortex cannot treat this feature as \eqref{eq:tension_theta0} gives the different behavior, $T_k={\sin^2(\pi k/N)\over \sin^2(\pi/N)}\sim k^2 T_1$ with $k\ll N$, which may be called as the ``sine-squared'' law. 
Even though this behavior is different from that of $4$d confinement of pure YM theory, we note that quantitative features can be changed under the $T^2$ compactification, regardless of whether the adiabatic continuity scenario works or not. 
We also point out that the vortex thickness is completely neglected in our semiclassical description. 
Some of previous studies suspect that the Casimir scaling may be attributed to the finite thickness of vortex~\cite{Greensite:2003bk}.

Let us compare this situation with other analytically calculable regimes of $4$d YM theory with adjoint matters, which are thought to be smoothly connected to pure YM dynamics. 
First, we consider softly-broken $\mathcal{N}=2$ Seiberg-Witten theory, which gives a model of confinement by Abelian monopole condensation~\cite{Seiberg:1994rs}. 
Since the Weyl group $S_N$ is completely Higgsed, the fundamental string splits into $N-1$ different confining strings~\cite{Douglas:1995nw}. 
Let $T_k$ be the smallest string tension in the $k$-box antisymmetric representation, then they obey the ``sine'' law, $T_k={\sin(\pi k/N)\over \sin (\pi/N)}T_1\sim k T_1$ when $k\ll N$. 
We note, however, that the $N$-ality rule for generic representations is violated because of the absence of $W$-bosons in its effective description. 

Next, we consider the YM theory on small $\R^3 \times S^1 $ adding the double-trace deformation or adjoint fermions with periodic boundary condition~\cite{Unsal:2007vu,Unsal:2007jx,Unsal:2008ch,Shifman:2008ja, Davies:2000nw}.
In this case, dilute gas of monopole-instantons~\cite{Kraan:1998kp, Kraan:1998sn, Lee:1998bb, Lee:1997vp} provides the semiclassical description of confinement. 
String tensions of deformed YM theory are studied in details in Ref.~\cite{Poppitz:2017ivi}, and they obey the square-root of Casimir scaling, $T_k=\sqrt{k(N-k)\over N-1}T_1\sim \sqrt{k}T_1$, for $k\ll N$, where $T_k$ is the minimal string in the $k$-box antisymmetric representation. 
In this model, $\mathbb{Z}_N$ subgroup of the Weyl group $S_N$ can be understood as the $0$-form center symmetry. 
As a result, the fundamental string tension is unique, but there are several confining strings for a higher irreducible representation, and the $N$-ality rule is violated. 
These behaviors are different from what we expect for non-Abelian confinement of pure YM theory or $\mathcal{N}=1$ SYM theory, but one can expect that these theories are smoothly connected. 

To summarize, each effective theory gives a prediction in its own valid regime, and different quantitative properties can be connected via smooth crossover even though dynamics need to be rearranged under the crossover. 
Important thing is that the center-vortex description on small $\R^2 \times T^2$ with 't~Hooft flux gives the same vacuum structure with that of $4$d YM theory,  
and we conjecture that they are adiabatically connected without phase transitions.

%-----------------------------------------------------
\subsection{'t Hooft anomaly and compactification of Yang-Mills theory}
\label{sec:anomaly_compact}

In this section, we discuss topological aspects of the $T^2$ compactification with the 't~Hooft flux. 
It is now well known that the $4$d YM theory has an 't~Hooft anomaly related to $\mathbb{Z}_N^{[1]}$ and $CP$ symmetry at $\theta=\pi$~\cite{Gaiotto:2017yup, Gaiotto:2017tne}. It requires the spontaneous breakdown of $CP$ at $\theta=\pi$ assuming that the YM theory shows confinement at any values of $\theta$. 
We have just seen that the $2$d effective theory with center vortices show the same phenomena, and let us demystify why this happens. 

We start with discussion on the anomaly of $4$d $SU(N)$ YM theory. 
Introducing the background $\mathbb{Z}_N$ $2$-form gauge field $B=B_{4\mathrm{d}}$, we have seen that the topological charge becomes fractional as \eqref{eq:Qtop_tHooft}. 
Under the $2\pi$ shift of the $\theta$ angle, the partition function $Z_\theta[B_{4\mathrm{d}}]$ with $B_{4\mathrm{d}}$ is transformed as 
\begin{equation}
    Z_{\theta+2\pi}[B_{4\mathrm{d}}]=\exp\left({\im N\over 4\pi}\int_{M_2\times T^2} B_{4\mathrm{d}}\wedge B_{4\mathrm{d}}\right) Z_\theta[B_{4\mathrm{d}}]. 
    \label{eq:anomaly_theta_period}
\end{equation}
This mild violation of the $2\pi$ periodicity plays a crucial role, because the presence of $CP$ symmetry at $\theta=\pi$ uses the fact that we can identify $\theta=-\pi$ with $\theta=\pi$ up to the $2\pi$ periodicity. 
At $\theta=0$, the $CP$ transformation is a good symmetry even with the background $B_{4\mathrm{d}}$, 
\begin{equation}
    CP:Z_{\theta=0}[B_{4\mathrm{d}}]\to Z_{\theta=0}[B_{4\mathrm{d}}], 
\end{equation}
but, at $\theta=\pi$, the partition function transforms as 
\begin{align}
    CP:Z_{\theta=\pi}[B_{4\mathrm{d}}]&\to Z_{\theta=-\pi}[B_{4\mathrm{d}}]\nonumber\\
    &=\exp\left(-{\im N\over 4\pi}\int_{M_2\times T^2} B_{4\mathrm{d}}\wedge B_{4\mathrm{d}}\right) Z_{\theta=\pi}[B_{4\mathrm{d}}]. 
\end{align}
For even $N$, there is no local counter term that eliminates this phase, so this is a genuine 't~Hooft anomaly between $\mathbb{Z}_N^{[1]}$ and $CP$ at $\theta=\pi$. 
For odd $N$, there is a local counter term that eliminates this phase, so there is no 't~Hooft anomaly at $\theta=\pi$, but we can obtain a slightly weaker condition on the ground state by the global inconsistency~\cite{Gaiotto:2017yup, Kikuchi:2017pcp, Tanizaki:2017bam, Karasik:2019bxn, Tanizaki:2018xto, Cordova:2019jnf,Cordova:2019uob}. 
In both cases, the most plausible scenario for $4$d YM theory is that confinement occurs at any values of $\theta$ and $CP$ is spontaneously broken at $\theta=\pi$ to match the 't~Hooft anomaly and/or global inconsistency. 

We then move on to the discussion of the $T^2$ compactification~\cite{Yamazaki:2017dra} (see also Refs.~\cite{Tanizaki:2017qhf, Tanizaki:2017mtm, Dunne:2018hog, Furusawa:2020qdz}). 
In order to emphasize the role of the 't~Hooft flux, we here reintroduce the 't~Hooft twist $n_{34}$ as a variable in $\mathbb{Z}_N$. 
The periodic compactification corresponds to $n_{34}=0$, while the center-vortex effective theory is obtained for $n_{34}=1$. 
For any choice of $n_{34}$, $2$d effective theory enjoys $\left( \Z_N^{[1]}\right)_{\rm 2d} \times \Z_N^{[0]} \times \Z_N^{[0]}$, and we can introduce their background gauge fields:
\begin{itemize}
    \item $B_{2\diff}$: $2$-form gauge field for $\mathbb{Z}_N^{[1]}$, which couples to $W_{\calR}(C)$ inside $M_2$. 
    \item $A_3$: $1$-form gauge field for one of $\mathbb{Z}_N^{[0]}$, which couples to $P_3$. 
    \item $A_4$: $1$-form gauge field for another $\mathbb{Z}_N^{[0]}$, which couples to $P_4$. 
\end{itemize}
We note that these gauge fields live on $M_2$, and thus they are independent of the $T^2$ coordinates. 
We can nicely summarize these information into the $4$d $\mathbb{Z}_N$ $2$-form gauge field as 
\begin{equation}
    B_{4\mathrm{d}}=B_{2\mathrm{d}}+A_3\wedge{\diff x_3\over L_s}+A_4\wedge {\diff x_4\over L_s}+{2\pi n_{34}\over N}{\diff x_3\wedge \diff x_4\over L_s^2}. 
    \label{eq:4d2d_gauge}
\end{equation}
We can confirm that this is a $\mathbb{Z}_N$ $2$-form gauge field by noting that $\exp\left(\im\int_{X_2} B_{4\diff}\right)\in \mathbb{Z}_N$ for any closed $2$-manifolds $X_2\subset M_4$. 
Indeed, this can be regarded as a configuration of the $\mathbb{Z}_N$ background gauge field for $4$d $\mathbb{Z}_N^{[1]}$ symmetry.

We can evaluate the $4$d topological action using \eqref{eq:4d2d_gauge} as follows: 
\begin{align}
    &{N\over 4\pi}\int_{M_2\times T^2}B_{4\diff}\wedge B_{4\diff}\nonumber\\
    =\,& {N\over 2\pi}\int_{M_2\times T^2}\left(B_{2\diff}\wedge {2\pi n_{34}\over N}{\diff x_3\diff x_4\over L_s^2}+A_3\wedge{\diff x_3\over L_s}\wedge A_4\wedge {\diff x_4\over L_s}\right)\nonumber\\
    =\,& n_{34}\int_{M_2} B_{2\diff}-{N\over 2\pi}\int_{M_2}A_3\wedge A_4. 
\end{align}
Let us denote the partition function with the 't~Hooft twist $n_{34}$ as $Z^{(n_{34})}_{\theta}[B_{2\diff},A_3,A_4]$, 
then the relation~\eqref{eq:anomaly_theta_period} becomes 
\begin{equation}
    Z^{(n_{34})}_{\theta+2\pi}[B_{2\diff},A_3,A_4]
    =\exp\left( \im\, n_{34}\int_{M_2} B_{2\diff}-{\im N\over 2\pi}\int_{M_2}A_3\wedge A_4\right)
    Z^{(n_{34})}_{\theta}[B_{2\diff},A_3,A_4]. 
\end{equation}
This shows that we must take the nontrivial twist $n_{34}\not=0$ in order to have the nontrivial relation between the confinement and the $\theta$ periodicity.  

When we take the periodic compactification, $n_{34}=0$, the $2$d $1$-form symmetry does not couple to the $4$d $\theta$ angle, and the confinement of the Wilson loop does not lead to the multi-branch structure of confining vacuum. 
Even though there still exists a mixed anomaly between the $0$-form center symmetry and the $\theta$ periodicity, the center symmetry is spontaneously broken as we have seen in Sec.~\ref{sec:periodic_T2}, and the anomaly matching is satisfied in a trivial manner.

The story is totally different for the twisted compactification, $n_{34}=1$. The following discussion applies for any twist $n_{34}$ that satisfies $\mathrm{gcd}(n_{34},N)=\pm 1$. 
In these cases, the $4$d 't~Hooft anomaly is kept intact as much as possible under the $T^2$ compactification, and there remains the mixed 't~Hooft anomaly between $2$d $\mathbb{Z}_N^{[1]}$ symmetry and the $\theta$ periodicity. 
The fractional $\theta$ dependence must appear if the Wilson loop obeys the area law to satisfy the anomaly matching. 
Moreover, at $\theta=\pi$, this leads to the mixed 't~Hooft anomaly (or global inconsistency) between $(\mathbb{Z}_{N}^{[1]})_{2\rmd}$ and the $CP$ symmetry, which requires the doubly degenerate ground states at $\theta=\pi$. 
As the Wilson loops play the role of the domain-wall defect connecting these vacua, the degeneracy of ground states implies deconfinement for one of the Wilson loops at $\theta=\pi$, as we have seen in Fig.~\ref{fig:tensions}. 
In our semiclassical analysis in Sec.~\ref{sec:center_vortex_theory}, both of these phenomena are caused by center vortices, and this fact corroborates the close connection between confinement and the fractional $\theta$ dependence.

%-------------------------------------------------------
\subsection{Quantum mechanics with compactification of another direction}
\label{sec:QM_T3R}

In this section, we consider another compactification by taking $M_2=\mathbb{R}\times S^1$, and discuss the quantum mechanics by regarding $\mathbb{R}$ as the time direction. 
This is useful to understand the Hamiltonian picture of the center-vortex induced semiclassical confinement. 
Moreover, this clarifies the relation between the center-vortex theory discussed in Sec.~\ref{sec:center_vortex_theory} and the work~\cite{Yamazaki:2017ulc} by M.~Yamazaki and K.~Yonekura. 

The spacetime manifold in this section is given by 
\begin{equation}
    M_4=\mathbb{R}\times (S^1)_A\times \underbrace{(S^1)_B\times (S^1)_C}_{n_{BC}=1}, 
\end{equation}
with
\begin{equation}
    L_s=L_B=L_C\ll L_l=L_A\, (\,\ll \Lambda^{-1}). 
\end{equation}
We note that the new $S^1$ direction is much larger than the original $T^2$, but it is still much smaller than $\Lambda^{-1}$. 
Let us denote the holonomy along the new $S^1$ direction as $P_A$.

We note that the holonomies, $P_B=P_3$ and $P_C=P_4$, along $T^2$ are already determined as \eqref{unbroken}, $P_B=S$ and $P_C=C$. 
The flatness condition shows $P_B P_A=P_A P_B$ and $P_C P_A=P_A P_C$, and thus we obtain 
\begin{equation}
    P_A=\rme^{2\pi \im m/N}\bm{1}, 
\end{equation}
with $m=0,1,\ldots, N-1$. 
Therefore, the center symmetry that acts on $P_A$ is spontaneously broken at the classical level, even though center symmetries acting on $P_B, P_C$ are unbroken.
We denote these classically center-broken states as $|m\rangle$:
\begin{equation}
    {1\over N}\tr(P_A)|m\rangle = \rme^{2\pi \im m/N} |m\rangle. 
    \label{eq:QM_centerbroken}
\end{equation}

An important question is if this broken center symmetry is restored by quantum effects. 
Here, the center vortex, or the fractional instanton, again plays an important role. 
In Sec.~\ref{sec:center_vortex_theory}, we have reviewed the numerical study~\cite{Gonzalez-Arroyo:1998hjb, Montero:1999by, Montero:2000pb} of self-dual configurations on $\mathbb{R}^2\times T^2$ with 't~Hooft flux, but there are also numerical studies on fractional instantons in $\mathbb{R}\times T^3$~\cite{GarciaPerez:1989gt, GarciaPerez:1992fj, Itou:2018wkm}. 
The transition amplitude of one fractional (anti-)instanton gives 
\begin{equation}
    \langle n|\exp(-T\widehat{H}_{\mathrm{YM}}) | m \rangle\sim \delta_{nm}+TL_l K \rme^{-S_\mathrm{I}/N}\left(\rme^{\im \theta/N} \delta_{n,m+1}+\rme^{-\im \theta/N} \delta_{n,m-1}\right), 
\end{equation}
where $\widehat{H}_{\mathrm{YM}}$ is the Hamiltonian of YM effective quantum mechanics and the indices in the Kronecker delta is understood in mod $N$. 
We can understand the second term as the contribution of the center vortex when $L_A$ is large. 
To see it, let us consider a single center vortex on a cylinder $M_2=\mathbb{R}\times S^1$. By moving $P_A$ across the center vortex, $P_A$ gets an extra phase, $\rme^{\pm 2\pi\im /N}$, as a result of the commutation relation between the center vortex and the Wilson loop. 
This shows that the center vortex gives the transition amplitude from $|m\rangle$ to $|m\pm 1\rangle$ as represented by the second term.

The eigenstate of this Hamiltonian is given by 
\begin{equation}
    \widetilde{|k\rangle}={1\over \sqrt{N}}\sum_{m=0}^{N-1}\rme^{2\pi \im km/N}|m\rangle. 
    \label{eq:correct_vac}
\end{equation}
We can explicitly check that this is an eigenstate as follows:
\begin{align}
    &\quad\exp(-T\widehat{H}_{\mathrm{YM}})\widetilde{|k\rangle} \nonumber\\
    &={1\over \sqrt{N}}\sum_{n,m}\rme^{2\pi \im km/N}|n\rangle \langle n|\exp(-T \widehat{H}_{\mathrm{YM}})|m\rangle\nonumber\\
    &\sim{1\over \sqrt{N}}\sum_{n,m}\rme^{2\pi \im km/N}|n\rangle \left\{\delta_{nm}+TL_l K\rme^{-S_\mathrm{I}/N}\left(\rme^{\im \theta/N} \delta_{n,m+1}+\rme^{-\im \theta/N} \delta_{n,m-1}\right)\right\}\nonumber\\
    &={1\over \sqrt{N}}\sum_n \rme^{2\pi\im kn/N}|n\rangle \left(1+2TL_l K \rme^{-S_\mathrm{I}/N}\cos{\theta-2\pi k\over N}\right)\nonumber\\
    &\sim \exp\left(2T L_l K \rme^{-S_\mathrm{I}/N}\cos{\theta-2\pi k\over N}\right)\widetilde{|k\rangle}. 
\end{align}
This also confirms that the energy eigenvalue $L_l E_k(\theta)$ is given by \eqref{eq:multi_branch}, and thus we obtain the same expression for both $L_l\ll\Lambda^{-1}$ and $L_l\gg \Lambda^{-1}$. 
This suggests that we do not encounter the phase transition for the size of the new compactified direction, $(S^1)_A$. 
The ground state is unique at generic values of $\theta$, and there is a level crossing at $\theta=\pi$ (see also Refs.~\cite{Witten:1982df, Cox:2021vsa}). 
We can readily check that 
\begin{equation}
    \widetilde{\langle k'|}{1\over N}\tr(P_A)\widetilde{|k\rangle}=\delta_{k',k+1}, 
    \label{eq:UA_element}
\end{equation}
and thus the ground state expectation value vanishes, $\widetilde{\langle k|}{1\over N}\tr(P_A)\widetilde{|k\rangle}=0$. The center symmetry is restored by fractional instantons. 
This also shows that the two-point function of $\tr(P_A)$ shows the exponential decay along the imaginary-time direction, and the exponent is given by $L_l(E_{1}(\theta)-E_0(\theta))=L_l(E_{0}(\theta+2\pi)-E_0(\theta))$, which is nothing but the string tension $L_l T_1(\theta)$ of the fundamental Wilson loop. 

In the work~\cite{Yamazaki:2017ulc} by Yamazaki and Yonekura, a different limit has been considered:
\begin{equation}
    L_A,L_B\ll L_C\, (\,\ll \Lambda^{-1}), 
\end{equation}
while the 't~Hooft flux is still inserted along the $BC$ directions, $n_{BC}=1$. 
They show that the effective theory on $\mathbb{R}\times (S^1)_C$ is basically given by the $\mathbb{Z}_N$-twisted $\mathbb{C}P^{N-1}$ sigma model that has been discussed in~\cite{Dunne:2012ae, Dunne:2012zk, Misumi:2014jua, Misumi:2016fno}, while the $\mathbb{C}P^{N-1}$ target space has several singularities. 
The fractional instantons in $4$d Yang-Mills theory with the 't~Hooft twist can be mapped to the fractional instantons of $2$d $\mathbb{Z}_N$-twisted $\mathbb{C}P^{N-1}$ model, and we can perform the semiclassical analysis without infrared divergence in the same way. 
Therefore, it would be natural to expect that these two regimes are adiabatically connected, and they provide a promising way of infrared regularization while keeping the confinement property of $4$d Yang-Mills theory.

%--------------------------------------------------------
%--------------------------------------------------------
\section{Supersymmetric Yang-Mills theory on \texorpdfstring{$M_2 \times T^2$}{M2xT2} with 't Hooft flux}
\label{sec:SYM}

Let us add the single Weyl fermion $\lambda$ in the adjoint representation, and we consider the $T^2$ compactified theory with the 't~Hooft flux. 
This theory has $\mathcal{N}=1$ supersymmetry at the massless point, and the Lagrangian of $\mathcal{N}=1$ super Yang-Mills (SYM) theory is given by 
\begin{align}
S_{\mathrm{SYM}}=S_{\mathrm{YM}} + {2\im\over g^2}\int \tr(\overline{\lambda} \overline{\sigma}^\mu  D_{\mu}  \lambda) \;,
\label{Lag}
\end{align} 
where   $ D_{\mu} \lambda   = \partial_\mu  \lambda     + \im\, [a_{\mu}, \lambda] $.

\subsection{Discrete chiral symmetry and anomaly matching}

The classical Lagrangian~\eqref{Lag} has an Abelian chiral symmetry, $\lambda\to \rme^{\im \alpha }\lambda$ and $\overline{\lambda}\to \overline{\lambda}\rme^{-\im \alpha }$. 
However, this classical symmetry is absent due to the Adler-Bell-Jackiw (ABJ) anomaly, and the path integral measure is transformed as 
\begin{equation}
    \Diff \overline{\lambda}\Diff \lambda\to \Diff \overline{\lambda}\Diff \lambda\exp\left({\im}{2N\alpha\over 8\pi^2}\int \tr(F^2)\right).
\end{equation} 
Since the topological charge is quantized to integers in closed spacetimes, there still exists the discrete chiral symmetry,
\begin{equation}
    (\mathbb{Z}_{2N})_\chi:\lambda \to \rme^{{2\pi\over 2N}\im}\lambda,\quad \overline{\lambda}\to \overline{\lambda}\rme^{-{2\pi\over 2N}\im}. 
\end{equation}
The $\mathbb{Z}_2$ subgroup of the chiral symmetry $(\mathbb{Z}_{2N})_\chi$ is identical to the fermion parity, and thus it cannot be spontaneously broken as long as the rotational symmetry is unbroken. 

Computation of the Witten index, $\tr(-1)^F=N$, shows that there are at least $N$ vacua, and it is a natural guess to conclude it comes from the spontaneous chiral symmetry breaking~\cite{Witten:1982df}
\begin{equation}
    (\mathbb{Z}_{2N})_\chi \xrightarrow{\mathrm{SSB}} \mathbb{Z}_2, 
    \label{eq:chiralSSB_SYM}
\end{equation}
and these $N$ vacua are specified by the chiral condensate,
\begin{equation}
    \bigl\langle \tr(\lambda \lambda)\bigr \rangle_k\sim \Lambda^3 \rme^{\im (\theta-2\pi k)/N},
    \label{eq:SYM_condensate}
\end{equation}
where $k=0,1,\ldots, N-1$. 
The fractional $\theta$ dependence of the chiral condensate is required for consistency with the spurious symmetry, $\lambda\to \rme^{\im \alpha}\lambda$, $\overline{\lambda}\to \overline{\lambda}\rme^{-\im \alpha}$, and $\theta\to \theta+2N\alpha$. 
This has various supporting evidences, and we here comment on the recent result of anomaly matching~\cite{Shimizu:2017asf, Komargodski:2017smk} as it is crucial for the following discussion on the compactified theory. 

Let us again introduce the $\mathbb{Z}_N$ $2$-form gauge field $B_{4\diff}$ as a background gauge field for the $4$d $1$-form symmetry $\mathbb{Z}_N^{[1]}$. 
Performing the discrete chiral transformation, $\lambda\to \rme^{{2\pi\over 2N}\im}\lambda$ and $\overline{\lambda}\to \overline{\lambda}\rme^{-{2\pi\over 2N}\im}$, the fermion path-integral measure changes as 
\begin{align}
    \Diff \overline{\lambda}\Diff \lambda
    &\to \Diff \overline{\lambda}\Diff \lambda
    \exp\left({\im}{2N\,{2\pi\over 2N}\over 8\pi^2}\int_{M_4} \tr\Bigl((\tilde{F}-B_{4\diff})^2\Bigr)\right)\nonumber\\
    &= \exp\left(-{\im N\over 4\pi}\int_{M_4} B_{4\rmd}\wedge B_{4\rmd}\right)\Diff \overline{\lambda}\Diff \lambda.
\end{align}
Here, the term that involves the dynamical gauge fields $\tilde{F}$ disappears due to the integer quantization of the $U(N)$ topological charge. 
As a result, we find that there is a mixed 't~Hooft anomaly between $\mathbb{Z}_{N}^{[1]}$ and $(\mathbb{Z}_{2N})_\chi$ that is characterized as 
\begin{equation}
    (\mathbb{Z}_{2N})_\chi: Z_{\mathrm{SYM}}[B_{4\diff}]\to 
    \exp\left(-{\im N\over 4\pi}\int_{M_4} B_{4\rmd}\wedge B_{4\rmd}\right) Z_{\mathrm{SYM}}[B_{4\diff}]. 
    \label{eq:anomaly_SYM_a}
\end{equation}
As a consequence, if the Wilson loop shows the area law for SYM, its discrete chiral symmetry should be maximally broken as \eqref{eq:chiralSSB_SYM} to satisfy the anomaly matching condition.

The presence of  't~Hooft anomaly~\eqref{eq:anomaly_SYM_a} says that the SYM partition function $Z[B_{4\diff},A_\chi]$ with both the $\mathbb{Z}_N$ $2$-form gauge field $B_{4\diff}$ and the $(\mathbb{Z}_{2N})_\chi$ gauge field $A_\chi$ cannot be gauge invariant. 
However, if we introduce the $5$d bulk topological action, 
\begin{equation}
    S_{5\diff, M_5}[B_{4\diff}, A_\chi]={2\pi\over N}\int_{M_5} {2N\over 2\pi}A_\chi\wedge {N^2\over 8\pi^2}B_{4\diff}\wedge B_{4\diff},
    \label{eq:anomaly_SYM}
\end{equation}
with $\partial M_5=M_4$, then the following combination, 
\begin{equation}
    Z_{\mathrm{SYM}}[B_{4\diff},A_\chi]\exp\left(\im\, S_{5\diff,M_5}[B_{4\diff},A_\chi]\right),
\end{equation}
is gauge invariant by anomaly inflow. 
The information of 't~Hooft anomaly is completely summarized into the $5$d topological action~\eqref{eq:anomaly_SYM}. 
We note that this $5$d topological action is quantized in $2\pi/N$, and it is gauge invariant under the large gauge transformations mod $2\pi$ on spin manifolds so that $\exp(\im\, S_{5\diff,M_5})$ is well defined when $M_5$ is a closed spin manifold.

Let us now consider the $T^2$ compactification: $M_4=M_2\times T^2$. The 't~Hooft flux on $T^2$ is given by $n_{34}\in \mathbb{Z}_N$, and we keep its dependence for a while.  
The $4$d $\mathbb{Z}_N$ $1$-form symmetry splits  into $(\mathbb{Z}_N^{[1]})_{2\diff}\times \mathbb{Z}_N^{[0]}\times \mathbb{Z}_N^{[0]}$, 
and $B_{4\diff}$ is given by \eqref{eq:4d2d_gauge} using the gauge fields $B_{2\diff},A_3,A_4$ on $M_2$ and the 't~Hooft twist. 
We can easily compute the 't~Hooft anomaly of its $2$d effective theory by computing the $5$d topological action with this setup. 
We take the $5$d bulk as $M_5=M_3\times T^2$ with some $M_3$ that satisfies $\partial M_3=M_2$, and we extend $2$d gauge fields to $M_3$. 
It gives the $3$d topological action on $M_3$ as follows:
\begin{align}
    S_{5\diff, M_3\times T^2}[B_{4\diff}, A_\chi]
    &={2N^2\over 8\pi^2}\int_{M_3\times T^2} A_\chi\wedge B_{4\diff}\wedge B_{4\diff}\nonumber\\
    &={2N^2\over 8\pi^2}\int_{M_3\times T^2} A_\chi\wedge 2\left({2\pi n_{34}\over N} B_{2\diff} -A_3\wedge A_4\right)\wedge {\diff x_3\diff x_4\over L^2}\nonumber\\
    &={2N n_{34}\over 2\pi}\int_{M_3} A_\chi\wedge B_{2\diff}-{2N^2\over (2\pi)^2}\int_{M_3}A_\chi\wedge A_3\wedge A_4. 
    \label{eq:anomaly_SYM_T2}
\end{align}
The first term describes the mixed anomaly between $(\mathbb{Z}_N^{[1]})_{2\diff}$ and $(\mathbb{Z}_{2N})_\chi$, and the second one describes the mixed anomaly between $\mathbb{Z}_N^{[0]}\times \mathbb{Z}_N^{[0]}$ and $(\mathbb{Z}_{2N})_\chi$. 
As in the case of pure YM theory, the first term exists only if we take the nontrivial twist, $n_{34}\not=0$. 

When $n_{34}=0$, the result crucially depends on the choice of the fermion boundary condition along $T^2$. 
When the boundary condition maintains supersymmetry, the perturbative effective potential~\eqref{pot} is canceled between the gluon and gluino contributions, and thus holonomies have the flat direction at the perturbative level. 
At the nonperturbative level, we get $N$ degenerate vacua as the partition function in this case is identical to the Witten index. This is consistent with the spontaneous chiral symmetry breaking. 
Therefore, the anomaly matching for the second term of \eqref{eq:anomaly_SYM_T2} is satisfied by chiral symmetry breaking, while the center symmetry is unbroken. 
This is an interesting situation, but we do not go into its details in this paper. 

When we take the supersymmetry-breaking boundary condition with $n_{34}=0$, then the fermions do not have the zero modes and the perturbative potential~\eqref{pot} prefers the center-broken vacua. 
Therefore, the 't~Hooft anomaly is trivially satisfied by spontaneous breaking of center symmetry. 

In the next section, we consider the case with a nontrivial 't~Hooft flux, $n_{34}=1$. 
In this case, there is a mixed anomaly between $(\mathbb{Z}_N^{[1]})_{2\rmd}$ and $(\mathbb{Z}_{2N})_\chi$, which means that the Wilson loop $W_{\calR}(C)$ in $M_2$ and the chiral condensate operator $\tr(\lambda\lambda)$ have the $\mathbb{Z}_N$ mutual statistics. 
The same 't~Hooft anomaly appears in the $1$-flavor charge-$N$ massless Schwinger model~\cite{Anber:2018jdf, Anber:2018xek, Armoni:2018bga, Misumi:2019dwq, Honda:2021ovk}, and the anomaly matching concludes both the spontaneous chiral symmetry breaking and the deconfinement of the Wilson loop. 
We shall see how the center-vortex theory reproduces the consequence of anomaly matching condition in the semi-classical manner. 

\subsection{Center vortex and discrete chiral symmetry breaking}

As we have done in Sec.~\ref{sec:QM_T3R}, let us compactify three directions, $M_4=\mathbb{R}\times (S^1)_A\times (S^1)_B \times (S^1)_C$, with 
\begin{equation}
    L_s=L_B=L_C\ll \Lambda^{-1}, 
\end{equation}
and we introduce the nontrivial 't~Hooft flux, $n_{BC}=1$, along small $T^2=(S^1)_B\times (S^1)_C$. 
We consider the case where $(S^1)_A$ is much larger than small $T^2$, 
\begin{equation}
    L_l=L_A\gg L_s.
\end{equation}
In Sec.~\ref{sec:QM_T3R}, $L_l\ll \Lambda^{-1}$ is also assumed, but we have shown that there is no phase transition for $L_l$. Therefore, we no longer have to impose this extra assumption. 
As we have obtained in \eqref{unbroken}, holonomies along $B,C$ directions are given by 
$P_B=S$ and $P_C=C$. 
We set the boundary condition for the adjoint fermion $\lambda$ as
\begin{align}
    \lambda(\bm{x},x_3+L_s,x_4)&=S^{-1} \lambda(\bm{x},x_3,x_4) S,\nonumber\\
    \lambda(\bm{x},x_3,x_4+L_s)&=C^{-1} \lambda(\bm{x},x_3,x_4)C, 
    \label{eq:twistedBC_gluino}
\end{align}
which is identical to that of gauge fields~\eqref{eq:twistedBC_gauge}. 
As a result, both the gauge field and gluino have mass gap, given in \eqref{gluonspectrum}.

We regard $\mathbb{R}$ as the imaginary-time direction, and then the classical vacua can be characterized by the holonomy $P_A$ along $(S^1)_A$. 
Since $P_A$ has to commute with $P_B=S$ and $P_C=C$, we have $P_A=\rme^{2\pi \im m/N}\bm{1}$ with some $m=0,1,\ldots, N-1$, and we denote these $N$ classical vacua as $|m\rangle$ as we have done in \eqref{eq:QM_centerbroken}. 
In the case of pure YM theory, the center vortex gives the tunneling amplitude from $|m\rangle $ to $|m\pm 1\rangle$, which resolves the degeneracy of $N$ ground states. 
Under the presence of the massless fermion, this does not happen because of the fermionic zero modes. 
The Atiyah-Singer index theorem tells that the index for the adjoint Dirac operator with the center vortex configuration is given by
\begin{equation}
    \mathrm{Index}(\im \slashed{D}_{\mathrm{adj}})=2NQ_{\mathrm{top}}=\pm 2, 
\end{equation}
so there are two fermionic zero modes. 
As a consequence, 
\begin{equation}
    \langle m\pm 1| \exp(-T\widehat{H}_{\mathrm{SYM}}) |m\rangle =0 
    \label{tunneling}
\end{equation}
for SYM, and there are $N$ ground states even at the quantum level. 
This is consistent with the fact that the Witten index is given by $\tr(-1)^F=N$ for the $\mathcal{N}=1$ $SU(N)$ SYM theory.

Although the center vortex does not contribute to the tunneling, it gives the chiral condensate. 
Indeed, as the chiral condensate operator $\tr(\lambda \lambda)$ has the axial charge $2$, two zero modes associated with the center vortex are soaked up by this operator (see also \cite{Cohen:1983fd}).  Therefore,
\begin{equation}
    \langle m'| \tr(\lambda \lambda)|m\rangle \sim K' \rme^{-S_\mathrm{I}/N}\rme^{\im \theta/N}\delta_{m',m+1}, 
\end{equation}
where $K'$ is some factor that comes from the fluctuation determinant both of gluon and gluino and from the zero-mode wave function. 
By dimensional analysis, we can estimate $K'\sim 1/L_s^3$ but  careful analysis will be required to determine its details. 
Since the first coefficient of the beta function is now given by ${11\over 3}N-{2\over 3}N=3N$, 
we get $\rme^{-S_\mathrm{I}/N}\sim (\Lambda L_s)^3$ so that 
\begin{equation}
    \langle m+1| \tr(\lambda \lambda)|m\rangle \sim \Lambda^3\rme^{\im \theta/N}.  
\end{equation}
We can diagonalize the chiral condensate operator by using $\widetilde{|k\rangle}$ given in \eqref{eq:correct_vac}, and obtain 
\begin{align} 
\widetilde{\langle k |}     \tr (\lambda \lambda) \widetilde{|  k \rangle}
\sim \Lambda^3 \rme^{\im (\theta-2\pi k)/N}.   
\end{align}
Remarkably, we can reproduce the chiral condensate~\eqref{eq:SYM_condensate} in the semiclassical analysis using center vortices with small $L_s\ll \Lambda^{-1}$, and we here verify that $N$ degenerate vacua is a consequence of spontaneous chiral symmetry breaking. 
We also note that this is consistent with the requirement of anomaly matching condition. 

In order to see more details of 't~Hooft anomaly in this semiclassical description, let us analyze the Wilson loop. 
In the chiral basis $\widetilde{|k\rangle}$, the holonomy ${1\over N}\tr(P_A)$ is no longer diagonal, and instead we have \eqref{eq:UA_element}, 
\begin{align}
\widetilde{\langle k'|} {1\over N}\tr(P_A) \widetilde{|k\rangle}=\delta_{k',k+1}. 
\end{align}
Since $k$ determines the fractional phase of the chiral condensate, this shows the nontrivial commutation relation between the Wilson loop $\tr(P_A)$ and the chiral condensate $\tr(\lambda\lambda)$:
\begin{equation}
    \tr(\lambda \lambda) \tr(P_A) = 
    \rme^{-2\pi \im /N}\tr(P_A) \tr(\lambda \lambda). 
\end{equation}
This mutual statistics is exactly the consequence of the mixed 't~Hooft anomaly between $(\mathbb{Z}_N^{[1]})_{2\rmd}$ and $(\mathbb{Z}_{2N})_\chi$ in $2$d with $n_{34}=n_{BC}=1$, and the same algebra has been observed for massless charge-$N$ Schwinger model~\cite{Anber:2018jdf, Anber:2018xek, Armoni:2018bga, Misumi:2019dwq, Honda:2021ovk}. 
At low energies in this $T^2$ compactified setup, sufficiently large Wilson loops in $M_2$ can be identified as the generator of discrete chiral symmetry.

As the Wilson loop is identified with the chiral symmetry generator that is spontaneously broken, energy densities inside and outside the loop are the same. 
This means that the Wilson loop in $M_2$ shows the perimeter law, and the system is 
deconfined.\footnote{This can be understood semi-classically as follows.  Since the center-vortices have fermionic zero modes, they do not 
lead to confinement,  unlike the situation in  pure Yang-Mills  on $\mathbb R^2 \times T^2$.  
 This is similar to the fact that monopole-instantons in ${\cal N}=1$ SYM  on $\R^3 \times S^1$ do not generate confinement either due to  their fermionic zero modes.  On the other hand, on 
$\mathbb R^3 \times S^1$,  correlated events such as magnetic bion (without fermionic zero modes, but with a non-vanishing magnetic charge) lead to confinement. One may wonder if  correlated events of vortices  on $\mathbb R^2 \times T^2$  can generate confinement. The answer to this question  is negative. Correlated events without fermionic zero modes have trivial mutual statistics with Wilson loops, and they cannot disorder it. Let us emphasize that deconfinement in our $T^2$ compactified theory is required by the 't~Hooft anomaly matching, and it is not an artifact of our semiclassical center-vortex analysis.} 
This result may naively look in contradiction with adiabatic continuity. A version of adiabatic continuity asserts that as long as the  $\Z_N^{[0]} \times \Z_N^{[0]}$ part of the center-symmetry is unbroken, the dynamics of the theory on $\R^2 \times T^2$  is smoothly connected to infinite volume $\R^4$ limit.  
Indeed, chiral condensate matches exactly as expected. But the discussion of the Wilson loops is more subtle. 
Does the perimeter law of the Wilson loop in $M_2$ imply the breakdown of adiabatic continuity?

We claim that this is not the case. 
In the $4$d case, the perimeter law of the Wilson loop implies the appearance of topological order, and thus the number of vacua on closed spatial manifolds can be different between the confined and deconfined theories. 
This is why confined and deconfined theories with the $1$-form symmetry has to be separated by quantum phase transitions. 
In our $T^2$ compactified setup of SYM theory with the 't~Hooft flux, however, spontaneous breakdown of $(\mathbb{Z}_N^{[1]})_{2\rmd}$ does not lead extra vacua. 
The number of ground states are just $N$, which comes from the spontaneous chiral symmetry breaking, and this is a special feature of $2$d field theories with mixed anomaly between $0$-form and $1$-form symmetries. 
As the large-$T^2$ and small-$T^2$ theories have the same number of vacua, these two regimes can be adiabatically connected.

\subsection{Mass deformed theory} 

Let us now turn on   a soft mass term in Lagrangian, 
\begin{equation}
    \Delta {\cal L}_m =  \frac{m}{g^2}  \left( \tr (\lambda \lambda) + \tr(\overline{\lambda}\,\overline{\lambda})\right). 
\end{equation} 
The mass deformation breaks both ${\cal N}=1$ supersymmetry and 
$\Z_{2N}^{[0]}$ chiral symmetry softly. 
In the limit $m\to \infty$, we obtain the pure YM theory and thus we can consider how the semiclassical computations for SYM can be related to those of pure YM. 

In the massless theory, tunneling by the center vortex is prohibited due to the fermionic zero modes, but now those zero modes can be soaked up by the mass term. 
At the leading order of mass perturbation, the energy densities of different branches are given by 
\begin{align}
   E_k(\theta) &= -L_s^2 m\Bigl(\langle\tr(\lambda \lambda)\rangle_k+\mathrm{c.c.}\Bigr) \nonumber\\
   &\sim  -2   m  \Lambda (\Lambda L_s)^2 \cos  \frac{ \theta- 2 \pi k} {N},  
  \end{align} 
The string tensions  in the semi-classical domain take the form of  
 \begin{align}
T_\calR (\theta) = 2 (m\Lambda)(\Lambda L_s)^2 \left\{\cos\left({\theta\over N}\right)-\cos\left({\theta+2\pi |\mathcal{R}| \over N}\right)\right\}. 
\end{align}
We note that these have the same $\theta$ dependence with those of pure YM theory on $\mathbb{R}^2\times T^2$ obtained in \eqref{thetadep}. 
This is somewhat expected as we obtain the pure YM theory by taking the massive limit, $m\to \infty$, so this suggests that there is no phase transition as a function of the fermion mass in this semiclassical center-vortex theory. 

For $m= 0$, the string tensions vanish as we have discussed in the previous section. 
For small fermion mass, the structure of the string tension is proportional to $m$, which is similar to massive charge-$N$ Schwinger model.     
Note, however, that this is not the expected behavior on the $\R^4$ thermodynamic limit, where we expect the tension is proportional to $\Lambda^2$ even for $m=0$. 
This peculiarity is related to the fact that  the long distance limit of the massless theory on $M_2 \times T^2$ is a deconfined topological theory, and it is forced by the presence of 't~Hooft anomaly that involves $(\mathbb{Z}_N^{[1]})_{2\rmd}$.

\section{'t Hooft flux in QCD  with fundamental fermions}
\label{sec:QCD_fund}

Since $SU(N)$ YM theories with adjoint matters possess the $\mathbb Z_N^{[1]}$ symmetry, one can turn on a 2-form background field associated with it. 
We have constructed the semiclassical description of confinement based on center vortex using this property. 
However, once  fundamental fermions are introduced, the theory no longer has the $\mathbb Z_N^{[1]}$ symmetry, and naively, one cannot impose 't Hooft twisted  boundary conditions~\cite{tHooft:1981sps}. 
Does this imply that the semiclassical center-vortex theory is not applicable to QCD with fundamental quarks?

Fortunately, we shall see that there are several ways to circumvent this obstacle in the case of QCD, and it is possible to effectively introduce the 't~Hooft flux in the gluon sector. 
The key ingredient is that the symmetry group that acts properly on local gauge-invariant operators does not act properly on quark fields. 
In other words, the flavor symmetry group for the quark fields has an overlap with the $SU(N)$ gauge redundancy, and the correct symmetry group is the one divided by the common center, $\mathbb{Z}_N$. 
As a result, the correct symmetry group for QCD with $N_f$ fundamental massless quarks is given by~\cite{Tanizaki:2018wtg} 
\begin{equation}
    G={SU(N_f)_{\rmL}\times SU(N_f)_{\rmR}\times U(1)_{\rmq}\over \mathbb{Z}_N\times \mathbb{Z}_{N_f}}\, . 
    \label{eq:QCD_symmetry}
\end{equation}
Here, $U(1)_{\rmq}$ is the quark number symmetry, and the baryon number symmetry is given by $U(1)_{\rmB}=U(1)_{\rmq}/\mathbb{Z}_N$. 
This subtlety of the symmetry group is responsible for introducing the 't~Hooft flux under the presence of fundamental quarks. Below, we shall consider two different ways to introduce minimal 't Hooft flux, 
by activating  either $SU(N_f)_\rmV$ flavor symmetry when $N_f=N$, or  the $U(1)_\rmB$ magnetic flux 
background for general $N_f$~(see Fig.~\ref{fig:twist}~(b) and (c), respectively).

It turns out that the semiclassical center-vortex calculation for QCD with fundamental quarks is very similar to that of charge-$N$ Abelian Higgs model with massless charge-$N$ fermion, which is discussed in Appendix~\ref{sec:AbelianHiggs_masslessfermion}, and it would be useful to compare these two.

\subsection{\texorpdfstring{$SU(N_f)$}{SU(Nf)}-twisted QCD on \texorpdfstring{$T^2$}{T2} for \texorpdfstring{$N_f=N$}{Nf=N} flavors}
\label{sec:QCD_flavor_twist}

Let us consider the case when the number of color and flavor are the same, $N_f=N$. 
In this case, both the color gauge group $SU(N)_{\mathrm{guage}}$ and the flavor symmetry group $SU(N_f)_{\rmV}\subset SU(N_f)_\rmL\times SU(N_f)_\rmR$ have the common center subgroup $\mathbb{Z}_N$. 
Let us denote the quark field $\psi$ as the $N\times N_f$ matrix-valued field, on which $g\in SU(N)_{\mathrm{gauge}}$ acts from the left and $V\in SU(N_f)_\rmV$ acts from the right:
\begin{equation}
    \psi(x)\to g^\dagger \psi(x)  V. 
\end{equation}
The diagonal center element $(\omega,\omega)\in \mathbb{Z}_N\subset SU(N)_{\mathrm{gauge}}\times SU(N_f)_{\rmV}$ for $N_f=N$ acts trivially on the quark fields, so that gauge group and flavor symmetry group are combined as
\begin{equation}
    {SU(N)_\mathrm{gauge}\times SU(N)_\rmV\over \mathbb{Z}_N}\,. 
\end{equation}
Using this feature, when we consider the $T^2$ compactification of QCD with $N_f=N$ flavors, 
the 't~Hooft twist in the color sector can be imposed by a suitable choice of the flavor twist.

Let us consider the flavor-twisted boundary condition for QCD on $\mathbb{R}^2\times T^2$. 
For quark fields, we impose 
\begin{align}
\psi(\bm{x},x_3 +L_3, x_4)&= g_3(x_4)^\dagger  \psi(\bm{x},x_3,  x_4) \Omega_3^F  , \cr
\psi(\bm{x},x_3 , x_4 + L_4)&= g_4(x_3)^\dagger  \psi(\bm{x},x_3,  x_4)  \Omega_4^F , 
\label{fbc}
\end{align}
where $\Omega_3^F,\Omega_4^F\in SU(N)_V$ denotes the twisted boundary condition with the flavor symmetry, and $g_3(x_4), g_4(x_3)$ are the $SU(N)_{\mathrm{gauge}}$ transition functions. 
Let us choose $\Omega_3, \Omega_4$ as the shift and clock matrices, 
\begin{equation}
    \Omega_3^F=S,\quad \Omega_4^F=C, 
\end{equation}
then the 't~Hooft twisted boundary condition for the gauge sector is automatically selected for the quark fields $\psi$ being well-defined:
\begin{equation}
    \psi(\bm{x},L,L) = \left\{
    \begin{array}{cc}
     g_3(L)^\dagger \psi(\bm{x},0,L) \Omega_3^F 
    = g_3(L)^\dagger g_4(0)^\dagger \psi(\bm{x},0,0) \Omega_4^F \Omega_3^F\,, \\
     g_4(L)^\dagger \psi(\bm{x},L,0) \Omega_4^F 
    = g_4(L)^\dagger g_3(0)^\dagger \psi(\bm{x},0,0) \Omega_3^F \Omega_4^F\,.
    \end{array}
    \right.
\end{equation}
Since $\Omega_3^F \Omega_4^F=\rme^{2\pi\im/N}\Omega_4^F \Omega_3^F$ in the above choice, we must require 
\begin{equation}
    g_3(L)^\dagger g_4(0)^\dagger=\rme^{2\pi \im/N}g_4(L)^\dagger g_3(0)^\dagger, 
\end{equation}
which is nothing but the 't~Hooft twisted boundary condition with $n_{34}=1$. 
We can choose the gauge so that $g_3(x_4)=S$ and $g_4(x_3)=C$. 
For sufficiently small $T^2$, the classical action should be minimized and we get $F_{34}=0$. Then, holonomies and transition functions can be identified, and we obtain \eqref{unbroken}, $P_3=S$ and $P_4=C$. 
Let us recall that the remnant gauge group is the center subgroup $\mathbb{Z}_N$, so the non-Abelian part is completely gauge fixed. 

\subsubsection{Discrete 't~Hooft anomaly and chiral effective Lagrangian}

Before studying the dynamics of this $T^2$-compactified QCD, let us discuss its 't~Hooft anomaly that is preserved by the twisted boundary condition~\eqref{fbc} on small $T^2$. 
We note that the symmetry group of $2$d effective theory is given by 
\begin{equation}
    U(1)_\rmB\times (\mathbb{Z}_N)_\rmL \subset G, 
\end{equation}
where $U(1)_\rmB=U(1)_\rmq/\mathbb{Z}_N$ is the baryon number symmetry and $(\mathbb{Z}_N)_\rmL\subset SU(N)_\rmL$ denotes the discrete chiral symmetry. 
This is because the $2$d symmetry must commute with $S,C\in SU(N)_\rmV$ to be consistent with the twisted boundary condition~\eqref{fbc}, and only the Abelian part can satisfy this requirement. 
This suggests that we can concentrate on the discrete anomaly of $4$d QCD that involves $U(1)_\rmB$ and $(\mathbb{Z}_N)_\rmL$, and we should discuss whether it persists under the $T^2$ compactification. 
Indeed, the subgroup,
\begin{equation}
    {SU(N)_\rmV\over \mathbb{Z}_N}\times U(1)_\rmB \times (\mathbb{Z}_N)_\rmL \subset G,
\end{equation}
has the $4$d discrete 't~Hooft anomaly, which is captured by the $5$d topological action~\cite{Tanizaki:2018wtg},
\begin{equation}
    S_{5\rmd}={N\over (2\pi)^2}\int_{M_5} A_\chi\wedge \diff A_\rmB \wedge B_f, 
\end{equation}
where $B_f$ is the $\mathbb{Z}_N$ $2$-form gauge field as a part of the $SU(N)_\rmV/\mathbb{Z}_N$ gauge field, $A_\rmB$ is the $U(1)_\rmB$ gauge field, and $A_\chi$ is the $(\mathbb{Z}_N)_\rmL$ gauge field. 
The twisted boundary condition~\eqref{fbc} corresponds to
\begin{equation}
    \int_{T^2} B_f={2\pi \over N}, 
\end{equation}
and thus the $5$d topological action reduces to the nontrivial $3$d topological action:
\begin{align}
    S_{5\rmd}&={N\over (2\pi)^2}\int_{M_3\times T^2} A_\chi\wedge \diff A_\rmB \wedge B_f
    ={1\over 2\pi}\int_{M_3}A_\chi\wedge \diff A_\rmB. 
    \label{eq:anomaly_flavor_twist}
\end{align}
This is the $3$d symmetry-protected topological state with $U(1)_\rmB\times (\mathbb{Z}_N)_\rmL$, which is nothing but the symmetry group of $2$d effective theory. 
Since the $T^2$-compactified QCD with the flavor twist \eqref{fbc} can be regarded as the boundary of this $3$d symmetry-protected topological state, the anomaly matching requires the $N$ degenerate vacua by the discrete chiral symmetry breaking or the presence of massless spectrum.

Let us briefly examine the prediction of the chiral effective Lagrangian when the torus size is large, $L\gg \Lambda^{-1}$. 
In this case, because of the spontaneous chiral symmetry breaking $SU(N_f)_\rmL\times SU(N_f)_\rmR \xrightarrow{\mathrm{SSB}} SU(N_f)_\rmV$, the low-energy effective theory is described by the chiral Lagrangian, 
\begin{align}
    S_{\chi\mathrm{EFT}}={f_\pi^2 \over 2}\int_{M_4}\tr_{\rm f} [\diff U^\dagger \wedge \star \diff U]+\cdots,
    \label{eq:chiralLag}
\end{align}
where $U$ is the $SU(N_f)$-valued field and $f_\pi$ is the pion decay constant. 
Since $U\sim \overline{\psi_\rmL}\psi_\rmR$, the flavor twist~\eqref{fbc} in terms of clock and shift matrices translates for the chiral field as 
\begin{align}
U(\bm{x},x_3 +L, x_4)&= S^\dagger U(\bm{x},x_3,  x_4)S, \cr
U(\bm{x},x_3 , x_4+L)&= C^\dagger U(\bm{x},x_3,  x_4)C.
\label{ftbc}
\end{align}
In order to satisfy this boundary condition with the constant fields, $U$ is restricted to the center elements, 
\begin{equation}
    \langle U(\bm{x},x_3,x_4)\rangle=\rme^{2\pi \im k/N_f}\bm{1}_{N_f}, 
\end{equation}
with $k=0,1,\ldots, N_f-1$. Therefore, we get $N_f(=N)$ disconnected vacua. 
Because of the twisted boundary condition~\eqref{ftbc}, the $N_f^2-1$ massless Nambu-Goldstone bosons become massive of $O(1/{N_f L})$. 
To see this, one can decompose the pion field given by 
$U= \rme^{\im \Pi/f_\pi}$ as $ \Pi (\bm{x},x_3,x_4)= \sum_{\bm{p}\not=0} \Pi^{(\bm{p})} (\bm{x},x_3,x_4) J_{\bm{p}}$  by using  the basis \eqref{basis1}, but now  the basis is interpreted in terms of flavor. The twisted boundary condition for $\Pi^{(\bm{p})}$ is given by \eqref{tbc2}, which yields the mass spectrum \eqref{gluonspectrum} for pions.  
We now find that the anomaly matching is satisfied by $N$ gapped vacua with discrete chiral symmetry breaking.  
In the next section, we consider if the semiclassical center-vortex theory give the same result or not in the small-$T^2$ regime, $L=L_s\ll \Lambda^{-1}$.

\subsubsection{Semiclassical center-vortex computation with small torus}

In this section, we study the dynamics of $N_f$-flavor massless QCD with $N_f=N$ on small $T^2$ by imposing the $SU(N)_\rmV$-twisted boundary condition~\eqref{fbc}. 
Since the holonomies along $T^2$ directions are given by $P_3=S$ and $P_4=C$ due to the 't~Hooft flux in the gauge sector, the $2$d gauge fields on $M_2$ are completely gapped at the perturbative level. 
We note, however, that the quark fields have one massless components. 
As the quark fields obey 
\begin{equation}
    \psi(x_3+L_s,x_4)=S^\dagger \psi(x_3,x_4)S,\quad 
    \psi(x_3,x_4+L_s)=C^\dagger \psi(x_3,x_4)C, 
\end{equation}
the massless mode exists only for the identity component, i.e. ${1\over N}\tr(\psi)\bm{1}_N$, and other $N^2-1$ components acquire the perturbative mass gap, same as the one given in \eqref{gluonspectrum}   for gluons. 
In the case of SYM theory, the fermion obeys the identical boundary condition, but the gluino field is traceless, and thus there are no perturbatively massless modes. 
This gives the sharp contrast between QCD with $N_f=N$ fundamental quarks and SYM theory in this semiclassical analysis. 

Let us denote the perturbative massless quark field as 
\begin{equation}
    \Psi=L_s\tr(\psi), 
\end{equation}
and then the $2$d effective theory is given by the massless $2$d fermions,
\begin{equation}
    \mathcal{L}=\overline{\Psi}\slashed{\partial} \Psi, 
\end{equation}
where $\slashed{\partial}=\gamma^1\partial_1+\gamma^2\partial_2$ is the $2$d free Dirac operator with $4$d gamma matrices. 
We here note that $\Psi$ still couples to the $\mathbb{Z}_N$ $1$-form gauge field, and thus $\Psi$ itself is not gauge invariant while the baryon operator $\Psi^N$ is gauge invariant. 
This is important to recognize that the global $U(1)$ symmetry of this system is given by $U(1)_\rmB=U(1)_\rmq/\mathbb{Z}_N$ instead of $U(1)_\rmq$, and we can find the correct coefficient for the 't~Hooft anomaly~\eqref{eq:anomaly_flavor_twist}. 
Anyway, within the perturbation theory, we obtain $2$-component $2$d massless Dirac fermion as the low-energy effective theory. 

As a next step, let us perform the semiclassical analysis by taking into account the center vortex configurations. 
Since the center vortex carries the topological charge $Q_{\mathrm{top}}=\pm 1/N$, there are fermionic zero modes according to the index theorem,
\begin{equation}
    2N_f \mathrm{Index}(\slashed{D}_{\mathrm{fund}})=2NQ_{\mathrm{top}}=\pm 2.
\end{equation}
This means that the vertex operator for the center vortex is associated with the chiral operators $\overline{\Psi_\rmL}\Psi_R$, $\overline{\Psi_\rmR}\Psi_\rmL$ of perturbatively massless fermions:
\begin{equation}
    \mathcal{V}(\bm{x})\sim {1\over L_s}\rme^{-S_{\mathrm{I}}/N}\rme^{\im \theta/N} \overline{\Psi_\rmL}\Psi_\rmR(\bm{x}), \quad 
    \mathcal{V}^*(\bm{x})\sim {1\over L_s}\rme^{-S_{\mathrm{I}}/N}\rme^{-\im \theta/N} \overline{\Psi_\rmR}\Psi_\rmL(\bm{x}). 
\end{equation}
We now perform the dilute gas approximation. 
We should note that the total topological charge has to be an integer, and this gives 
\begin{align}
    &\quad \sum_{n,\overline{n}\ge 0}{1\over n!\, \overline{n}!}\left(\int \diff^2\bm{x}\calV \right)^n \left(\int \diff^2 \bm{x}\calV^*\right)^{\overline{n}}\delta_{n-\overline{n}\in N\mathbb{Z}}\nonumber\\
    &=\sum_{k=0}^{N-1}\exp\left[\int \diff^2\bm{x}{1\over L_s}\rme^{-S_{\mathrm{I}/N}}\left(\rme^{\im(\theta-2\pi k)/N}\overline{\Psi_\rmL}\Psi_\rmR+\rme^{-\im(\theta-2\pi k)/N}\overline{\Psi_\rmR}\Psi_\rmL\right)\right]. 
\end{align}
The dilute gas of center vortex gives the nonperturbative mass gap for $\Psi$ and $\overline{\Psi}$. 
Furthermore, the $2$d effective theory decomposes into $N$ distinct sectors, and the complex fermion mass acquires different phases on each sector. 
As a result, the fermion bilinear condensate is given by 
\begin{align}
    \langle \overline{\Psi_\rmL}\Psi_\rmR\rangle
    &\sim {1\over L_s}\rme^{-S_{\mathrm{I}}/N}\rme^{-\im (\theta-2\pi k)/N} \cr
    &\sim \Lambda^3 L_s^2 \rme^{-\im (\theta-2\pi k)/N}. 
\end{align}
Recalling that $\Psi=L_s \tr(\psi)$, this gives the correct behavior of $4$d chiral condensates. 
We obtain $N$ degenerate vacua as a consequence of the discrete chiral symmetry breaking of $(\mathbb{Z}_N)_\rmL$. 
This satisfies the anomaly matching condition, and, moreover, it reproduces the result of the chiral effective Lagrangian for large $T^2$. 
This strongly suggests that, even with fundamental quarks, we can achieve the adiabatic continuity by suitable choice of the boundary condition along $T^2$. 

We note, however, that it is somewhat accidental that we can reproduce the correct magnitude for the chiral condensate, and this coincidence comes from the fact that the leading coefficient of the beta function are the same between SYM theory and QCD with $N_f=N$ fundamental quarks. 
What is important is that qualitative features of vacua are the same between the large-$T^2$ and small-$T^2$ regimes, and thus strongly-coupled $4$d dynamics of QCD may be smoothly connected to the semiclassical description with center vortex without phase transitions.

\subsection{QCD on \texorpdfstring{$T^2$}{T2} with baryon-number magnetic flux}
\label{sec:QCD_baryon_flux}

So far, we have seen that our flavor-twisted boundary condition is useful for $N_f=N$ flavor QCD.
Here, we would like to consider another setup, so that the center-vortex theory is applicable to QCD for any number of flavors $N_f$. 

The idea is to use the baryon number symmetry, $U(1)_\rmB$. 
Let us introduce the minimal $U(1)_\rmB$ magnetic flux along small $T^2$ direction, 
\begin{equation}
    \int_{T^2}\diff A_\rmB=2\pi. 
\end{equation}
To be concrete, let us choose the specific $U(1)_\rmB$ gauge field as 
\begin{equation}
    A_\rmB={2\pi\over L^2}x_3\diff x_4. 
\end{equation}
We note that 
\begin{equation}
    A_\rmB(x_3+L, x_4)=A_\rmB(x_3,x_4)+{2\pi\over L}\diff x_4,  
\end{equation}
and thus there is a $U(1)_\rmB$-valued transition function, $\rme^{\im \alpha(x_4)}$, with the $2\pi$-periodic function $\alpha(x_4)={2\pi\over L}x_4$, when we identify $(x_3+L, x_4)\sim (x_3, x_4)$.

Let us describe the boundary condition for quark fields under this background $A_\rmB$. However, we should note that $U(1)_\rmB=U(1)_\rmq/\mathbb{Z}_N$, so the quark fields have fractional charges $\pm 1/N$ under $U(1)_\rmB$. 
As a result, the fractionalized transition function, $\rme^{\im \alpha(x_4)/N}$, appears in the connection formula, which may seem to be illegal at the first sight. 
Introducing the $SU(N)_\mathrm{gauge}$ transition functions, $g_3(x_4)$ and $g_4(x_3)$, we get 
\begin{align}
\psi(x_3 +L, x_4)&= \rme^{-\im \alpha(x_4)/N}g_3(x_4)^\dagger  \psi(x_3,  x_4), \cr
\psi(x_3 , x_4 + L)&= g_4(x_3)^\dagger  \psi(x_3,  x_4) .
\end{align}
Consistency of the fermion wave function requires that 
\begin{align}
    g_3(L)^\dagger g_4(0)^\dagger
    &=\rme^{\im (\alpha(L)-\alpha(0))/N} g_4(L)^\dagger g_3(0)^\dagger \nonumber\\
    &= \rme^{2\pi \im /N} g_4(L)^\dagger g_3(0)^\dagger. 
\end{align}
This is again nothing but the 't~Hooft twisted boundary condition with $n_{34}=1$ for the gauge sector. 
Turning on the baryon-number magnetic flux, the structure group is given by $(SU(N)_{\mathrm{gauge}}\times U(1)_\rmq)/\mathbb{Z}_N\simeq U(N)$ instead of the naive one, $SU(N)\times U(1)$, and the 't~Hooft flux is inserted through quotient by the common center subgroup.

Let us briefly discuss the implications of turning on the $U(1)_\rmB$ background for the large $T^2$ by using the chiral Lagrangian~\eqref{eq:chiralLag}, assuming spontaneous chiral symmetry breaking in $4$d.  
The chiral field $U$ itself does not transform under $U(1)_\rmB$.  
To describe the effect of the background baryon-number magnetic flux in the chiral Lagrangian,  we must minimally couple $A_{\rm B}$ to the Skyrmion current, $J_{\rm B}={1\over 24\pi^2}\tr_\rmf[(U^{\dagger}\diff U)^3]$,~\cite{Witten:1983tw}:
\begin{align}
\int_{M_4} A_{\rm B} \wedge J_{\rm B}=\int_{M_5}\diff A_\rmB\wedge J_\rmB. 
\label{min_coup}
\end{align} 
Here, we have introduced the auxiliary $5$d bulk, $M_5$, that satisfies $\partial M_5=M_4$ for convenience. 
Consider ${M_5} = M_3\times T^2$, and the boundary of $M_3$ is given by $ \partial M_3=M_2$.
The background $A_{\rm B}$ field correspond to a magnetic monopole whose flux is piercing $T^2$, and thus it gives the level-$1$ Wess-Zumino-Witten (WZW) term, 
\begin{align}
  \int_{M_3\times T^2}\diff A_\rmB\wedge J_B={1\over 12\pi}\int_{M_3}\tr_\rmf [(U^\dagger \diff U)^3]=:\Gamma_{\mathrm{WZW}}[U]. 
\end{align}
Therefore, the chiral Lagrangian is modified by $\Gamma_{\mathrm{WZW}}$, and $2$d effective theory on $M_2$ is described by $2$d level-1  WZW model as a consequence of $4$d perturbative 't~Hooft anomaly for $U(1)_\rmB$-$SU(N_f)_\rmL$-$SU(N_f)_\rmL$. 
Here, let us emphasize that $2$d effective theory is gapless due to the WZW term, $\Gamma_{\mathrm{WZW}}$. 
If the WZW term is absent, the $2$d theory is described by $SU(N_f)$ principal chiral model, which is believed to acquire the nonperturbative mass gap with the unique ground state. 
Therefore, we need to introduce the nonzero $U(1)_\rmB$ flux in order to make the $2$d theory gapless at the nonperturbative level by turning on $\Gamma_{\mathrm{WZW}}$. 

We must examine if this prediction by the chiral Lagrangian can be reproduced by the semiclassical analysis of QCD on small $T^2$. 
For this purpose, we first need to obtain the perturbative massless spectrum. 
Because of the 't~Hooft flux in the gauge sector, the gluons are gapped.  
To find the perturbative zero modes of the quark field, let us solve 
\begin{equation}
    \left[\gamma_3 \partial_3+\gamma_4\left(\partial_4+\im{1\over N}A_{\rmB,4}\right)\right]\psi(x_3,x_4)=0, 
    \label{eq:zeromode1}
\end{equation}
with the boundary condition, 
\begin{equation}
    \psi(x_3+L,x_4)=\rme^{-{2\pi \im\over NL}x_4}S^\dagger \psi(x_3,x_4),\quad 
    \psi(x_3,x_4+L)=C^\dagger \psi(x_3,x_4). 
\end{equation}
Since $(\im\gamma_3\gamma_4)^2=1$, the eigenvalue of $\im\gamma_3\gamma_4$ is given by $\pm 1$, and thus \eqref{eq:zeromode1} can be decomposed as 
\begin{align}
    \left[\partial_3\mp \im \left(\partial_4+\im{2\pi\over NL^2}x_3\right)\right]\psi_{\pm}(x_3,x_4)=0,
    \label{eq:zeromode2}
\end{align}
where $(\im\gamma_3\gamma_4)\psi_{\pm}=\pm  \psi_{\pm}$. 

We can immediately conclude that $\psi_-$ does not have zero modes by noticing the following equation,
\begin{align}
    &\quad \int_{T^2}\diff^2x \left|\left\{\partial_3\mp \im \left(\partial_4+\im{2\pi\over NL^2}x_3\right)\right\}\psi_{\pm}\right|^2\nonumber\\
    &= \int_{T^2}\diff^2x \left(\Bigl|\partial_3 \psi_{\pm}\Bigr|^2
    +\left|\left(\partial_4+\im{2\pi\over NL^2}x_3\right)\psi_{\pm}\right|^2
    \mp{2\pi\over NL^2}\Bigl|\psi_{\pm}\Bigr|^2\right). 
\end{align}
The left-hand-side is $0$ for zero modes. 
For $\psi_{-}$, however, each term on the right-hand-side is positive semi-definite, and thus the zero-mode equation can be satisfied only if $\psi_{-}=0$, which means that zero mode is absent. 
From now on, let us solve the zero modes of $\psi_+$, and we neglect the spin degrees of freedom for a while. 
For $\psi_+$, general solutions of \eqref{eq:zeromode2} take the form of 
\begin{equation}
    \psi_+(x_3,x_4)=\exp\left(-{\pi\over NL^2}x_3^2\right) u(z), 
\end{equation}
where $u(z)$ is an $N$-component holomorphic function of the dimensionless complex coordinate, $z=(x_3-\im x_4)/L$. The boundary condition for $u(z)$ is given by 
\begin{align}
    u(z+1)=\rme^{{\pi\over N}+{2\pi\over N}z}S^\dagger u(z),\quad 
    u(z-\im)= C^\dagger u(z). 
\end{align}
From the second condition, we find the Fourier series, 
\begin{equation}
    u_\ell(z)=\rme^{{2\pi (\ell-1)\over N}z}\sum_{n\in \mathbb{Z}}c_{n,\ell}\, \rme^{2\pi n z}, 
\end{equation}
where $\ell=1,2,\ldots, N$ is the color label. Substituting this into the first condition, we find 
\begin{align}
    c_{n,\ell-1}&=\rme^{2\pi n+{2\pi (\ell-1)\over N}-{\pi\over N}}c_{n,\ell} \quad (\ell=2,\ldots, N),\nonumber\\
    c_{n-1,N}&=\rme^{2\pi n-{\pi\over N}}c_{n,1}. 
\end{align}
From this expression, we obtain $c_{n+1,1}=\rme^{-\pi N(2n+1)}c_{n,1}$, and it can be solved as $c_{n,1}=\rme^{-\pi N n^2}c_{0,1}$. This gives the convergent sum in the Fourier expansion of $\psi_+$, and thus we find the unique zero mode. 
If we repeat the same computation for $\psi_-$, it turns out that the Fourier sum is divergent, and we can confirm that there is no zero mode for $\psi_-$ in the concrete manner. 
Recalling that $\psi_+$ has $2$ spin components, we obtain a single $2$d Dirac fermion for each flavor since these $2$ spin components have opposite $2$d chirality. 
To see this, we note that $2$d and $4$d chiralities for $\psi_+$ are related by $\im \gamma_1\gamma_2=(\im\gamma_1\gamma_2)(\im \gamma_3\gamma_4)=-\gamma_5$. 
As a result, we obtain $2$d $N_f$-flavor massless Dirac fermion as a $2$d effective theory in this perturbative analysis. 

By non-Abelian bosonization, $2$d $N_f$-flavor massless Dirac fermions can be mapped to $2$d level-$1$ $U(N_f)$ WZW model~\cite{Witten:1983ar}, 
\begin{equation}
    S={1\over 8\pi}\int_{M_2}\tr_\rmf(\diff \tilde{U}\wedge \star \diff \tilde{U}^\dagger)+{1\over 12\pi}\int_{M_3}\tr_\rmf[(\tilde{U}^\dagger \diff \tilde{U})^3], 
\end{equation}
where $\tilde{U}$ is a $U(N_f)$-valued field. 
The result is already close to that of the chiral Lagrangian prediction for large $T^2$, and we would like to resolve the  difference between $U(N_f)$ and $SU(N_f)$ WZW models. 
By taking into account the center vortex, we expect that the $U(1)$ axial symmetry is explicitly broken while the $SU(N_f)_\rmL\times SU(N_f)_\rmR$ chiral symmetry is kept intact. 
To be consistent with these requirements and with spurious symmetry, the center-vortex vertex should take the following form,
\begin{equation}
    \Delta S\sim -{1\over L^2}\rme^{-S_{\mathrm{I}}/N}\left(\rme^{\im \theta/N}(\det \tilde{U})^{1/N}+\rme^{-\im \theta/N}(\det \tilde{U}^\dagger)^{1/N}\right). 
\end{equation}
Here, the branch label $k$ is eliminated because these are now connected by the Abelian part of $U(N_f)$ field, $\tilde{U}$, and thus discrete vacua does not appear. 
In other words, the branch label is compensated by the ambiguity of taking the $1/N$-th power of the $U(1)$-valued field $\det(\tilde{U})$. 
By formally regarding the number of color $N$ as a large number, the center-vortex vertex can be approximated as 
\begin{equation}
    \Delta S\sim {\Lambda^2(\Lambda L)^{{5N-2N_f\over 3N}}\over N^2}\left(\im \ln \bigl(\det(\tilde{U})\bigr)-\theta\right)^2, 
    \label{eq:WV_formula}
\end{equation}
and we can obtain the same functional form for the $\eta'$ mass in the large-$N$ QCD with fixed numbers of flavor. 
Indeed, if we neglect the correction of $\mathcal{O}(N_f/N)$ in the coefficient, then the comparison between \eqref{eq:multi_branch} and \eqref{eq:WV_formula} reproduces the Witten-Veneziano formula~\cite{Witten:1979vv,Veneziano:1979ec} that relates the YM topological susceptibility and the $\eta'$ mass. 

We would like to emphasize that the center vortex has resolved the issue of the perturbative analysis. 
Within perturbation theory, there are $N_f$ massless fermions so the $2$d central charge is given by $N_f$, while the $SU(N_f)_1$ WZW model has the central charge $N_f-1$. 
Since the center vortex gives a mass to the $U(1)$ part of $U(N_f)$-valued field, these two descriptions now match with each other.

\section{Concluding remarks and outlooks}

In this paper, we derived the novel semiclassical description of confinement based on the center vortices by putting $4$d gauge theories on small $\mathbb{R}^2\times T^2$ with the 't~Hooft flux. 
These $T^2$ compactifications preserve the 't~Hooft anomaly of $4$d gauge theories, and thus $2$d effective theories and the original $4$d gauge theories are constrained by the same anomaly matching condition. 
We have seen that analytic computations of the semiclassical center-vortex theory give the prediction consistent with the expected behaviors of $4$d strong dynamics.  

We find ourselves in an exciting situation, as we now have two different compactifications of $4$d gauge theories down to small $\R^3 \times S^1$ (with center-stabilization)~\cite{Unsal:2007vu,Unsal:2007jx,Unsal:2008ch,Shifman:2008ja, Davies:2000nw} and to small $\R^2 \times T^2$ (with 't Hooft flux), both of which provide semiclassical descriptions of $4$d gauge theories. We expect that both of them are adiabatically connected to the $\R^4$ limit since we can reproduce the nonperturbative properties such as confinement, chiral symmetry breaking, and the multi-branch structure of vacua.

On small $\mathbb{R}^3\times S^1$, the Polyakov loop in the small $S^1$ direction behaves as an adjoint Higgs field. 
Adding suitable deformations, such as the double-trace deformation or several adjoint fermions with the periodic boundary condition, the $0$-form center symmetry is stabilized and the expectation value of the Polyakov loop induces the adjoint Higgsing, $SU(N)\to U(1)^{N-1}$. 
Then, the dilute gas of monopole-instantons (or magnetic bions in some cases) gives the semiclassical description of confinement. 
In this case, monopole-instantons are exact BPS solutions of the self-dual YM equation and they carry fractional topological charge $Q_{\mathrm{top}}=1/N$. 
This construction has been used in the last fifteen years to address nonperturbative properties of gauge theories, and examples include deformed YM theory, adjoint QCD~\cite{Unsal:2007vu,Unsal:2007jx,Unsal:2008ch,Shifman:2008ja}, and fundamental QCD with the flavor-twisted boundary condition~\cite{Poppitz:2013zqa,Iritani:2015ara,Cherman:2016hcd,Cherman:2017tey} (see also \cite{Kouno:2012zz,Kouno:2013mma,Kouno:2015sja}). 
They provide a realization of the idea of adiabatic continuity. 

It is quite suggestive that both semiclassical descriptions rely on the topological defects with the fractional topological charge, $Q_{\mathrm{top}}=1/N$, although their dimensions are different in $4$d spacetime. 
Moreover, both defects carry magnetic charges so that their liberation causes confinement. 
However, we have to note that there is a difference in detail about their magnetic charges. 
The center vortex has nontrivial mutual statistics with the Wilson loop, and this is why its liberation gives the correct $N$-ality structure in string tensions. On the other hand, monopoles have magnetic charges in root lattice so they only have trivial mutual statistics with Wilson loops. This tells that these two cases have different microscopic mechanisms for confinement despite certain similarities. 
Still, we should be able to make some connections between them by considering the decompactification of one $S^1$ component of small $T^2$ assuming that the confinement always occurs in this process.  
This assumption is violated for pure YM theory, while this is a valid scenario for softy-broken $\mathcal{N}=1$ SYM theory. 
It would be an interesting future study to find the explicit connection between these two semiclassical descriptions of color confinement with some continuous deformations. 

There are multiple interesting open directions to pursue about the semiclassical center-vortex theory. 
Lastly, let us pick up some of them with brief comments:

\begin{itemize}
\item {\bf Analytic solution of the center-vortex configuration:} 
As discussed in Sec.~\ref{sec:center_vortex_theory}, the center vortex on $\mathbb{R}^2\times T^2$ is constructed only numerically~\cite{Gonzalez-Arroyo:1998hjb, Montero:1999by, Montero:2000pb}, and it satisfies the self-dual equation and has $Q_{\mathrm{top}}=1/N$. 
Since the center vortex plays a vital role in our semiclassical descriptions of confinement, it is desirable to have analytic solutions for studying its properties in more detail. 
Given the importance of the problem, it would be nice to devote an effort towards finding either the BPS solutions or their existence proof in nontrivial 't Hooft flux background for fractional topological charge configurations.

\item  {\bf Chiral gauge theory dynamics:} 
Some chiral theories (especially, chiral quiver theories~\cite{Douglas:1996sw,Strassler:2001fs,Shifman:2008cx,Sulejmanpasic:2020zfs})  possess the $\Z_N^{[1]}$ symmetry, so we can turn on the 't~Hooft flux background under $T^2$ compactifications.  
We can apply the setup of this paper to study the nonperturbative dynamics of such $4$d chiral gauge theories. 

\item {\bf Two-index QCD-like theories:} 
Using the baryon number $U(1)_{\rm B}$ magnetic flux background, it should also be possible to explore the dynamics of  QCD with antisymmetric and symmetric representation Dirac fermions (QCD(AS/S)).   
These theories possess, at most, a $\Z_2^{[1]}$ 1-form symmetry, and the use of baryon number background is necessary to impose minimal 't Hooft flux. 
For QCD with bi-fundamental fermions (QCD(BF)), since there is a  $\Z_N^{[1]}$ 1-form symmetry, one can use standard construction.  This may shed new insights into the nonperturbative large-$N$ orbifold and orientifold equivalences between the one-flavor QCD(AS/S/BF) and  ${\cal N}=1$ SYM and the multi-flavor generalization \cite{Armoni:2003gp, Unsal:2006pj} 
\end{itemize}

%%%%%%%%%%   ACKNOWLEDGMENTS   %%%%%%%%%%

\acknowledgments
Y.~T. appreciates useful discussions during the virtual conference ``Paths to Quantum Field Theory'' organized at Durham University on 23-27 August 2021, and especially thanks Tin~Sulejmanpasic for the wonderful invitation.  
The work of Y.~T. was supported by Japan Society for the Promotion of Science (JSPS) KAKENHI Grant-in-Aid for Research Activity Start-up, 20K22350.
The work of M.~\"U. was supported by U.S. Department of Energy, Office of Science, Office of Nuclear Physics under Award Number DE-FG02-03ER41260. 
%---------------------------------------------------------------------------------------

\appendix

\section{Review on several perspectives of 't Hooft flux}
\label{app:tHooft_flux}

In this appendix, we summarize some basic properties of 't~Hooft flux in several viewpoints. 
After giving the definition on generic manifolds using the language of principal bundles, we specialize to the case of $2$-torus $T^2$. 
We discuss it in the continuum formulation first, and explain the same result from the lattice gauge theory. 

\subsection{Principal \texorpdfstring{$SU(N)/\mathbb{Z}_N$}{SU(N)/ZN} bundle and 't~Hooft flux}
\label{sec:definition_tHooft_flux}

Let us assume that we are interested in the $\mathfrak{su}(N)$ gauge theory coupled to adjoint matter fields. 
In this case, we should make a choice if the global structure of the gauge group is whether $SU(N)$ or $PSU(N)=SU(N)/\mathbb{Z}_N$.\footnote{To be more precise, there are other choices $SU(N)/\mathbb{Z}_K$, where $\mathbb{Z}_K\subset \mathbb{Z}_N$ is a subgroup of the center group, and we may also have different choices of discrete theta angles for the same gauge group. 
In this subsection, we neglect those subtleties for simplicity, and it is sufficient for the purpose of this paper.} 
Here, 't~Hooft flux becomes the key to understand these differences in the path-integral quantization. 

First, let us put $SU(N)$ gauge theory with a matter $\phi$ in an irreducible representation $\rho$ on a compact Euclidean manifold $M$. 
To take care of the global data correctly, we introduce a good open cover $\{U_i\}$ of $M$, i.e. $\cup_i U_i=M$, and $U_i$, $U_{ij}=U_i\cap U_j$, etc. are diffeomorphic to the open ball. 
Then, the gauge field $a$ is the collection of the following mathematical data:
\begin{itemize}
    \item $a_i$ is an $\mathfrak{su}(N)$-valued $1$-form on $U_i$.
    \item $g_{ij}:U_{ij}=U_i\cap U_j\to SU(N)$ is a transition function. We set $g_{ji}=g_{ij}^{-1}$. 
    \item On $U_{ij}$, $a_j$ and $a_i$ are related by the gauge transformation with $g_{ij}$: \begin{equation}
    a_j=g_{ij}^{-1}a_i g_{ij}-\im g^{-1}_{ij}\diff g_{ij}.
    \label{eq:connection_gauge}
    \end{equation}
    \item On $U_{ijk}=U_{i}\cap U_{j}\cap U_{k}$, the transition functions $g_{ij}$, $g_{jk}$, and $g_{ki}$ satisfy the cocycle condition, 
    \begin{equation}
        g_{ij}g_{jk}g_{ki}=1. 
        \label{eq:cocycle}
    \end{equation}
\end{itemize}
In this sense, the gauge field $a$ is locally a $1$-form, but it is not necessarily globally well-defined as a $1$-form field, and the subtlety is taken into account by transition functions. 

Similarly, the matter field $\phi$ is not a globally defined function, 
but it has to be regarded as a section of the associated bundle: 
\begin{itemize}
    \item $\phi_i$ is a function on $U_i$.
    \item On $U_{ij}$, $\phi_j$ and $\phi_i$ are related by the gauge transformation, 
    \begin{equation}
        \phi_j=\rho(g^{-1}_{ij})\phi_i. 
        \label{eq:connection_matter}
    \end{equation}
\end{itemize}
We note that the cocycle condition \eqref{eq:cocycle} is a sufficient condition for the well-definedness of $\phi$.
To see this, we consider a triple overlap $U_{ijk}$, and we cyclically relate $\phi_i, \phi_j$ and $\phi_k$ as follows:
\begin{equation}
    \phi_i=\rho(g_{ki}^{-1})\phi_k=\rho(g_{ki}^{-1})\rho(g_{jk}^{-1})\phi_j=\rho(g_{ki}^{-1})\rho(g_{jk}^{-1})\rho(g_{ij}^{-1})\phi_i. 
\end{equation}
As a result, we obtain 
\begin{equation}
    \left[1-\rho((g_{ij}g_{jk}g_{ki})^{-1})\right]\phi_i=0. 
\end{equation}
Since $\rho$ is an irreducible representation, this is true for arbitrary $\phi_i$ if and only if 
\begin{equation}
    \rho(g_{ij}g_{jk}g_{ki})=1. 
    \label{eq:single_valued}
\end{equation}
Since $\rho(1)=1$, we have shown that \eqref{eq:cocycle} is indeed a sufficient condition. 

If $\phi$ is in a defining (or, fundamental) representation, then $\phi$ is valued in $\mathbb{C}^N$, and $\rho(g)\phi=g\cdot \phi$ when $g$ is realized as a unitary $N\times N$ matrix. 
For this case, the cocycle condition \eqref{eq:cocycle} is the necessary and sufficient condition for the single-valuedness of $\phi$. 
However, this is not necessarily true for other representations. 
Especially for the adjoint representation, the condition \eqref{eq:single_valued} only requires that 
\begin{equation}
    g_{ij}g_{jk}g_{ki}=\exp\left({2\pi \im\over N}n_{ijk}\right)
    \label{eq:tHooft_flux_def}
\end{equation}
for some integers $n_{ijk}$ modulo $N$. 
This additional degree of freedom $n_{ijk}$ is nothing but the 't~Hooft flux. 
When the matter field $\phi$ is in the adjoint representation, we have to make a choice whether we impose \eqref{eq:cocycle} or we relax it as \eqref{eq:tHooft_flux_def} without violating locality and unitarity of quantum field theories. 
They correspond to $SU(N)$ and $SU(N)/\mathbb{Z}_N$ gauge theories, respectively. 
More physically, those theories have different set of genuine line operators~\cite{Aharony:2013hda}. 

Let us discuss properties of $\{n_{ijk}\}$ to identify its physical degrees of freedom. 
We first note that $n_{ijk}$ is totally antisymmetric in its labels $i,j,k$: $n_{ijk}=n_{jki}=n_{kij} \bmod N$ and $n_{ijk}=-n_{jik} \bmod N$. This can be seen by massaging the definition of 't~Hooft flux, \eqref{eq:tHooft_flux_def}. 
Cyclic property is obtained by multiplying $g_{ij}^{-1}$ from the left and $g_{ij}$ from the right of \eqref{eq:tHooft_flux_def}, then 
\begin{equation}
    \exp\left({2\pi \im\over N}n_{ijk}\right)=g_{ij}^{-1}\exp\left({2\pi \im\over N}n_{ijk}\right)g_{ij}=g_{jk}g_{ki}g_{ij}=\exp\left({2\pi\im\over N}n_{jki}\right). 
\end{equation}
By taking the inverse of \eqref{eq:tHooft_flux_def}, we find 
\begin{equation}
    \exp\left(-{2\pi\im\over N}n_{ijk}\right)=g_{ki}^{-1}g_{jk}^{-1}g_{ij}^{-1}=g_{ik}g_{kj}g_{ji}=\exp\left({2\pi\im\over N}n_{ikj}\right), 
\end{equation}
so $n_{ijk}$ flips its sign under odd permutations. 

Next, we must identify the gauge redundancy for $n_{ijk}$. 
We note that the connection formulas \eqref{eq:connection_gauge} (and \eqref{eq:connection_matter} for adjoint matters) are invariant under the transformation,
\begin{equation}
    g_{ij}\mapsto \exp\left({2\pi\im\over N}\lambda_{ij}\right)g_{ij},
\end{equation}
with $\lambda_{ij}\in \mathbb{Z}_N$ and $\lambda_{ji}=-\lambda_{ij} \bmod N$. In order to achieve invariance of the modified cocycle condition \eqref{eq:tHooft_flux_def}, $n_{ijk}$ must transform as  
\begin{equation}
    n_{ijk}\mapsto n_{ijk}+(\delta \lambda)_{ijk}, 
\end{equation}
where we have introduced the derivative $\delta$ by 
\begin{equation}
    (\delta \lambda)_{ijk}\equiv\lambda_{ij}-\lambda_{ik}+\lambda_{jk}. 
\end{equation}
This is the $\mathbb{Z}_N$ gauge symmetry acting on the transition functions, and $\{n_{ijk}\}$ is nothing but the gauge field for that gauge symmetry. This is called $\mathbb{Z}_N$ one-form gauge symmetry, and $\{n_{ijk}\}$ is a realization of corresponding $\mathbb{Z}_N$ two-form gauge field. 

By construction, the $\mathbb{Z}_N$ two-form gauge field obeys the flatness condition, 
\begin{equation}
    (\delta n)_{ijk\ell}\equiv n_{ijk}-n_{ij\ell}+n_{ik\ell}-n_{jk\ell}=0 \bmod N. 
    \label{eq:flatness_2form_gauge}
\end{equation}
We note that $\delta \delta \lambda=0$, so the  left-hand-side is gauge invariant and it is regarded as a field strength of the discrete gauge field. 
In order to see this flatness condition, we consider a quadruple overlap $U_{ijk\ell}=U_i\cap U_j\cap U_k \cap U_\ell$. We first compute 
\begin{equation}
    g_{ij}g_{jk}g_{k\ell}g_{\ell i}=(g_{ij}g_{jk}g_{ki})(g_{ik}g_{k\ell}g_{\ell i})=\exp\left({2\pi \im \over N}(n_{ijk}+n_{ik\ell})\right). 
\end{equation}
Since this belongs to the center element, we can multiply $g_{ij}^{-1}$ from the left and $g_{ij}$ from the right without changing the result. Therefore, 
\begin{align}
\exp\left({2\pi\im\over N}(n_{ijk}+n_{ik\ell})\right)&= g_{ij}^{-1}(g_{ij}g_{jk}g_{k\ell} g_{\ell i}) g_{ij}=g_{jk}g_{k\ell} g_{\ell i} g_{ij}\nonumber\\
&=(g_{jk}g_{k\ell} g_{\ell j})(g_{j\ell}g_{\ell i}g_{ij})=\exp\left({2\pi\im\over N}(n_{jk\ell}+n_{j\ell i})\right). 
\end{align}
This gives \eqref{eq:flatness_2form_gauge}. 
Mathematically, this claims that the gauge-equivalence class of $\{n_{ijk}\}$ belongs to the second $\mathbb{Z_N}$-valued cohomology, $[\{n_{ijk}\}]\in H^2 (M;\mathbb{Z}_N)$. 
$[\{n_{ijk}\}]\not = 0 $ characterizes the obstruction to lifting an $SU(N)/\mathbb{Z}_N$ principal bundle to an $SU(N)$ principal bundle. 

This flatness condition, $\delta n=0 \bmod N$, means that $4$d $SU(N)$ and $SU(N)/\mathbb{Z}_N$ gauge theories have the same contents for local, dynamical excitations. 
To see this, it is convenient to introduce one adjoint Higgs field, and assume that the gauge symmetry is Higgsed as $SU(N)\to U(1)^{N-1}$. 
In this regime, we can construct 't~Hooft-Polyakov monopoles, whose magnetic charges belong to the root lattice. 
For $SU(N)$ gauge theories, these are the minimal magnetic charges allowed by Dirac quantization, because the fundamental quark can be introduced at least as test particles by Wilson lines. 
For $SU(N)/\mathbb{Z}_N$ gauge theories, however, the fundamental Wilson loop is not a genuine line operator, and there can be fractionally charged monopoles belonging to the weight lattice. 
If $\delta n\not =0 \bmod N$, the magnetic defect is nothing but those fractionally charged monopoles. 
The flatness condition claims that those extra monopoles can be introduced only as test particles by 't~Hooft lines and do not exist as dynamical excitations, and we can see that local dynamics is the same for both $4$d $SU(N)$ and $SU(N)/\mathbb{Z}_N$ gauge theories. 

\subsection{'t~Hooft flux and \texorpdfstring{$\mathbb{Z}_N$}{ZN} two-form gauge fields}
\label{sec:tHooft_flux_gauging}

Here, we explain the fact that introduction of 't~Hooft flux can be understood as the coupling of a topological $\mathbb{Z}_N$ two-form gauge theory to $SU(N)$ Yang-Mills theory, following Refs.~\cite{Kapustin:2014gua,Kapustin:2013qsa}. 
This is already suggested in the previous subsection, as we have seen that 't~Hooft flux is characterized by $[\{n_{ijk}\}]\in H^2(M,\mathbb{Z}_N)$. 

The starting point is the following topological action, 
\begin{equation}
    S_{\mathrm{top}}[G,B^{(2)},B^{(1)}]=\im \int_M {1\over 2\pi}G\wedge (N B^{(2)}-\diff B^{(1)}), 
    \label{eq:ZN_two_form_TQFT}
\end{equation}
where $G$ is an $\mathbb{R}$-valued $2$-form field, $B^{(2)}$ is a $U(1)$ $2$-form gauge field, and $B^{(1)}$ is a $U(1)$ $1$-form gauge field. 
Here, $G$ is an auxiliary field introduced as a Lagrange multiplier and its equation of motion imposes
\begin{equation}
    NB^{(2)}=\diff B^{(1)}. 
    \label{eq:ZN_two_form_EoM}
\end{equation}
This $2$-form gauge field $B^{(2)}$ turns out to play the role of the 't~Hooft flux $[\{n_{ijk}\}]$ in the previous subsection. 
Let us discuss the gauge redundancy of this theory. There are both $0$-form and $1$-form gauge transformations, 
\begin{equation}
    B^{(2)}\mapsto B^{(2)}+\diff \Lambda^{(1)},\quad B^{(1)}\mapsto B^{(1)}+N\Lambda^{(1)}+\diff \Lambda^{(0)}. 
\end{equation}
Here, the $1$-form gauge transformation parameter $\Lambda^{(1)}$ is a $U(1)$ $1$-form gauge field, and the $0$-form (i.e. ordinary) gauge transformation parameter $\Lambda^{(0)}$ is a $2\pi$-periodic scalar field. 

Let us consider physical observables of this theory~\eqref{eq:ZN_two_form_TQFT}. 
There are surface and loop operators as nontrivial observables. 
The surface operator is given by
\begin{equation}
    U_k(\Sigma)=\exp\left(\im k \int_{\Sigma}B^{(2)}\right)
\end{equation}
for $k\in \mathbb{Z}$ and closed $2$-manifolds $\Sigma\subset M$. 
The level $k$ should be quantized to integers for the $1$-form gauge invariance. 
Moreover, because of the equation of motion~\eqref{eq:ZN_two_form_EoM}, $\int_\Sigma B^{(2)}\in {2\pi\over N}\mathbb{Z}$ due to the Dirac quantization for $B^{(1)}$, so the label $K$ is identified in mod $N$, $k\sim k+N$. 

Next, we construct the loop operator. We note that the Wilson loop $\exp\left(\im \int_C B^{(1)}\right)$ is not a physical loop operator because it violates the $1$-form gauge invariance. 
Instead, the 't~Hooft loop for $B^{(1)}$ is well-defined as a gauge-invariant loop operator, 
\begin{equation}
    H(C),
\end{equation}
which is defined as a defect operator and imposes the boundary condition for $B^{(1)}$ so that $\int_{S^2_*}\diff B^{(1)}=2\pi$ for small spheres linking to $C$. 
This can be easily seen by performing the path integral for $B^{(1)}$ in \eqref{eq:ZN_two_form_TQFT} before doing it for $G$. 
The equation of motion for $B^{(1)}$ tells $\diff G=0$, and, moreover, summing up topological sectors for $B^{(1)}$ imposes $\int_\Sigma G\in 2\pi \mathbb{Z}$ for any closed $2$-manifolds $\Sigma \subset M$. 
Therefore, $G$ can be regarded as the $U(1)$ gauge field strength, 
\begin{equation}
    G=\diff A, 
\end{equation}
and $S_{\mathrm{top}}$ becomes the level-$N$ $BF$ theory,
\begin{equation}
    S_{\mathrm{top}}=\im \int_{M}{N\over 2\pi} \diff A \wedge B^{(2)}. \label{eq:levelN_BF}
\end{equation}
With this description, the 't~Hooft loop is expressed as the Wilson loop of the dual $U(1)$ gauge field $A$, 
\begin{equation}
    H(C)=\exp\left(\im \int_C A\right). 
\end{equation}
As in the case of the surface operator, we also have the identification, $H(C)^{k+n}=H(C)^k$. 

We can deform the topological action~\eqref{eq:ZN_two_form_TQFT} by a discrete $\theta$ parameter, 
\begin{equation}
    S_{\mathrm{top},p}=\im \int_M\left[{1\over 2\pi}G\wedge (N B^{(2)}-\diff B^{(1)})+{Np\over 4\pi} B^{(2)}\wedge B^{(2)}\right]. 
    \label{eq:ZN_TQFT_discreteTheta}
\end{equation}
For simplicity, let us assume that $M$ admits a spin structure. 
When $p$ is an integer, this is invariant under the following gauge transformations, 
\begin{align}
    & B^{(2)}\mapsto B^{(2)}+\diff \Lambda^{(1)},\quad B^{(1)}\mapsto B^{(1)}+N\Lambda^{(1)}+\diff \Lambda^{(0)}, \nonumber\\
    & G\mapsto G-p\, \diff \Lambda^{(1)}.
\end{align}
We can repeat the above discussions with general $p$, and the contents of gauge-invariant loop and surface operators are changed. For details, see \cite{Kapustin:2014gua,Kapustin:2013qsa}. 

We can couple the above topological theories to $SU(N)$ Yang-Mills theory to introduce the 't~Hooft flux, or to consider the $SU(N)/\mathbb{Z}_N$ Yang-Mills theory. 
For this purpose, it is convenient to consider the $U(N)$ Yang-Mills theory, 
\begin{equation}
    S_{\mathrm{YM},U(N)}={1\over g^2}\int_M |F(\tilde{a})|^2+{\im \theta\over 8\pi^2}\int_M \tr[F(\tilde{a})^2], 
    \label{eq:U(N)_YM}
\end{equation}
where $\tilde{F}:=F(\tilde{a})=\diff \tilde{a}+\im\,\tilde{a}\wedge \tilde{a}$ for $U(N)$ gauge field $\tilde{a}$ and $|\tilde{F}|^2=\tr(\tilde{F}\wedge \star \tilde{F})$. 
This is obviously different both from $SU(N)$ and $SU(N)/\mathbb{Z}_N$ Yang-Mills theories for several reasons. 
There are massless photons in this theory described by $\tr(\tilde{a})$, and the topological sectors are not characterized by instantons on closed manifolds as there are extra monopole sectors. 
We can resolve these discrepancy by introducing $U(1)$ $2$-form gauge field $B$ and we impose the following constraint,
\begin{equation}
    NB=\tr(\tilde{F}).
    \label{eq:ZN_two_form_PSU(N)}
\end{equation}
Here, we simply denote $B$ for $B^{(2)}$. We note that $\tr(\tilde{F})$ can be identified as the $U(1)$ field strength, and \eqref{eq:ZN_two_form_PSU(N)} corresponds to \eqref{eq:ZN_two_form_EoM}. 
When performing the path integral $\int \Diff \tilde{a}$, the configuration space should be restricted to satisfy \eqref{eq:ZN_two_form_PSU(N)}. 

By introducing $B$, we can additionally require the $1$-form gauge invariance. The $U(N)$ $0$-form and $U(1)$ $1$-form gauge transformations are given by 
\begin{equation}
    \tilde{a}\mapsto g^{-1}\tilde{a} g-\im g^{-1}\diff g +\Lambda^{(1)}\bm{1},\quad 
    B\mapsto B+\diff \Lambda^{(1)}, 
\end{equation}
where $g$ is a $U(N)$-valued function and $\Lambda^{(1)}$ is a $U(1)$ gauge field. The $1$-form gauge invariance is the key to eliminate the massless photon $\tr[\tilde{a}]$ from the physical spectrum as in the case of would-be Nambu-Goldstone bosons in Higgs mechanism. 
We also have the gauge redundancy of the gauge transformation parameters as 
\begin{equation}
    \Lambda^{(1)}\mapsto \Lambda^{(1)}+\diff \lambda,\quad g\mapsto \rme^{-\im \lambda}g, 
\end{equation}
where $\lambda$ is a $2\pi$-periodic scalar. 
The action~\eqref{eq:U(N)_YM} for $U(N)$ YM theory is no longer gauge invariant, so we need to change it. 
We can easily achieve it by replacing $\tilde{F}$ by $\tilde{F}-B$ everywhere in \eqref{eq:U(N)_YM}, and we obtain 
\begin{equation}
    S_{\mathrm{YM},p}[\tilde{a},B]={1\over g^2}\int_M |\tilde{F}-B|^2+{\im \theta\over 8\pi^2}\int_M (\tilde{F}-B)^2+{\im Np\over 4\pi}\int_M B^2.
    \label{eq:YMaction_tHooft_flux}
\end{equation}
The last term is the discrete $\theta$ parameter discussed in \eqref{eq:ZN_TQFT_discreteTheta}. 
In this theory, the fundamental Wilson line is no longer a genuine line operator, and it becomes a boundary of the topological surface operator,
\begin{equation}
    W(\partial \Sigma,\Sigma)=\tr\left[\mathcal{P}\exp\left(\im \int_{\partial \Sigma}\tilde{a}\right)\right]\exp\left(-\im \int_\Sigma B\right). 
\end{equation}
Instead, we have magnetically charged genuine lines, and they are dyonic for nonzero values of $p$~\cite{Aharony:2013hda}:
\begin{equation}
    HW^{p}(C). 
\end{equation}
Therefore, these theories are physically distinguished even though local dynamics are identical.

\subsection{'t Hooft flux on \texorpdfstring{$T^2$}{T2} and its minimal action configurations}
\label{sec:tHooft_flux_2dtorus_continuum}

Let us consider a $2$-dimensional torus $T^2$, and the coordinate $(x,y)\in T^2$ is subject to the identification $x\sim x+L_x$, $y\sim y+L_y$. 
In the case of torus, instead of working with the good open cover as presented in Appendix~\ref{sec:definition_tHooft_flux}, it is more convenient to work explicitly with the coordinate~\cite{vanBaal:1982ag}. 

In gauge theories, the fields at $(x,y)$ and at $(x+L_x,y)$ should be identified up to gauge transformations. 
Let us denote the gauge-transformation function as $g_x(y)$, then 
\begin{align}
    & a(x=L_x,y)=g_x(y)^{\dagger}a(x=0,y)g_x(y)-\im g_x(y)^{\dagger}\diff g_x(y),\\
    & \phi(x=L_x,y)=\rho(g_x(y)^{\dagger})\phi(x=0,y). 
\end{align}
The similar relation exists for fields at $(x,y)$ and $(x,y+L_y)$, and we denote the transition function as $g_y(x)$:
\begin{align}
    & a(x,y=L_y)=g_y(x)^{\dagger}a(x,y=0)g_y(x)-\im g_y(x)^{\dagger}\diff g_y(x),\\
    & \phi(x,y=L_y)=\rho(g_y(x)^{\dagger})\phi(x,y=0). 
\end{align}
When $\rho$ is the adjoint representation, the consistency condition requires that~\cite{tHooft:1979rtg} 
\begin{equation}
    g_x(L_y)^{\dagger}g_y(0)^{\dagger}=g_y(L_x)^{\dagger}g_x(0)^{\dagger}\exp\left({2\pi\im\over N}n\right). 
    \label{eq:tHooft_twist_continuum}
\end{equation}
This label $n\in\mathbb{Z}_N$ is nothing but the 't~Hooft flux $[\{n_{ijk}\}]$ discussed in the previous section. 
We note that this label $n$ is already $\mathbb{Z}_N$ one-form gauge invariant under $g_x(y)\to \exp({2\pi \im \over N}\lambda_x)g_x(y)$, and $g_y(x)\to \exp({2\pi \im \over N}\lambda_y)g_y(x)$. 

In $2$d, we can perform $SU(N)$ gauge transformations so that the transition functions $g_x(y)$ and $g_y(x)$ become independent of coordinates as $SU(N)$ is simply-connected, $\pi_1(SU(N))=0$. 
As a result, the minimal 't~Hooft flux, $n=1$, can always be explicitly realized by the constant matrices, 
\begin{equation}
g_x(y)=S,\quad g_y(x)=C, 
\end{equation}
where $C$ and $S$ are clock and shift matrices in $SU(N)$. For $SU(3)$, they are given as 
\begin{equation}
    C=\begin{pmatrix}
    1& & \\
     & \omega & \\
     & & \omega^2
    \end{pmatrix}, \quad S=\begin{pmatrix}
    0 & 1 & 0\\
    0 & 0 & 1\\
    1 & 0 & 0
    \end{pmatrix}, 
\end{equation}
with $\omega=\exp\left({2\pi \im\over3}\right)$, 
and generalization to $SU(N)$ is straightforward. It satisfies 
\begin{equation}
    S C = CS \exp\left({2\pi \im \over N}\right), 
\end{equation}
and the nontrivial 't~Hooft flux on $T^2$ is obtained. 

Next, we shall see that the 't~Hooft flux in $2$d is compatible with the flat connection, $F=\diff a+\im a\wedge a=0$. 
Moreover, we can see that such a gauge configuration is unique up to gauge transformations. 
To show this, we first solve $F=0$ as 
\begin{equation}
    a=-\im V^\dagger \diff V
\end{equation}
with some $V:[0,L_x]\times [0,L_y]\to SU(N)$. Performing the gauge transformation with $V^\dagger$, we obtain $a'=0$ and the transition functions are replaced as 
\begin{equation}
    g'_x(y)=V(0,y)g_x(y)V(L,y)^\dagger,\quad g'_y(x)=V(x,0)g_y(x)V(x,L)^\dagger. 
\end{equation}
We note that new transition functions, $g'_x(y), g'_y(x)$, also satisfy \eqref{eq:tHooft_twist_continuum}. 
Furthermore, by using the connection formula for $a=-\im V^\dagger\diff V$, we see that $\partial_y g'_x(y)=0$ and $\partial_x g'_x(y)=0$, so the new transition functions must be constant.
Then, by performing constant gauge transformations, we can always set $g'_x(y)=S$ and $g'_y(x)=C$ while keeping $a'=0$. 
This is the result, which was to be proven.

Let us compute Wilson loops around nontrivial cycles for this classical solution with the 't~Hooft flux. 
They are given by 
\begin{align}
    P_x(y)&=g_x(y)\, \mathcal{P}\rme^{\im \int_0^{L_x} a_x(x,y)\diff x}=S,\nonumber\\
    P_y(x)&=g_y(x)\, \mathcal{P}\rme^{\im \int_{0}^{L_y}a_y(x,y)\diff y}=C. 
    \label{eq:unbroken_hol_con}
\end{align}
We note that transition functions have to be multiplied so that $P_x$ and $P_y$ transform in the adjoint representation under gauge transformations, and $\tr(P_{x})$ and $\tr(P_y)$ are gauge invariant. 
Since the classical solution is given by $a=0$ after gauge transformations, these Wilson lines become identical to the transition functions.

\subsection{Lattice viewpoint of 't Hooft flux on \texorpdfstring{$T^d$}{Td}}
\label{sec:tHooft_flux_2dtorus_lattice}

Let us discuss Wilson's lattice gauge action for $SU(N)$ gauge theory with the 't~Hooft background on the hypertorus $T^d$. 
We first consider with general spacetime dimensions $d$, and we will set $d=2$ to find the classical solutions and confirm the results in the previous section, Appendix~\ref{sec:tHooft_flux_2dtorus_continuum}. 

The Wilson lattice action is given by 
\begin{equation}
    S_{\mathrm{W}}=\sum_{x}\sum_{\mu\not=\nu}\tr\left(\bm{1}_N-\tilde{U}_\mu(x)\tilde{U}_{\nu}(x+\hat{\mu})\tilde{U}^\dagger_{\mu}(x+\hat{\nu})\tilde{U}^\dagger_\nu(x)\right), 
    \label{eq:Wilson_lattice}
\end{equation}
where $x=(x_1,\ldots,x_d)$ denotes the lattice sites, $\mu,\nu$ specify directions, $1,\ldots,d$, and $\hat{\mu},\hat{\nu}$ are unit vectors along those directions. 
We here set the lattice constant to be $1$ and denote $N_\mu$ be the size of the torus along the $\mu$ direction, i.e. $x_\mu=0,1,\ldots, N_\mu-1$ and we identify $x_\mu\sim x_\mu+N_\mu$. 
The link variables $\tilde{U}_\mu(n)$ are given by
\begin{equation}
    \tilde{U}_\mu(x)=\mathcal{P}\exp\left(\int_{x}^{x+\hat{\mu}}a\right),
\end{equation}
and they obey the twisted boundary condition, 
\begin{equation}
    \tilde{U}_\mu(x+N_\nu \hat{\nu})=g_{\nu}^\dagger(x) \tilde{U}_\mu(x) g_{\nu}(x+\hat{\mu}). 
    \label{eq:lattice_twistedbc}
\end{equation}
Here, we note that the arguments of $\tilde{U}_\mu(x)$ in \eqref{eq:Wilson_lattice} are restricted to $x_\nu=0,1,\ldots, N_\nu$ for $\nu\not=\mu$ and $x_\mu=0,1,\ldots,N_\mu-1$. 
Therefore, we can set $x_\nu=0$ in \eqref{eq:lattice_twistedbc}, so we can assume that $g_\nu(x)$ does not depend on $x_{\nu}$. 
If there is a matter field $\phi$ in the representation $\rho$, it satisfies 
\begin{equation}
    \phi(x+N_\nu \hat{\nu})=\rho(g^\dagger_{\nu}(x)) \phi(x), 
\end{equation}
and we assume that $\rho$ has trivial $N$-ality. 
The transition functions specify the 't~Hooft flux $n_{\mu\nu}\in \mathbb{Z}_N$ on the torus,
\begin{equation}
    g_\mu^\dagger(x+N_\nu \hat{\nu})g_{\nu}^\dagger(x)=g_{\nu}^\dagger (x+N_\mu \hat{\mu})g_{\mu}^\dagger(x) \exp\left({2\pi \im\over N} n_{\mu\nu}\right). 
\end{equation}
This is the lattice realization of the 't~Hooft twisted boundary condition~\eqref{eq:tHooft_twist_continuum}. 

We can redefine the link variables so that they obey the periodic boundary condition, and denote them as $U_\mu(x)$. For $x_\mu=N_\mu-1$, we relate it to  $\tilde{U}_\mu(x)$ as 
\begin{equation}
    U_\mu(x_\mu=N_\mu-1,x_{\nu\not=\mu})=\tilde{U}_\mu(x_\mu=N_\mu-1,x_{\nu\not=\mu})g^\dagger_\mu(x_{\nu}). 
\end{equation}
When $x_\mu=0,1,\ldots,N_\mu-2$, we identify $U_\mu(x)$ and $\tilde{U}_\mu(x)$. 
For a link variable $\tilde{U}(x)$ with $x_\mu=N_\mu$ in \eqref{eq:Wilson_lattice}, we first relate it to the link variable $\tilde{U}(x)$ with $0\le x_\mu\le N_\mu-1$ using the boundary condition~\eqref{eq:lattice_twistedbc} and then rewrite it with the periodic link variable $U(x)$. 
After this manipulation, the Wilson action~\eqref{eq:Wilson_lattice} becomes 
\begin{equation}
     S_{\mathrm{W}}=\sum_{x}\sum_{\mu\not=\nu}\tr\left(\bm{1}_N-\rme^{-{2\pi \im\over N}B_{\mu\nu}(x)} U_\mu(x)U_{\nu}(x+\hat{\mu})U^\dagger_{\mu}(x+\hat{\nu})U^\dagger_\nu(x)\right), 
\end{equation}
where $B_{\mu\nu}(x)=n_{\mu\nu}$ at $(x_\mu,x_\nu)=(N_\mu-1, N_\nu-1)$ and $B_{\mu\nu}(x)=0$ otherwise. 
In this expression, the 't~Hooft twist is introduced as the specific realization of the background gauge field for $\mathbb{Z}_N$ $1$-form symmetry of $SU(N)$ Yang-Mills theory, and its effect can be seen very explicitly. 
This corresponds to \eqref{eq:YMaction_tHooft_flux} (with $p=0$) in the continuum formulation given in Appendix~\ref{sec:tHooft_flux_gauging}. 
 
\begin{figure}[t]
\vspace{-2.2cm}
\begin{center}
\includegraphics[width = 1.0\textwidth]{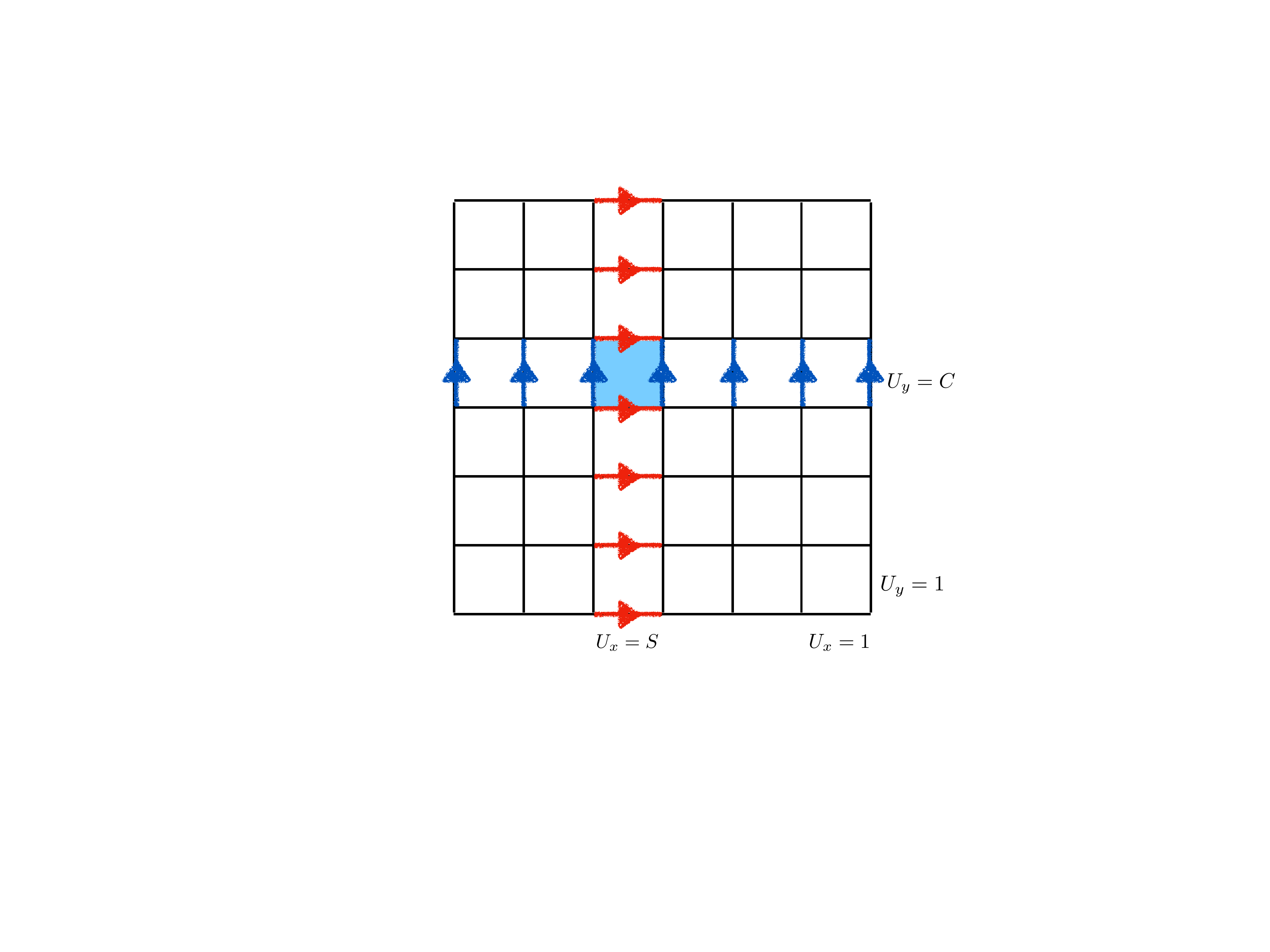}
\vspace{-3.8cm}
\caption{Turning on one unit of 't Hooft flux in the small $T^2$, the classical moduli space parametrized in terms of two Polyakov loops gets lifted.  The  classical minimum  of the system is shown in the figure. As a result, Polyakov loops in the $x$ and $y$ direction becomes the non-commuting pair $P_x=S$ and $P_y= C$. This configuration respects $ \Z_N^{[0]} \times \Z_N^{[0]}$ part of the center-symmetry at the classical level. 
    }
\label{fig:fluxmin}
\end{center}
\end{figure}

Let us consider the case $d=2$, and we set $n_{xy}=1$. 
By regarding $T^2$ as the rectangle with the periodic boundary condition, the 't~Hooft flux $B_{xy}=1$ is inserted on the plaquette at the top-right corner. 
Other plaquettes are not affected by the twisted boundary condition in the above realization. 
In order to minimize the classical action $S_{\mathrm{W}}$, we can set 
\begin{equation}
    U_{\mu\nu}(x,y)=1
\end{equation}
for $(x,y)\not=(N_x-1,N_y-1)$, while 
\begin{equation}
U_{\mu\nu}(N_x-1,N_y-1)=\rme^{2\pi\im/N}. 
\end{equation}
This can be solved in terms of the link variables as 
\begin{equation}
    U_x(x\not=N_x-1,y)=1,\quad U_y(x,y\not=N_y-1)=1,
\end{equation}
and 
\begin{equation}
    U_x(N_x-1,y)=S,\quad U_y(x,N_y-1)=C. 
\end{equation}
We can always perform the gauge transformation to get this gauge configuration from arbitrary minimal action configurations with the 't~Hooft flux, and there is no remnant of gauge transformations except the center elements $\mathbb{Z}_N$. 
This is the lattice derivation of the results in Appendix~\ref{sec:tHooft_flux_2dtorus_continuum}, and it is shown in Fig.~\ref{fig:fluxmin}. 
In the figure, we introduce the 't~Hooft flux in the light-blue shaded plaquette, and the red and blue arrowed links denote $U_x=S$ and $U_y=C$, respectively. 
As a result, holonomy along each direction is given by 
\begin{align}
    P_x(y)&=\mathcal{P}\prod_{x'} U_x(x',y)=S, \nonumber\\
    P_y(x)&=\mathcal{P}\prod_{y'} U_y(x,y')=C, 
\end{align}
which reproduces \eqref{eq:unbroken_hol_con}.

\section{\texorpdfstring{$2$}{2}d Abelian-Higgs model with charge \texorpdfstring{$N$}{N}}
\label{sec:2dAbelianHiggs}

In this appendix, we consider $2$d $U(1)$ gauge theory coupled to charge-$N$ complex scalar $\Phi$. The Lagrangian is given by 
\begin{equation}
    \mathcal{L}={1\over 2e^2}|f_{12}|^2+|(\p_\mu +\im N a_\mu)\Phi|^2+\lambda(|\Phi|^2-v^2)^2-{\im \theta\over 2\pi}f_{12}, 
    \label{eq:Lagrangian_AbelianHiggs}
\end{equation}
with $f_{12}=\p_1 a_2-\p_2 a_1$. 
Here, mass dimensions of fields and coupling constants are given by $[a_\mu]=1$ so that $[a]=[a_\mu \diff x^\mu]=0$, $[\Phi]=0$, $[e^2]=2$, $[v]=0$, and $[\lambda]=2$. 
This system has $\mathbb{Z}_N$ $1$-form symmetry, which acts on the Wilson loops, $W(C)=\exp(\im \oint_C a)$. 

We consider the case with the wine-bottle potential, $v^2>0$, so that the system is in the Higgs regime. 
The bosonic potential is minimized by setting 
\begin{equation}
    \Phi=v\, \rme^{\im \varphi},
    \label{eq:AbelianHiggs_classical_minima}
\end{equation}
where $\varphi$ is a $2\pi$-periodic scalar field. 
When $v$ is sufficiently large, the classical equation of motion of $a_\mu$ is then given by 
\begin{equation}
    N a=-\diff \varphi,
    \label{eq:AbelianHiggs_classical_eom}
\end{equation}
and $U(1)$ gauge field is Higgsed to $\mathbb{Z}_N$ gauge field. 
Within the perturbation theory, the Wilson loop obeys the perimeter law, so the system is in the deconfined phase. 
However, it is usually quite difficult to break the $1$-form symmetry in $2$d spacetime, unless there is a requirement by 't~Hooft anomaly~\cite{Gaiotto:2014kfa, Misumi:2019dwq}. 
This suggests that we are missing some important nonperturbative effects in the above perturbative discussion, and we should take it into account to get the correct physics. 

In the following, we first discuss the vortex configuration and its topological nature in Appendix~\ref{sec:AbelianHiggs_vortex}. 
In Appendix~\ref{sec:AbelianHiggs_DVGA}, we perform the dilute vortex gas approximation, and we find the confinement of Wilson loops and the multi-branch structure of $\theta$ vacua. 
In Appendix~\ref{sec:AbelianHiggs_masslessfermion}, we add a charge-$N$ massless Dirac fermion, and we observe the presence of $N$ vacua due to the spontaneous chiral symmetry breaking and also the perimeter law of Wilson loops. 

\subsection{Vortex and its fractional \texorpdfstring{$\theta$}{theta} dependence}
\label{sec:AbelianHiggs_vortex}
 
As a nontrivial semiclassical configuration in $2$d, we have a vortex configuration which is a point-like defect in the Euclidean spacetime. 
When we are sufficiently far away from the vortex, it is a good approximation to stay in the classical minima~\eqref{eq:AbelianHiggs_classical_minima}. 
Let us take the polar coordinate $(r,\phi)$ of the spacetime, where the vortex center locates at the origin, and then the vortex is characterized by the winding number $n$ by the identification, 
\begin{equation}
    \varphi=n\phi. 
\end{equation}

Let us comment on important properties of this vortex configuration. 
We note that we can use the classical equation of motion~\eqref{eq:AbelianHiggs_classical_eom} when we are sufficiently far away from the vortex center. 
Using the Stokes theorem, we can see that there should be nonzero field strength near the vortex center because 
\begin{equation}
    \int_{\mathbb{R}^2}\diff a=\int_{S^1_{\infty}}a=-{n\over N}\int \diff \phi=-{2\pi n\over N}. 
    \label{eq:fractional_charge_vortex}
\end{equation}
That is, the single vortex does not necessarily satisfy the Dirac quantization of $U(1)$ gauge fields, and the $\theta$ dependence is fractionalized. 
This also shows that the Wilson loop has a complex phase if it surrounds the vortex: 
\begin{equation}
W(C)=\exp\left(\im \int_C a\right)=\exp\left(-{2\pi\im\over N}n\right), 
\end{equation}
when $C$ surrounds the vortex with the winding number $n$. This phase fluctuation due to the vortex is important to have the area law of the Wilson loop, as we shall see in the next subsection.

 \subsection{Dilute vortex gas and confinement}
 \label{sec:AbelianHiggs_DVGA}
 
Let us denote the Euclidean action of the vortex as $S_v$. As we have just seen in \eqref{eq:fractional_charge_vortex}, the minimal vortex carries the fractional topological charge, and thus the Boltzmann weight for the single vortex configuration is given by 
\begin{equation}
    \exp(-S_v) \exp(\pm\im\, \theta/N), 
\end{equation} 
where the $\pm$ sign for the $\theta$ dependence is for the vortex with the winding number $\pm1$.
In order to obtain the partition function within the semi-classical approximation, we assume that the interaction between vortices are negligible, or, in other words, we consider the dilute gas of vortices. 
 
When computing the partition function on closed $2$-manifolds, we have to note that the topological charge must be integer because of the Dirac quantization,
\begin{equation}
     \int_M{\diff a\over 2\pi}\in \mathbb{Z}. 
\end{equation}
Therefore, the numbers of vortices and anti-vortices, $n,\overline{n}$, should satisfy the constraint,
\begin{equation}
    n-\overline{n}\in N\mathbb{Z}. 
    \label{eq:restriction_vortex_number}
\end{equation}
As a result, the $\theta$ dependence of the partition function is given by 
\begin{align}
    Z(\theta)&=\sum_{n,\overline{n}\ge 0}{V^{n+\overline{n}}\over n! \overline{n}!}\rme^{-(n+\overline{n})S_v}\rme^{\im (n-\overline{n})\theta/N}\delta_{n-\overline{n}\in N\mathbb{Z}}\nonumber\\
    &=\sum_{k=0}^{N-1}\exp\left[-V\left(-2\rme^{-S_v}\cos\left({\theta-2\pi k\over N}\right)\right)\right]. 
\end{align}
Therefore, the vacua has $N$-branch structure, and the ground-state energies are given by 
\begin{equation}
    E_k(\theta)=-2\rme^{-S_v}\cos\left({\theta-2\pi k\over N}\right). 
\end{equation}
The true ground state is determined by choosing the branch label $k$ with the minimal energy density, and thus the ground-state energy is 
\begin{equation}
    E(\theta)=\min_{k} E_k(\theta). 
\end{equation}
Especially when $-\pi<\theta<\pi$, the label $k=0$ is chosen, and when $\theta$ goes across $\pi$, there is the first-order phase transition from $k=0$ to $k=1$. 

Now, let us compute the expectation value of the Wilson loop. 
Here, we assume $|\theta|<\pi$ and the ground state is given by the $k=0$ brunch. 
Let us rewrite the Wilson loop as 
\begin{equation}
    W(C)^q=\exp\left(2\pi\im\,q\int_{D}{\diff a \over 2\pi}\right), 
\end{equation}
where $D$ is the $2$d surface with $\partial D=C$. Therefore, $W^q(C)$ shifts the $\theta$ angle inside the loop $C$ as $\theta\to \theta+2\pi q$. 
Therefore, the leading behavior of the Wilson loop is given by 
\begin{equation}
    \bigl\langle W(C)^q\bigr\rangle=\exp\bigl(-(E_{-q}(\theta)-E_0(\theta))\mathrm{area}(D)\bigr). 
\end{equation}
Therefore, the string tension $T_q$ of the probe charge $q$ is given by 
\begin{equation}
    T_q=E_{-q}(\theta)-E_{0}(\theta)=E_0(\theta+2\pi q)-E_0(\theta). 
\end{equation}
When $q\not\in N\mathbb{Z}$ and $|\theta|<\pi$, the string tension shows $T_q>0$ and the Wilson loops are confined. 
When $q$ is a multiple of $N$, $T_q=0$ and it can be understood as a consequence of string breaking by  $\Phi$ quanta.

\subsection{Adding massless fermion}
\label{sec:AbelianHiggs_masslessfermion}

Let us introduce one massless Dirac fermion with $U(1)$ charge $N$, 
\begin{equation}
    \overline{\psi}\gamma^{\mu}(\partial_\mu+\im N a_\mu)\psi,
\end{equation}
in the Abelian-Higgs model~\eqref{eq:Lagrangian_AbelianHiggs}. 
Under the $U(1)_A$ rotation, $\psi\to \rme^{\im \alpha \gamma_3}\psi$ and $\overline{\psi}\to \overline{\psi}\rme^{\im \alpha \gamma_3}$, the classical Lagrangian is invariant, but the fermion path-integral measure $\Diff \overline{\psi}\Diff \psi$ does not have this invariance, and it shifts the $\theta$ angle by $\theta\to \theta+2N\alpha$. 
Because of the $2\pi$ periodicity of $\theta$, the $\mathbb{Z}_{2N}$ subgroup of $U(1)_A$ generates the symmetry of the quantum system, so this system has the discrete axial symmetry. 
We note that $\mathbb{Z}_2\subset \mathbb{Z}_{2N}$ gives the fermion parity, so it cannot be broken spontaneously as long as the Lorentz symmetry is unbroken.\footnote{If we consider the charge-$N$ Schwinger model, the fermion parity is a part of the $U(1)$ gauge redundancy, so the correct axial symmetry should be identified as $(\mathbb{Z}_{2N})/\mathbb{Z}_2\simeq \mathbb{Z}_N$ instead of $\mathbb{Z}_{2N}$. In other words, there are no fermionic gauge-invariant local operators in the Schwinger model. In our case, however, the presence of bosonic field $\Phi$ and fermionic field $\psi$ changes the story, and the fermionic parity is a global symmetry. Indeed, it is part of the $U(1)$ symmetry, $\psi\to \rme^{\im \alpha }\psi$, $\Phi\to \Phi$, and the charged gauge-invariant operator is given by $\Psi\sim\Phi^*\psi$. } 

We can show that $\mathbb{Z}_N$ $1$-form symmetry and $\mathbb{Z}_{2N}$ axial symmetry has the mixed 't~Hooft anomaly (see the next subsection~\ref{sec:AbelianHiggs_anomaly}). 
If the system is gapped, the axial symmetry should be spontaneously broken in order to satisfy the anomaly matching constraint,
\begin{equation}
    \mathbb{Z}_{2N}\xrightarrow{\mathrm{SSB}} \mathbb{Z}_2. 
\end{equation}
Within perturbation theory, chiral symmetry is unbroken and the anomaly matching is satisfied by gapless fermions. 
We can easily observe this by noting that the classical bosonic vacua~\eqref{eq:AbelianHiggs_classical_minima} satisfy $N a=-\diff \varphi$, and thus the effective Lagrangian is given by 
\begin{equation}
    \mathcal{L}_{\mathrm{eff}}=\overline{\Psi}\gamma^\mu \partial_\mu \Psi, 
\end{equation}
where $\Psi=\rme^{-\im \varphi}\psi\sim \Phi^*\psi$ is the gauge-invariant Dirac fermion. In the bosonic case, however, the drastic difference is caused by the vortex, so let us again examine the dilute vortex gas picture. 

Since the vortex has the fractional $\theta$ dependence, it must be associated with two fermionic zero modes:
\begin{equation}
    \rme^{-S_v-\im \theta/N}\overline{\Psi}_R\Psi_L,\quad 
    \rme^{-S_v+\im \theta/N}\overline{\Psi}_L\Psi_R. 
\end{equation}
We note that this vertex is invariant under the spurious $U(1)$ axial symmetry, $\psi\to \rme^{\im \alpha \gamma_3}\psi$, $\overline{\psi}\to \overline{\psi}\rme^{\im \alpha \gamma_3}$, and $\theta\to \theta-2N\alpha$. 
By summing up the vortex gas with the Dirac quantization constraint, we have 
\begin{equation}
    Z=\sum_{k=0}^{N-1}\int \Diff \overline{\Psi}\Diff \Psi \exp\left(-\int \diff^2x\,\mathcal{L}_{\mathrm{eff},k}[\overline{\Psi},\Psi]\right),
\end{equation}
with 
\begin{equation}
    \mathcal{L}_{\mathrm{eff},k}=\overline{\Psi}\gamma^\mu \partial_\mu \Psi-\rme^{-S_v}\left(\rme^{-\im (\theta-2\pi k)/N}\overline{\Psi}_R\Psi_L+ 
    \rme^{\im (\theta-2\pi k)/N}\overline{\Psi}_L\Psi_R\right). 
\end{equation}
On each label $k$, the discrete axial symmetry looks to be explicitly broken, but the total system has the axial symmetry because of the summation over $k$. 
In this way, the system is gapped and the anomaly matching is satisfied by the spontaneous breaking, $\mathbb{Z}_{2N}\to \mathbb{Z}_2$. 

We note that the ground state energies become the same for all labels $k$. As a result, the string tension vanishes, and all the Wilson loops obey the perimeter law, 
\begin{equation}
    \langle W(C)^q\rangle \sim 1. 
\end{equation}
Under the presence of massless Dirac fermions, the system shows deconfinement because of the screening by massless fermions. 

\subsection{Anomaly matching and its semiclassical realization}
\label{sec:AbelianHiggs_anomaly}

In this subsection, we first discuss the (generalized) 't~Hooft anomaly and global inconsistency for the charge-$N$ Abelian-Higgs model, and explicitly show the semiclassical realization of anomaly in the dilute vortex gas picture. 

In order to observe the 't~Hooft anomaly, let us introduce the background $\mathbb{Z}_N$ two-form gauge field $B\in H^2(M,\mathbb{Z}_N)$. 
We can realize it as a pair of $U(1)$ two-form and one-form gauge fields $(B^{(2)}, B^{(1)})$, which satisfies the constraint,
\begin{equation}
    NB^{(2)}=\diff B^{(1)}. 
\end{equation}
This constraint has the one-form gauge invariance, 
\begin{equation}
    B^{(2)}\mapsto B^{(2)}+\diff \lambda^{(1)},\quad 
    B^{(1)}\mapsto B^{(1)}+N\lambda^{(1)}, 
\end{equation}
where the gauge transformation parameter $\lambda^{(1)}$ is another $U(1)$ gauge field. 
Under this one-form gauge transformation, we require that 
\begin{equation}
    a\mapsto a+\lambda^{(1)}. 
\end{equation}
In order to achieve the one-form gauge invariance of \eqref{eq:Lagrangian_AbelianHiggs}, we must replace the field strength as 
\begin{equation}
    \diff a\rightarrow \diff a-B^{(2)}, 
\end{equation}
and the covariant derivative as 
\begin{equation}
    (\partial_\mu+\im Na_\mu)\Phi\rightarrow (\partial_\mu+\im (N a_\mu-B^{(1)}_\mu))\Phi,
\end{equation}
so that the gauged action is given as 
\begin{align}
    S_{\mathrm{gauged}}&=\int \left({1\over 2e^2}|f-B^{(2)}|^2+|(\p_\mu +\im (N a_\mu-B^{(1)}_\mu)\Phi|^2+\lambda(|\Phi|^2-v^2)^2\right)\nonumber\\
    &\quad -{\im \theta\over 2\pi}\int(f-B^{(2)}). 
    \label{eq:Lagrangian_AbelianHiggs_gauged}
\end{align}
As a result, when we shift $\theta\to \theta+2\pi$, the path-integral weight changes as 
\begin{align}
    \exp\left(-S_{\text{gauged}}\right)&\to \exp\left(-S_{\text{gauged}}+{\im}\int_M (\diff a-B^{(2)})\right)\nonumber\\
    &=\exp\left(-S_{\text{gauged}}\right)\exp\left(-\im \int_{M}B^{(2)}\right). 
\end{align}
By performing the path integral of both sides, we obtain 
\begin{equation}
    Z_{\theta+2\pi}[B]=\exp\left(-\im \int_M B^{(2)}\right)Z_{\theta}[B]
\end{equation}
for the partition function $Z_{\theta}[B]$ with the background gauge field $B=(B^{(2)},B^{(1)})$.

Let us reproduce this relation using the dilute vortex gas picture. 
We can set 
\begin{equation}
    \exp\left(-\im \int_M B^{(2)}\right)=\rme^{-{2\pi \im\over N}m}
\end{equation}
for some integer $m$. The topological charge is no longer quantized to integers, but it takes the fractional values 
\begin{equation}
    Q_{\text{top}}={1\over 2\pi}\int_M (\diff a-B^{(2)})\in -{m\over N}+\mathbb{Z}. 
\end{equation}
We note that the vortex and anti-vortex have the topological charges $\pm 1/N$, and thus there have to be imbalance between the vortex and anti-vortex numbers, $n$ and $\overline{n}$:
\begin{equation}
    n-\overline{n}\in -m+N\mathbb{Z}. 
\end{equation}
The partition function is 
\begin{align}
    Z_\theta[B]&=\sum_{n,\overline{n}\ge 0}{V^{n+\overline{n}}\over n! \overline{n}!} \rme^{-(n+\overline{n})S_v}\rme^{\im (n-\overline{n})\theta/N}\delta_{n-\overline{n}\in -m+N\mathbb{Z}}\nonumber\\
    &=\sum_{k=0}^{N-1}\sum_{n,\overline{n}\ge 0}{V^{n+\overline{n}}\over n! \overline{n}!} \rme^{-(n+\overline{n})S_v}\rme^{\im (n-\overline{n})\theta/N}\rme^{-{2\pi\im\over N}(n-\overline{n}+m)k}\nonumber\\
    &=\sum_{k=0}^{N-1} \rme^{-{2\pi \im\over N}mk}\exp\left[-V\left(-2\rme^{-S_v}\cos{\theta-2\pi k\over N}\right)\right]. 
\end{align}
Under the $2\pi$ shift of $\theta$, $\theta\to \theta+2\pi$, the label $k$ should also be shifted as $k\to k'=k+1$, which reproduces the anomaly:
\begin{align}
    Z_{\theta+2\pi}[B]&=\sum_{k=0}^{N-1}\rme^{-{2\pi \im\over N}mk}\exp\left[-V\left(-2\rme^{-S_v}\cos{\theta-2\pi (k-1)\over N}\right)\right]\nonumber\\
    &=\sum_{k'=0}^{N-1}\rme^{-{2\pi \im\over N}m(k'+1)}\exp\left[-V\left(-2\rme^{-S_v}\cos{\theta-2\pi k'\over N}\right)\right]\nonumber\\
    &=\rme^{-{2\pi \im\over N}m}Z_{\theta}[B]. 
\end{align}
With a massless fermion, the discrete axial transformation can be regarded as the $2\pi$ shift of $\theta$, and we obtain the same anomalous phase.

\section{Another viewpoint of the area law based on center vortex}
\label{app:area_law}

In this section, we give another derivation of the string tension~\eqref{thetadep} for the Wilson loops from a slightly different viewpoint. 
The derivation here is based on the canonical ensemble of center vortex, while the derivation in Sec.~\ref{sec:DCVGA} is based on the grand canonical ensemble. 
Of course, these are equivalent, but it would be useful to have both perspectives.

We consider the situation, where a large Wilson loop $W_{\calR}(C)$ is inside $M_2$, and both $M_2$ and the loop $C$ are sufficiently large. 
We denote the total area (volume) of $M_2$ as $V$, and $\calA$ is the area surrounded by the loop $C$. 
Let us fix the total number of center vortex and anti-vortex in $M_2$ as $\mathcal{N}$. 
%One may wonder why this is possible since the vortex and anti-vortex can be pair created or annihilated.
We point out that our derivation is limited to the regime where the dilute gas description is valid. 
Since $g^2N \ll 1$ at the scale of compactification, the density of vortices is small,  and controlled by $\rme^{-S_\mathrm{I}/N} = \rme^{-{8 \pi^2}/{g^2N}}$. 
%Hence, a weak coupling analysis is justified. 
%so such pair creation or annihilation do not give major contributions. 
What becomes rather subtle in this derivation is the effect of global topology of $M_2$. 
When we decompose $\mathcal{N}=n+\overline{n}$, where $n$ and $\overline{n}$ are the number of vortex and anti-vortex, then we should have $n-\overline{n}\in N\mathbb{Z}$ on closed manifolds. 
% When $N$ is even, this also constrains that $\mathcal{N}$ has to be even. 
Let us circumvent such complications by taking $M_2$ as an open $2$-manifold, such as a disk, then we do no have constraints by the topological charge.

As we have done in Sec.~\ref{sec:DCVGA}, let $n_1$ and $n_2$ be the numbers of vortices inside and outside the loop $C$, and $\overline{n}_1$ and $\overline{n}_2$ be the ones for anti-vortices, then $n_1+n_2+\overline{n}_1+\overline{n_2}=\mathcal{N}$. 
The probability that $\mathcal{N}$ vortices  are distributed into the above  $n_1, n_2, \overline{n}_1, \overline{n}_2$  is 
\begin{align}
P_{\mathcal{N}}(n_1, n_2, \overline{n}_1, \overline{n}_2) =  {{\mathcal{N}!}\over {n_1!\, n_2!\,\overline{n}_1!\, \overline{n}_2! }}\, p_1^{n_1}\, p_2^{n_2}\, \overline{p}_1^{\overline{n}_1}\,  \overline{p}_2^{\overline{n}_2},
\end{align}
where  $p_i$ and $\overline{p}_i$ are determined by the relative areas, 
\begin{align}
p_1= \overline{p}_1= \frac{1}{2} \frac{{\cal A}}{V}, \qquad p_2= \overline{p}_2= \frac{1}{2}\left(1-\frac{{\cal A}}{V}\right), 
\end{align}
which give the probability of  a single chosen configuration to be in one of the four classes. Obviously, this satisfies the normalization,
\begin{align}
\sum_{\substack{n_1, n_2, \overline{n}_1, \overline{n}_2 \\
n_1+ n_2+ \overline{n}_1+\overline{n}_2 =\mathcal{N}}}  P_\mathcal{N}(n_1, n_2, \overline{n}_1, \overline{n}_2) = (p_1 + p_2 +  \overline{p}_1 + \overline{p}_2)^{\mathcal{N}}= 1.  
\end{align}
Inside the loop $C$, the $n_1+ \overline{n}_1$ vortex  and anti-vortex  configurations contribute to  the Wilson loop   $W_{\cal R} (C)$   as 
 $ \rme^{\im \frac{2 \pi |\calR|}{N} (n_1-\overline{n}_1)} $, and the configurations outside do not contribute.   
The average of the Wilson loop for fixed $\mathcal{N}$ takes the following form,
\begin{align}
\langle W_{\cal R}(C) \rangle & = \sum_{\substack{n_1, n_2, \overline{n}_1, \overline{n}_2 \\
n_1+ n_2+ \overline{n}_1+\overline{n}_2 =\mathcal{N}}}  P_{\mathcal{N}}(n_1, n_2, \overline{n}_1, \overline{n}_2) \, \rme^{\im \frac{ \theta+  2 \pi |\calR|}{N}  (n_1-\overline{n}_1)} 
\rme^{\im \frac{ \theta}{N}  (n_2-\overline{n}_2)} 
\cr 
&=  (p_1 \, \rme^{\im \frac{ \theta+  2 \pi |\calR|}{N}  } + \overline{p}_1\, \rme^{-\im \frac{  \theta+ 2 \pi |\calR|}{N}  }
+ p_2 \,\rme^{\im \frac{ \theta}{N}  } + \overline{p}_2\, \rme^{-\im \frac{ \theta}{N}  })^\mathcal{N}
\cr
& =  \left(  1 +  \frac{{\cal A}}{V} \Big(\cos  \frac{ \theta+ 2 \pi |\calR|}{N}    -\cos  \frac{ \theta}{N} \Big) \right)^\mathcal{N}. 
\label{fixedN}
\end{align}
We take the infinite volume limit $V\to \infty$, while keeping the density of vortices fixed as 
\begin{align}
\rho =   \frac{\mathcal{N}}{ V} \sim L_s^{-2}  \rme^{-S_\mathrm{I}/N}   \sim  \Lambda^2  (\Lambda L_s)^{\frac{5}{3}} . 
\label{density}
\end{align}
We can rewrite \eqref{fixedN} and take the $\mathcal{N}\to \infty$ limit as
\begin{align}
\langle W_{\cal R}(C) \rangle  
&=  \left(  1 +  \frac{ \rho {\cal A}}{\mathcal{N}} 
\Big(\cos  \frac{ \theta+ 2 \pi |\calR|}{N}    -\cos  \frac{ \theta}{N} \Big) \right)^{\mathcal{N}}
\cr
&\to \exp\left[\rho {\cal A} \Big(\cos  \frac{ \theta+ 2 \pi |\calR|}{N}    -\cos  \frac{ \theta}{N} \Big) \right]. 
\end{align}
This gives the area law of the Wilson loop with nonzero $N$-alities, and reproduces the formula~\eqref{thetadep}. 
In this derivation, since the effect of global topology of $M_2$ is eliminated by making $M_2$ an open $2$-manifold, the appearance of multi-branch structure is hidden. 
This comes from the special feature of $2$d field theories with $1$-form symmetry, as the Hilbert space can be decomposed by the charge of $1$-form symmetry~\cite{Pantev:2005rh, Hellerman:2006zs, Hellerman:2010fv,Tanizaki:2019rbk, Komargodski:2020mxz, Cherman:2020cvw, Honda:2021ovk}.
We note that the $4$d YM theory on $\mathbb{R}^4$ does not have such decomposition property. 
This is an emergent feature for small $T^2$ compactification, which is valid as long as the dynamics that involves the compactified direction is negligible.

\bibliographystyle{utphys}
\bibliography{./QFT.bib,./refs.bib}
\end{document}